\documentclass{aa}  
\usepackage{graphicx}
\usepackage{float}
\usepackage{rotating}
\usepackage[section]{placeins}
\usepackage{amsmath}
\usepackage{mathastext}
\newcommand{\angstrom}{\mbox{\normalfont\AA}}
\usepackage{xcolor}
\usepackage{url}
\usepackage{csvsimple}
\usepackage{pgfplotstable}
\usepackage{longtable}
\usepackage[varg]{txfonts}
\usepackage[version=4]{mhchem}
\usepackage{isotope}
\usepackage{units} % use either units or siunitx
\usepackage{siunitx}
\usepackage{comment}
\usepackage{threeparttablex}
\usepackage{geometry}

\usepackage{lscape}
\usepackage{booktabs}
\usepackage{array}
\pgfplotsset{compat=1.17}
\usepackage{verbatim}
\usepackage{multicol} % required for letting additional caption span over all cols
\usepackage{multirow}
\usepackage{colortbl}

\makeatletter
\renewcommand*\aa@pageof{, page \thepage{} of \pageref*{LastPage}}

\makeatletter
\newcommand\footnoteref[1]{\protected@xdef\@thefnmark{\ref{#1}}\@footnotemark}
\makeatother

\usepackage[breaklinks=true, colorlinks=true, linkcolor=blue, citecolor=blue, urlcolor=blue]{hyperref} %% to avoid \citeads line fills

\usepackage{natbib,twoopt} % turn in-text citations into clickers
\bibpunct{(}{)}{;}{a}{}{,}  %% natbib format for A&A and ApJ
\begin{document}

   \title{Unveiling the chemical fingerprint of phosphorus-rich stars}
   \subtitle{I. In the infrared region of APOGEE-2}
   
   \author{Maren Brauner\inst{1}\fnmsep\inst{2}
          \and
          Thomas Masseron\inst{1}\fnmsep\inst{2}
          \and
          D. A. García-Hernández\inst{1}\fnmsep\inst{2} 
          \and
          Marco Pignatari\inst{3}\fnmsep\inst{4}\fnmsep\inst{5}\fnmsep\inst{6}\fnmsep\inst{7}
          \and
          Kate A. Womack\inst{5}
          \and \\
          Maria Lugaro\inst{3}\fnmsep\inst{4}\fnmsep\inst{8}\fnmsep\inst{9}
          \and
          Christian R. Hayes\inst{10} % ORCID 0000-0003-2969-2445
          }

   \institute{Instituto de Astrofísica de Canarias, C/Via Láctea s/n, E-38205 La Laguna,      Tenerife, Spain\\
              \email{maren.brauner@iac.es}
        \and
             Departamento de Astrofísica, Universidad de La Laguna, E-38206 La Laguna, Tenerife, Spain
        \and
            Konkoly Observatory, Research Centre for Astronomy and Earth Sciences, Eötvös Loránd Research Network (ELKH), H-1121 Budapest, Konkoly Thege M. út 15-17, Hungary
        \and
            CSFK, MTA Centre of Excellence, Budapest, Konkoly Thege Miklós út 15-17, H-1121, Hungary
        \and
            E. A. Milne Centre for Astrophysics, University of Hull, Hull, HU6 7RX, UK
        \and
            Joint Institute for Nuclear Astrophysics - Center for the Evolution of the Elements
        \and
            The NuGrid Collaboration, http://www.nugridstars.org
        \and
            ELTE Eötvös Loránd University, Institute of Physics, Budapest 1117, Pázmány Péter sétány 1/A, Hungary
        \and
            School of Physics and Astronomy, Monash University, VIC 3800, Australia
        \and
            NRC Herzberg Astronomy and Astrophysics Research Centre, 5071 West Saanich Road, Victoria, B.C., Canada, V9E 2E7
            \\
             %\thanks{The university of heaven temporarily does not
             %        accept e-mails}
             }

   \date{Received; accepted }

% \abstract{}{}{}{}{} 
% 5 {} token are mandatory
 
  \abstract
  % context heading (optional), leave it empty if necessary  
   {The origin of phosphorus, one of the essential elements for life on Earth, is currently unknown. Prevalent models of Galactic chemical evolution (GCE) systematically underestimate the amount of P compared to observations, especially at low metallicities. The recently discovered P-rich ([P/Fe]$~\gtrsim \unit[1.2]{dex}$) and metal-poor ([Fe/H]$~\simeq \unit[-1.0]{dex}$) giants further challenge the GCE models, calling current theories on stellar nucleosynthesis into question.
   }
   {Since the observed low-mass giants are not expected to produce their high P contents themselves, our primary goal is to find clues on their progenitor or polluter. By increasing the number of known P-rich stars, we aim to narrow down a statistically reliable chemical abundance pattern that defines these peculiar stars. In this way, we place more robust constraints on the nucleosynthetic mechanism that causes the unusually high P abundances. In the long term, identifying the progenitor of the P-rich stars may contribute to the search for the source of P in our Galaxy.}
   {We performed a detailed chemical abundance analysis based on the high-resolution near-infrared (H band) spectra from the latest data release (DR17) of the APOGEE-2 survey. Employing the BACCHUS code, we measured the abundances of 13 elements in the inspected sample, which is mainly composed of a recent collection of Si-enhanced giants. We also analyzed the orbital motions and compared the abundance results to possible nucleosynthetic formation scenarios, and also to detailed GCE models. These models were produced with the OMEGA+ chemical evolution code, using four different massive star yield sets to investigate different scenarios for massive star evolution.
   }
   {We enlarged the sample of confirmed P-rich stars from 16 to a group of 78 giants, which represents the largest sample of P-rich stars to date. The sample includes the first detection of a P-rich star in a Galactic globular cluster. Significant enhancements in O, Al, Si, and Ce, as well as systematic correlations among the studied elements, unveil the unique chemical fingerprint of the P-rich stars. In contrast, the high [Mg/Fe] and [(C+N)/Fe] found in some of the P-rich stars with respect to P-normal stars is not confirmed over the full sample because of the current uncertainties. Strikingly, the strong overabundance in the $\alpha$-element Si is accompanied by normal Ca and S abundances. This is at odds with current stellar nucleosynthesis models of massive stars. Our analysis of the orbital motion showed that the P-rich stars do not belong to a locally specific population in the Galaxy. In addition, we confirm that the majority of the sample stars are not part of binary systems.
   }
   {}
   

   \maketitle
%
%________________________________________________________________
%
\section{Introduction\label{Introduction}}
\noindent

Phosphorus (P) is a crucial component in the building blocks of life on Earth \citep[see, e.g.,][]{gulick1955,butusov13}. However, the Galactic origin of this critical element is only sparsely explored. To trace the evolution of P over time, observations of long-lived stars covering a broad range of metallicities are required. The difficulty of obtaining observations like this is that the \ion{P}{i} absorption features, located in the near-ultraviolet (NUV) and near-infrared (NIR) J- and H-band spectral regions, are intrinsically weak and consequently hard to detect \citep[e.g.,][]{caffau11,roederer14}. Some high-resolution observations in the NIR, intended for determining the P content in stellar atmospheres, have been carried out by \citet{caffau11} using the single-object spectrographs CRIRES at VLT\footnote{CRyogenic high-resolution InfraRed Echelle Spectrograph at the Very Large Telescope} and by \citet{caffau16,caffau19} using GIANO at the TNG\footnote{Telescopio Nazionale Galileo}, spanning a total metallicity range of $\unit[-1.0]{dex}$ < [Fe/H] < $\unit[0.3]{dex}$. Complementary observations in the NUV region enabled \citet{roederer14} and \citet{jacobson14} to derive the P abundances of 13 late-type stars down to [Fe/H]$~\approx \unit[-3.3]{dex}$. The most recent observation by \citet{nandakumar22}, who used the NIR spectrograph IGRINS\footnote{Immersion GRating INfrared Spectrometer at Gemini South}, provided P measurements of 38 K-type giants in the metallicity range $\unit[-1.2]{dex}$ < [Fe/H] < $\unit[0.4]{dex}$. Detailed reviews of the few estimations of P in stars can be found in \citet{roederer14,maas19} and \citet{nandakumar22}. \\
In stellar nucleosynthesis, the odd-Z element P, which has only one stable isotope ($\isotope[31]{P}$),
is mostly produced via neutron capture on isotopes of Si during the convective C-burning shell 
in massive stars \citep[e.g.,][]{woosley:02,pignatari:16}. The produced P is ejected into the interstellar medium (ISM) by the following core-collapse supernovae (CCSNe).
The nucleosynthesis triggered by the CCSN explosion itself is not expected to contribute significantly to the final P yields \citep[e.g.,][]{woosley:95}. In Type Ia supernovae (SNeIa), the thermonuclear explosion of a carbon-oxygen white dwarf (WD), the production of P is also irrelevant \citep{leung18}. Asymptotic giant branch (AGB) stars are likewise not expected to produce significant amounts of P \citep{Karakas16}.\\
By taking into account the nucleosynthetic yields from different stellar sources, Galactic chemical evolution (GCE) models replicate the evolution of elements in the Galaxy as a function of metallicity and time \citep[see][and references therein]{matteucci:21}. In particular, the understanding of the chemical evolution of P in the Galaxy has been challenging since the earliest GCE simulations.
In general, while the [P/Fe]\footnote{\label{Bracket}$\left[\frac{X}{Fe}\right]=\log\left(\frac{N_X}{N_H}\right)_{*}-\log\left(\frac{N_X}{N_H}\right)_{\odot}-\left[\frac{Fe}{H}\right]$\\ with~$\left[\frac{Fe}{H}\right]=\log\left(\frac{N_{Fe}}{N_H}\right)_{*}-\log\left(\frac{N_{Fe}}{N_H}\right)_{\odot}$ and $N_X$ the number of atoms of element X.} abundance ratio can be roughly reproduced for stars at solar metallicity \citep{cescutti12}, it becomes difficult to produce enough P to match observations in metal-poor stars because of the secondary-like production of this element in massive stars \citep[see, e.g.,][]{goswami:00}. Several authors \citep[e.g.,][]{caffau11,roederer14,jacobson14,maas19} reported a major discrepancy between models and observations, concluding that GCE models are not able to quantitatively reproduce the behavior of [P/Fe] at subsolar metallicities. In particular, \citet{cescutti12} demonstrated that the P yields from \citet{woosley:95} and \citet{kobayashi06} have to be multiplied by a factor of approximately 3 to match the observations from \citet{caffau11}. The recent observation by \citet{nandakumar22} confirmed the conclusions made by \citet{cescutti12}. \citet{nandakumar22} showed that the behavior of P is more similar to that of the $\alpha$-elements produced in CCSNe, such as Mg, Si and S, than to other odd-Z elements. This finding supports the hypothesis from \citet{clayton03}, who proposed that in contrast to \citet{woosley:95}, CCSNe indeed contribute to the P production. More recently, \cite{ritter:18} showed that a merger between the convective C shell and the convective O shell in massive stars may boost the production of P in the CCSN progenitor and in the final CCSN ejecta. The following impact on the GCE is uncertain, and it depends, among other things, on the frequency and the nature of the shell mergers. However, \cite{ritter:18} showed that they could potentially explain the typical [P/Fe] ratio in stars in the MW and even reach a much larger P enrichment. In summary, the exact production sites of P is still unclear, causing a mismatch between GCE models and observations. \\  
Through the advent of multi-object spectroscopic surveys, such as the Apache Point Observatory Galactic Evolution Experiment (APOGEE; \citealp[e.g.,][]{majewski17APOGEE1}), it became possible to perform an abundance census in the infrared on a vast number of stars. The APOGEE-2 data release 14 (DR14; \citealp{abolfathi18APOGEEDR14}) provided NIR H-band spectra of more than 263\,000 stars in the Milky Way and allowed the discovery of 16 metal-poor ([Fe/H]$~\simeq \unit[-1.0]{dex}$) giants with high and unexplained amounts of P ([P/Fe]$~\ga \unit[1.2]{dex}$) \citep{masseron20a,masseron20b}. The GCE models, which already fail to reproduce the usual Galactic P abundances, are even less capable of accounting for the overabundances of these P-rich stars. \\
According to \citet{masseron20a}, the P-rich stars also present high amounts of O, Mg, Al, Si, and Ce. Because they are low-mass giants, it is not expected that the P-rich stars produced the enhanced elements themselves \citep{masseron20a}. An extragalactic origin of these stars was also discarded by \citet{masseron20a}, and therefore, the search for the source of P is now focused on finding the progenitor that polluted the ISM out of which the now observed P-rich stars were born. By comparing their chemical abundance pattern with current theoretical models of stellar nucleosynthesis, \citet{masseron20a,masseron20b} were able to rule out a number of possible formation scenarios, among others, novae and super-AGB stars. \citet{masseron20a} also discussed the aforementioned contribution of stellar rotation and/or nucleosynthesis in convective-reactive regions in massive stars such as C-O shell mergers \citep{ritter:18} as possible alternatives to explain the peculiar chemistry of P-rich stars. \\
In the present work, we built an enlarged sample of P-rich star candidates and performed a line-by-line abundance analysis on this new sample using the Brussels Automatic Code for Characterizing High accUracy Spectra (BACCHUS). The analysis includes 13 elements and is based on the NIR spectra from the latest APOGEE-2 data release (DR17). We found a total of 78 P-rich stars, the highest number of known P-rich stars to date, allowing us to report a statistically reliable chemical fingerprint of these peculiar stars. This new sample can guide us in the search for their progenitor by placing more robust constraints on the nucleosynthetic channel that produced the unusual abundance pattern. Furthermore, we studied the orbital motion of the P-rich stars in the Galaxy and discuss nucleosynthetic scenarios that have not been considered until now or have been suggested recently, such as nonthermal nucleosynthesis. Finally, we present four updated GCE models aiming to improve our understanding of the P production in the Milky Way. 

\section{Spectral data and sample\label{DataSample}}
\subsection{ APOGEE-2 survey\label{APOGEE}}
We used the spectral data provided by the APOGEE survey \citep[e.g.,][]{majewski17APOGEE1} and its extension APOGEE-2 \citep{blanton17SDSS}, which are both part of the Sloan Digital Sky Survey (SDSS-IV). The latter combines observations from the Northern and Southern hemispheres, using the $\unit[2.5]{m}$ Sloan Foundation Telescope at the Apache Point Observatory \citep{gunn06Telescope1} and the $\unit[2.5]{m}$ Irénée du Pont Telescope at Las Campanas Observatory \citep{bowen73Telescope2}, respectively, alongside the corresponding APOGEE twin spectrographs \citep{wilson19APOSpec}. \\
The latest data release (DR17; \citealp{abdurrouf22APOGEEDR17}) delivered high-resolution (R~$\approx 22\,500$) and spectra with a high signal-to-noise ratio S/N (>~$\unit[100]{/pixel}$) in the NIR H-band ($\lambda=\SI{1.51}{\micro m}-\SI{1.70}{\micro m}$) for more than $650\,000$ targets in the Galaxy. The reduction procedure of the DR17 data is mostly the same as described in \citet{nidever15} and \citet{holtzman15}, including various updates made in previous data releases \citep[see][and references therein]{abdurrouf22APOGEEDR17}. \\
To automatically obtain stellar parameters and chemical abundances, the APOGEE Stellar Parameters and Chemical Abundance Pipeline (ASPCAP; \citealp{garciaperez16ASPCAP}) is used. This tool is based on the optimization code FERRE \citep{allendeprieto06FERRE} and MARCS model atmospheres \citep{gustafsson08,jonsson20}. In DR17, as an update to previous data releases, ASPCAP employs the synthetic spectra generated by the code Synspec \citep{hubeny11}. This change was made to enable the use of new NLTE population calculations for the elements Na, Mg, K, and Ca performed by \citet{osorio20}. Another update was made by adopting an improved line list with respect to DR16 \citep{smith21}. \\
In short, the ASPCAP workflow is as follows: First, the MARCS atmospheres and line lists are combined to generate the grid of synthetic spectra \citep[see, e.g.,][]{zamora15} using the Synspec code. The grid is then used to identify the best-matching model compared to the observed spectrum. Finally, stellar parameters and chemical abundances are derived from the best match.

\subsection{Sample selection\label{Sample}}
\citet{masseron20a,masseron20b} showed that their 16 P-rich stars are also significantly enriched in Si and Al, suggesting that they might be part of a larger population of Si- or Al-rich stars \citep[e.g.,][]{fernandeztrincado19,schiavon17}. Silicon displays many strong H-band absorption lines, which are accessible in a wide metallicity range, and therefore, it is one of the most precisely determined elements from APOGEE-2 spectra \citep{jonsson20}. Thus, the sample of Galactic Si-rich ([Si/Fe]~$\ga \unit[+0.5]{dex}$) giants previously identified by \citet{fernandeztrincado19,fernandeztrincado20} served as a starting point for us to build a sample of potentially P-rich stars. After dropping a suspected member of a binary system included in the work of \citet{fernandeztrincado20} (discussed in Sect. \ref{binary}), we ended with a group of 53 Si-rich stars that constitute the first subsample of P-stars candidates. To our analysis, we also added the 16 P-rich and metal-poor ($-1.26$ < [Fe/H] < $-0.75$) giants reported by \citet{masseron20a,masseron20b} for homogeneity and consistency reasons; that is, to update their chemical abundances using the latest APOGEE-2 data release (DR17).
Another subsample of 18 stars was obtained by applying the merely arbitrary conditions on the P abundance ratio [P/Fe]~>$~0.8$ and on the overall metallicity [M/H]~<$~ -0.6$ to the targets included in the Value Added Catalog (VAC) called BACCHUS Analysis of Weak Lines in APOGEE spectra (BAWLAS\footnote{In this work, an earlier version of the VAC was used. The final version is available at \url{https://www.sdss.org/dr17/data_access/value-added-catalogs/}}; \citealp{hayes22}). This VAC explored weak and blended elemental species, such as Na, P, and S,  of approximately 120\,000 giants with high S/N (>$\unit[150]{px^{-1}}$) spectra in the APOGEE-2 DR17. The calculation of their chemical abundances was performed by means of the same code as employed in the present work (BACCHUS; \citealp{masseron2016}), which is presented in Sect. \ref{BACCHUS}. By merging these three subgroups, a total of 87 P-rich star candidates with overall metallicities $-1.5 <$ [M/H] $ < -0.6$ were found to be the input of our chemical abundance analysis. \\
To clearly detect (and visualize) possible chemical overabundances, a sample of field stars with similar stellar parameters as the P-star candidates was also selected to constitute a reference group. The detailed selection procedure for this reference group (or
background sample) is described in Appendix \ref{BackgroundSelection}. Finally, the reference group of field stars comprises about 5500 stars.

\section{Abundance calculations\label{Calculation}}
\subsection{BACCHUS\label{BACCHUS} code}
The elemental abundances of interest were derived by means of the latest version ($\sim$v67) of BACCHUS \citep{masseron2016}. As its basic functionality, BACCHUS computes synthetic spectra for a range of abundances by combining MARCS model atmosphere grids especially built for APOGEE \citep{gustafsson08, jonsson20} with the 1D-LTE turbospectrum radiative transfer code v19.1.3 \citep{plez12}, incorporating a spherical radiative transfer and the APOGEE-2 DR17 line list version 20180901t20\footnote{\url{https://data.sdss.org/sas/dr17/apogee/spectro/speclib/linelists/turbospec/}} \citep{smith21,shetrone15}. Molecular line lists were taken from \citet{li15CO} for CO, from \citet{sneden14CN} for CN, and from \citet{brooke16OH} for OH. The generated synthetic spectra were then compared to those observed in each spectral line, allowing the measurement of abundance values on a line-by-line basis. \\
The input data processed by BACCHUS consist of the APOGEE-2 DR17 spectra downloaded from the SDSS-IV Science Archive Server\footnote{\url{https://data.sdss.org/sas/dr17/apogee/spectro/redux/dr17/stars/}} together with the calibrated parameters and abundances derived by the ASPCAP pipeline \citep{garciaperez16ASPCAP}. In case of duplicate entries in the ASPCAP data table, which are due to multiple observations of the same target in different fields, the entry corresponding to the highest S/N value was used. For ten stars, multiple spectra are available, but only five of them had more than one spectrum without flag (see Appendix \ref{BackgroundSelection} for a description of the flags). We do not expect a significant improvement of the measurement precision by combining them because all of these spectra have an S/N > 150. Missing entries in parameter values of the sample, for example, in T$_{\text{eff}}$ or log g, were replaced by the values provided by \citet{fernandeztrincado20}, which correspond to DR16. This had to be done for five stars. The absence of parameter values in DR17 can be ascribed to warnings in the ASPCAP quality flags. Therefore, we checked the spectra of the five affected stars and found that the warnings did not influence the abundance analysis. Nevertheless, we report the source of the parameters (DR16 or DR17) in our final abundance table. \\
To obtain the synthetic spectra, the MARCS model atmospheres grid in spherical geometry was interpolated at the calibrated values of T$_{\text{eff}}$, log g, and [M/H] from ASPCAP and the microturbulence velocity ($\mu_t$) from \citet{masseron19}, where $\mu_t$ is described as a function of log g. For the synthetic spectra, a Gaussian line profile was adopted to account for the line broadening through instrumental resolution, macroturbulence, and rotation. The corresponding convolution parameter (CONVOL) was determined for each spectrum by means of the Si lines because they are reliable in strength and clearness over the whole [M/H] and T$_{\text{eff}}$ range considered in this work. \\
From the synthetic spectra, the code identifies the continuum points to use for the normalization of the observed spectra. Additionally, a window of relevant pixels for the abundance determination is selected in each line. \\
In its latest version, BACCHUS includes five different methods that compare the observed spectrum with the synthetic spectra. Each method has its own set of quality flags in the form of integers, indicating whether the fit is of good quality (flag = 1) or if a fit is considered problematic (flag $\neq$ 1). The different methods are more or less beneficial depending on the specific line case \citep{hayes22}. For example, two of the five methods (called wln and int) are based on individual pixels, exactly at or around the line core, and are therefore suitable for examining heavily blended lines. The problem with using this type of methods is their sensitivity to adverse effects on the level of individual pixels. Instead, the remaining three methods (called eqw, syn, and chi2 or $\chi^2$) use multiple pixels of a line, including the wings. As a consequence, these methods are less sensitive to effects occurring in individual pixels, but may be affected by poorly fitted blends at the wings. One of the methods that uses multiple pixels, the $\chi^2$-method, minimizes the squared differences between synthetic and observed spectra. The choice of minimizing the squared differences adds weight to the core of the line, making the $\chi^2$-method less sensitive to blends at the wings compared to the eqw and syn methods, which apply a rather flat weighting over the line. We therefore chose the $\chi^2$-method to be the default method for our measurements because it has the best balance between strengths and weaknesses for our purpose and is therefore the most robust approach, especially in combination with a visual inspection intended to trace poorly fitted blends.

\subsection{Line selection and processing\label{LineSelecProcessing}}

\begin{figure}
  \resizebox{\hsize}{!}{\includegraphics{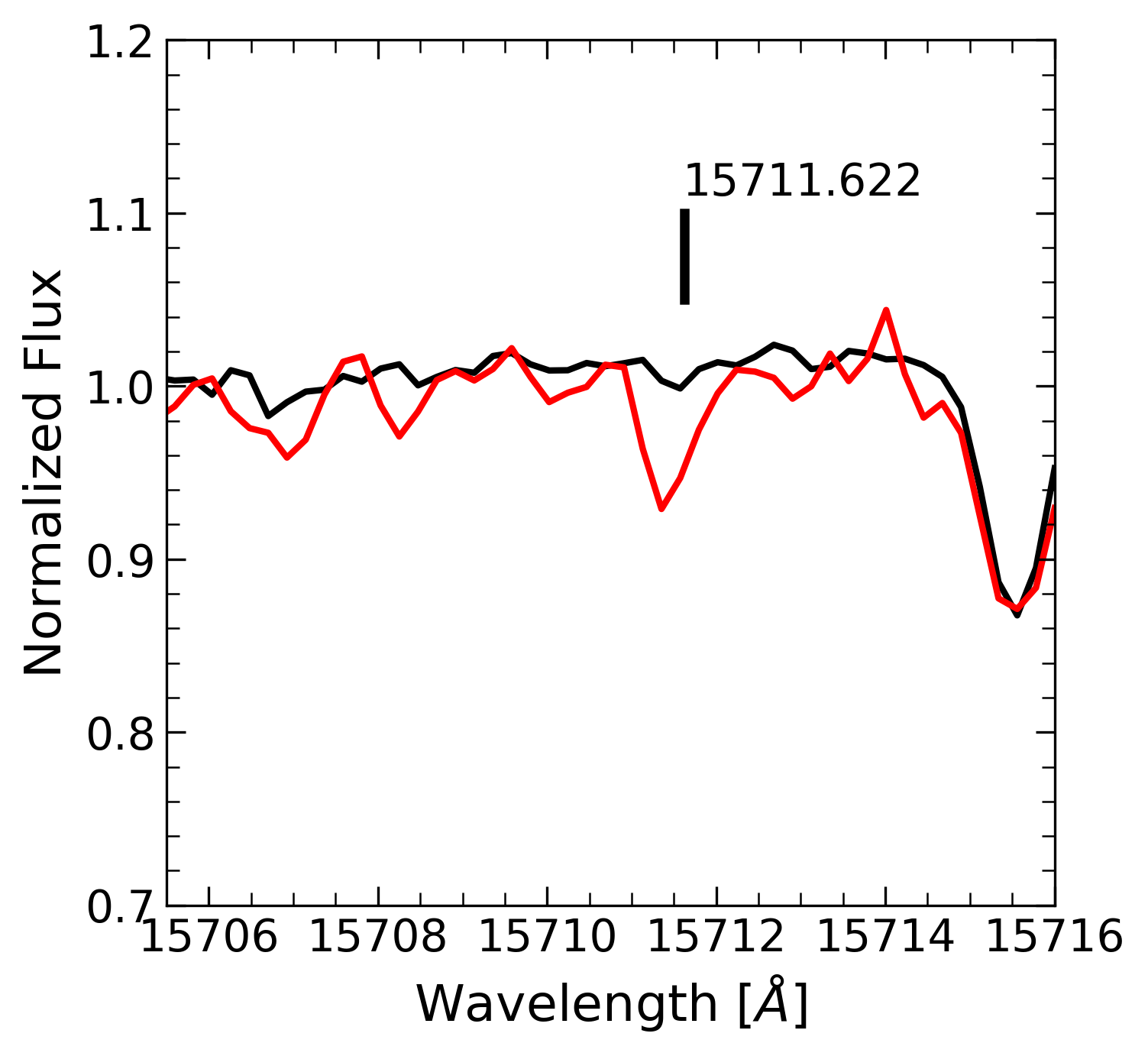}}
  \resizebox{\hsize}{!}{\includegraphics{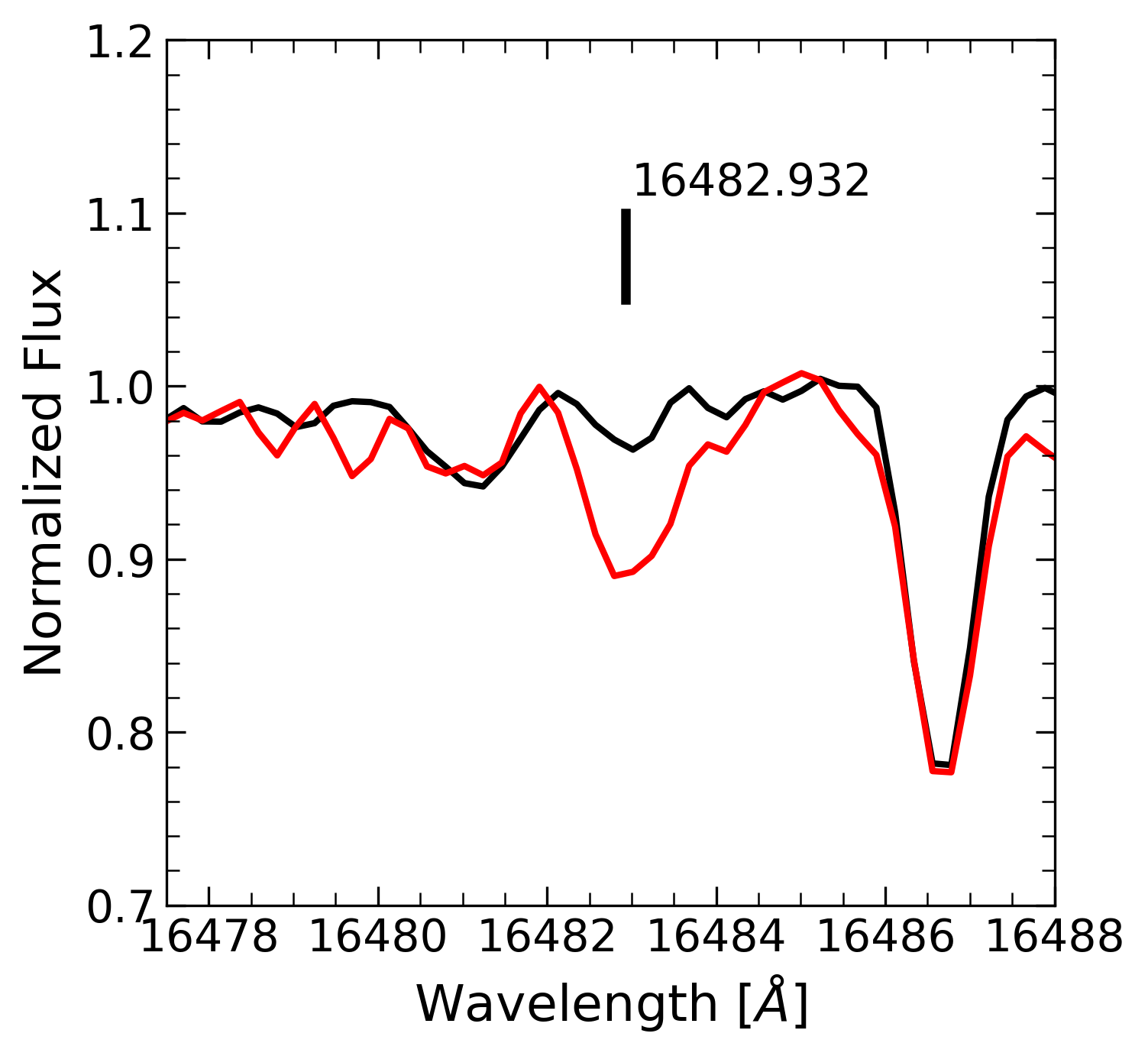}} 
  \caption{Spectra around the \ion{P}{i} lines at $\unit[15711.622]{\angstrom}$ (top panel) and $\unit[16482.932]{\angstrom}$ (bottom panel) of a P-rich (red line) and a P-normal (black line) star with similar stellar parameters: 2M18280709-3900094 (P-rich) with T$_{\text{eff}}$ = $\unit[4574.4]{K}$, log g = 1.77, [M/H]=$\unit[-0.95]{dex}$, and [P/Fe]=$\unit[1.42]{dex}$, and 2M16353092+2651591 (P-normal) with T$_{\text{eff}}$ = $\unit[4502.0]{K}$, log g = 1.74, [M/H]=$\unit[-0.92]{dex}$, and [P/Fe]<$\unit[0.48]{dex}$.}
  \label{fig:Plines}
\end{figure}

We started the BACCHUS iteration by fitting C, O, and N, aiming to properly subtract the corresponding molecular features because they mainly cause the blends in the lines of other species of interest. We then continued the iteration by determining the abundances of the remaining elements Na, Mg, Al, Si, S, Fe, Ce, Nd, P, and Ca. The wavelengths of the lines used in the fitting procedure are listed in Table \ref{tab:LL}. For each element, the BACCHUS code iterated automatically until it reached convergence. To accelerate this procedure, the ASPCAP abundances were used as the initial guess. When no abundance was provided by ASPCAP, the initial guess was fixed to a common abundance value of the sample stars. For self-consistency between the abundances, we carried out a total of three iterations over all the elements. \\
The next step consisted of an examination of the fits, where 
we required that none of the methods should be quality flagged for that line to be further processed. A visual inspection of the fitting output was also performed, rejecting suspicious fits that passed the intrinsic quality flagging of BACCHUS. In this procedure, a line was rejected when the abundance values derived by the different methods disagreed significantly or when the abundance of a certain line differed notably from the values obtained in the other lines of the element. We also paid special attention to possible blends, for example, in the case of P, by observing the quality of the fit of the CO line near $\unit[15724.00]{\angstrom}$. Since CO features may affect the \ion{P}{i} line at $\unit[16482.93]{\angstrom}$, a correctly modeled CO line near $\unit[15724.00]{\angstrom}$ indicates that the \ion{P}{i} is deblended from this feature. Furthermore, we visually monitored the potential blend in the \ion{P}{i} $\unit[15711.62]{\angstrom}$ line by a residual telluric feature. Figure \ref{fig:Plines} displays the spectra of a P-rich star compared to a P-normal star around the aforementioned lines of \ion{P}{i} that we used for the abundance determination. \\
After completing the quality selection, we proceeded with the post-processing of the line-by-line abundance measurements, given for each line in the form of $\epsilon(X)=\log\left(\frac{N_X}{N_H}\right)$ when extracted from the $\chi^2$-method. These values were converted into bracket notation\footnoteref{Bracket} using the solar zeropoints from \citet{Grevesse2007}. In the case of P, the value from \citet{Grevesse2007} ($\unit[5.36\pm 0.04]{dex}$) is slightly lower than the solar value from \citet{Asplund21} ($\unit[5.41\pm 0.03]{dex}$). The final abundance was then obtained by averaging the abundance over all selected lines. \\
To determine the abundances of the approximately 5500 reference group stars, an automated run of three iterations in the same form as for the sample was performed on a supercomputer. Since a visual inspection is not viable in this approach, the employed line list was updated (see Table \ref{tab:LL}), excluding the lines that were repeatedly problematic during the visual inspection of the sample fits. In doing this, we aimed to minimize the contribution of poorly fitted lines to the results (see Appendix \ref{LPvalidation} for the validation). \\
We complemented the results with an estimation of upper limits in cases where no valid measurement was obtained. For the sample stars, we estimated an upper limit by manually generating a set of synthetic spectra and visually choosing the best match for each line. The final upper limit for the element in question was then chosen to be the highest value out of all estimates in the lines. For the reference group, we applied the upper limit relations from \citet{hayes22}. These linear relations depend on T$_{eff}$ and act as cuts for distinguishing upper limits from real measurements. Unlike \citet{hayes22}, we applied the relations only on the usually most reliable line of each element to identify the upper limit. Because of the high quality of the APOGEE-2 spectra, the threshold t defined by \citet{hayes22} up to which an abundance should be detectable was imposed to be t = 1\% for all the elements. In other words, if the average $\chi^2$ abundance value was lower than the corresponding relation of the best line with t = 1\% \citep[see Table 7 of][]{hayes22}, the value was considered an upper limit and was not reported as a measurement.

\FloatBarrier
\subsection{Uncertainty estimation\label{Errors}}

\begin{table}
    \centering
    \caption{Parameters of the two stars used for the uncertainty estimation.}
    \begin{tabular}{l c c} \hline\hline
         & 2M16441013 & 2M18070782 \\
        \hline
        T$_{eff}$ [K] & 4727.5 & 4091.9 \\
        $\log g$ [dex] & 1.96 & 0.74 \\ 
        $\text{[M/H]}$ [dex] & –0.91 & –1.10\\
        $\mu_t$ [km/s] & 1.39 & 1.93 \\
        $\text{CONVOL}$ [km/s] & 14.23 &  13.78 \\ \hline
    \end{tabular}
    \label{tab:ParametersUncert}
\end{table}

\begin{table*}[htb!]
\tiny
\caption{Final abundance ratios of the studied sample (87 stars).}  
\label{table:AbusShort}      
\centering
\begin{tabular}{c c c c c c c c c c c c c c}        
\hline\hline                 
APOGEE-ID & [Fe/H] & [C/Fe] & [N/Fe] & [O/Fe] & [Na/Fe] & [Mg/Fe] & [Al/Fe] & [Si/Fe] & [P/Fe] & [S/Fe] & [Ca/Fe] & [Ce/Fe] & [Nd/Fe] \\    
\hline                        
   2M00044180 & -1.03 & 0.15   & 0.56   & 0.86 & 0.66+ & 0.54 & 0.80 & 0.58 & 1.17* &  0.22  & 0.34  & 0.61   & 0.55 \\      
   2M02175837 & -0.92 & -0.01  & 0.66   & 1.09 & 0.11* & 0.70 & 1.27 & 0.79 & 1.18* & 0.12* & 0.44  & 0.76   & 0.74*  \\
   2M04371861 & -0.70 & 0.16   & -0.47  & 0.54 & 0.46  & 0.40 & 0.53 & 0.29 & 0.77: & 0.21  & 0.28  & 0.22*: & 1.05+: \\
   2M04463289 & -0.89 & -0.12: & 0.59   & 0.90 & 0.16+ & 0.55 & 0.99 & 0.80 & 1.30 & 0.17  & 0.35  & 0.64   & 0.78 \\
   2M06263707 & -0.70 & 0.02   & 0.16   & 0.64 & 0.03+ & 0.44 & 0.52 & 0.30 & 0.87* & 0.03: & 0.25  & 0.12   & 0.56+  \\
   \ldots & \ldots & \ldots & \ldots & \ldots & \ldots & \ldots & \ldots & \ldots & \ldots & \ldots & \ldots & \ldots & \ldots \\
\hline
\end{tabular}
\tablefoot{Plus signs indicate upper limits, asterisks denote values that are based on one line only, and colons highlight results that have to be handled with caution because no clear abundance determination was possible. The full version of this table including stellar parameters and standard deviations is available in machine-readable form at the CDS.}
\end{table*}

We explored the influence of errors associated with the most important parameters used in the abundance derivation, that is, the systematic errors $\sigma_{param,[X/Fe]}$. The parameters and error intervals we considered, following the analysis of \citet{fernandeztrincado20}, are T$_{\text{eff}}\pm \unit[100]{K}$, $\log g \pm \unit[0.3]{dex}$ and $\mu_t \pm \unit[0.05]{km/s}$. For the overall metallicity, we adopted the error  $\text{[M/H]}\pm \unit[0.15]{dex}$, corresponding to the highest difference between literature and ASPCAP-derived [M/H] from \citet{garciaperez16ASPCAP}. In addition to these stellar parameters, we also varied the convolution parameter CONVOL. As mentioned before, this parameter accounts for several line-broadening effects and is determined based on the Si lines before iterating over the elements. When CONVOL is determined, BACCHUS provides an estimation of the corresponding error based on the root-mean-square error of the individual Si lines convolution that is used to refine this parameter. We used a typical value of this estimation as error interval, that is, $\text{CONVOL}\pm 1.5$. To perform the uncertainty analysis, we selected two typical stars from the sample, 2M16441013-1850478 and 2M18070782-1517393, with the stellar parameters given in Table \ref{tab:ParametersUncert}. We fixed their chemical composition to the $\epsilon(X)$ values obtained from the final iteration. Thereafter, we changed one of the parameters of interest and derived the $\epsilon(X)$ value of each element again, but without updating the chemical composition with these new $\epsilon(X)$. In this manner, we achieved an independent error analysis of all the elements. The abundances obtained in the + and $-$ run (e.g., T$_{\text{eff}}+\unit[100]{K}$ and T$_{\text{eff}}-\unit[100]{K}$, respectively) were then compared to the original abundance values of the stars, and the highest difference was stored in $\sigma_{param,[X/Fe]}$. The values obtained in the two runs can be found for each parameter in Table \ref{tab:PlusMinus}. \\
Similar to Eq. (1) from \citet{fernandeztrincado20}, the total uncertainty is given by the sum of the squared $\sigma_{param,[X/Fe]}$ for each element,
\begin{multline}
     \sigma^2_{[X/Fe]} = \sigma_{T_{eff},[X/Fe]}^2 + \sigma_{\log g,[X/Fe]}^2 + \\ + \sigma_{\mu_t,[X/Fe]}^2 + \sigma_{[M/H],[X/Fe]}^2 + \sigma_{CONVOL,[X/Fe]}^2. 
     \label{eq:totalsysterr}
\end{multline}
We found that the largest $\sigma_{param,[X/Fe]}$ is related to T$_{eff}$ and $\log g$ followed by $\text{CONVOL}$, meaning that uncertainties in these parameters have the highest impact on the final abundances. In contrast, deviations in $\text{[M/H]}$ and $\mu_t$ do not have a great influence. The final systematic errors $\sigma^2_{[X/Fe]}$ are also listed in Table \ref{tab:PlusMinus}. \\
Another estimation of the uncertainty is provided by the line-to-line random error or standard deviation $\sigma_{stdev,[X/Fe]}$, calculated for each element simultaneously to the final abundances. When the abundance is based on one spectral line only, the standard deviation was set to be a typical value, that is, the average $\sigma_{stdev,[X/Fe]}$ of that element over all the sample stars. The average $\sigma_{stdev,[X/Fe]}$ alongside $\sigma_{[X/Fe]}$ is given in Table \ref{tab:stdev+syst}. Both types of errors are on the same order of magnitude for nearly every element. In general, $\sigma_{[X/Fe]}$ is lower than or equal to $\unit[0.20]{dex}$, except for S in 2M18070782-1517393, which reflects the strong temperature dependence of the S lines (see Sect. \ref{alpha-elements}). The abundances of C, O, and N are also highly dependent on temperature because the measurements of these species are mainly based on molecular lines. This explains the relatively strong deviation toward higher values of $\sigma_{[X/Fe]}$ compared to $\sigma_{stdev,[X/Fe]}$. \\
For clarity, we provide a general error bar in the following figures, corresponding to the typical (average) $\sigma_{stdev,[X/Fe]}$. This is listed in Table \ref{tab:stdev+syst}. \\ \\

All parameters, abundances, and standard deviations obtained by the approach outlined in Sect. \ref{Calculation} for the entire sample described in Sect. \ref{Sample} are available in electronic form at the CDS. A preview including only the abundance results is given in Table \ref{table:AbusShort}.

%\FloatBarrier

\section{Sample analysis\label{Results}}

\subsection{Refining the sample: Outliers and special cases\label{Outliers}}

\begin{table*}[ht!]
\tiny
\caption{Abundance table of outliers, peculiar cases, and finally excluded stars.}   
\label{table:Outliers}      
\centering
\begin{tabular}{c c c c c c c c c c c c c c}        
\hline\hline                 
APOGEE-ID & [Fe/H] & [C/Fe] & [N/Fe] & [O/Fe] & [Na/Fe] & [Mg/Fe] & [Al/Fe] & [Si/Fe] & [P/Fe] & [S/Fe] & [Ca/Fe]  & [Ce/Fe] & [Nd/Fe] \\    \hline
   2M04371861 & -0.70 & 0.16 & -0.47  & 0.54 & 0.46  & 0.40 & 0.53 & 0.29 & 0.77: & 0.21 & 0.28  & 0.22*: & 1.05+: \\
   2M06330984 & -0.64 & 0.05 & 0.27 & 0.64 & 0.03* & 0.31 & 0.37 & 0.22 & 0.66: & 0.03: & 0.23  & 0.06* & 0.09+ \\
   2M06394177 & -0.88 & 0.16 & 0.31 & 0.76 & 0.25+ & 0.36 & 0.40 & 0.27 & 0.77: & 0.15 & 0.30  & 0.22+ & 1.23 \\
   2M14513934 & -1.39: & 0.51+ & 0.84: & 0.93+ & 1.22+ & -0.28 & 1.05 & 0.36 & 1.23+ & 0.75+ & 0.30+  & 0.60+ & None \\
   2M15170852 & -1.13: & 0.02: & 1.60+ & 1.72+ & 1.46+ & 0.64 & 0.11: & 0.74 & 1.02+ & 0.34: & 0.67+ & 1.15+ & 1.48+ \\
   2M16485945 & -0.89 & -0.04 & 0.43 & 0.67 & 0.15* & 0.34 & 0.27 & 0.26 & 0.51* & 0.25*: & 0.31 & 0.37 & 0.47* \\
   2M17073023 & -1.08 & -0.05 & 0.75 & 0.98 & 0.66+ & 0.63 & 0.94 & 0.60 & 0.72+ & 0.18* & 0.45 & 0.49 & 0.93+ \\
   2M22375002 & -1.12 & -0.71*: & 1.44 & 0.72+ & 0.82+ & -0.92*: & 1.10 & 0.57 & 0.51+: & 0.27 & 0.5 & 0.24+ & None: \\ \hline
   2M17171752 & -1.03 & 0.07 & -0.25+ & 0.41 & 0.58+ & 0.41 & 0.44 & 0.42 & 1.77+ & 0.35 & 0.40 & 0.73 & 0.98+ \\ \hline
   2M16231729 & -0.96 & -0.13: & 1.28 & 0.89 & 0.69+ & 0.42 & 0.74 & 0.45 & 1.00 & 0.18: & 0.27 & 0.44 & 0.91+ \\ \hline
\end{tabular}
\tablefoot{Plus signs indicate upper limits, asterisks denote values based on one line only, and colons highlight preliminary results that have to be further confirmed because no clear abundance determination was possible with the available data.}
\end{table*}

As described in Sect. \ref{Sample}, the sample is composed of three subgroups. Two of them potentially host P-rich stars, and the third is already proved to be P-rich. \\
The selection of 18 additional stars from the VAC \citep{hayes22} by applying rather arbitrary conditions ([P/Fe]~$>\unit[0.8]{dex}$ and without imposing ASPCAPFLAG = 0) aimed to uncover possible underestimates of their P abundance made in the catalog. Recalculating the abundances of these stars confirmed the results from the VAC in some cases, meaning that not all of the stars contained in this subgroup can be called P rich. Except for this expected reduction, some more stars are not P rich for various reasons. In the following, we point out each case separately and justify the rejection before we present the final chemical signature in Sect. \ref{EnhanceCorr}. \\
In total, 9 out of the initial group of 87 stars were excluded, leading to a final sample of 78 stars. This is still the largest number of P-rich giants to date. It is necessary here to clarify that we did not systematically search for P-rich stars in the APOGEE-2 DR17 with more than 650\,000 targets, as was done previously in DR14 \citep[263\,000 targets,][]{abolfathi18APOGEEDR14} to find the first set of P-rich stars \citep{masseron20a}. This means that our final number of 78 P-rich stars has to be regarded as a lower limit because we expect more undiscovered P-rich stars to be present in DR17. \\   

\begin{figure}
    \resizebox{\hsize}{!}{\includegraphics{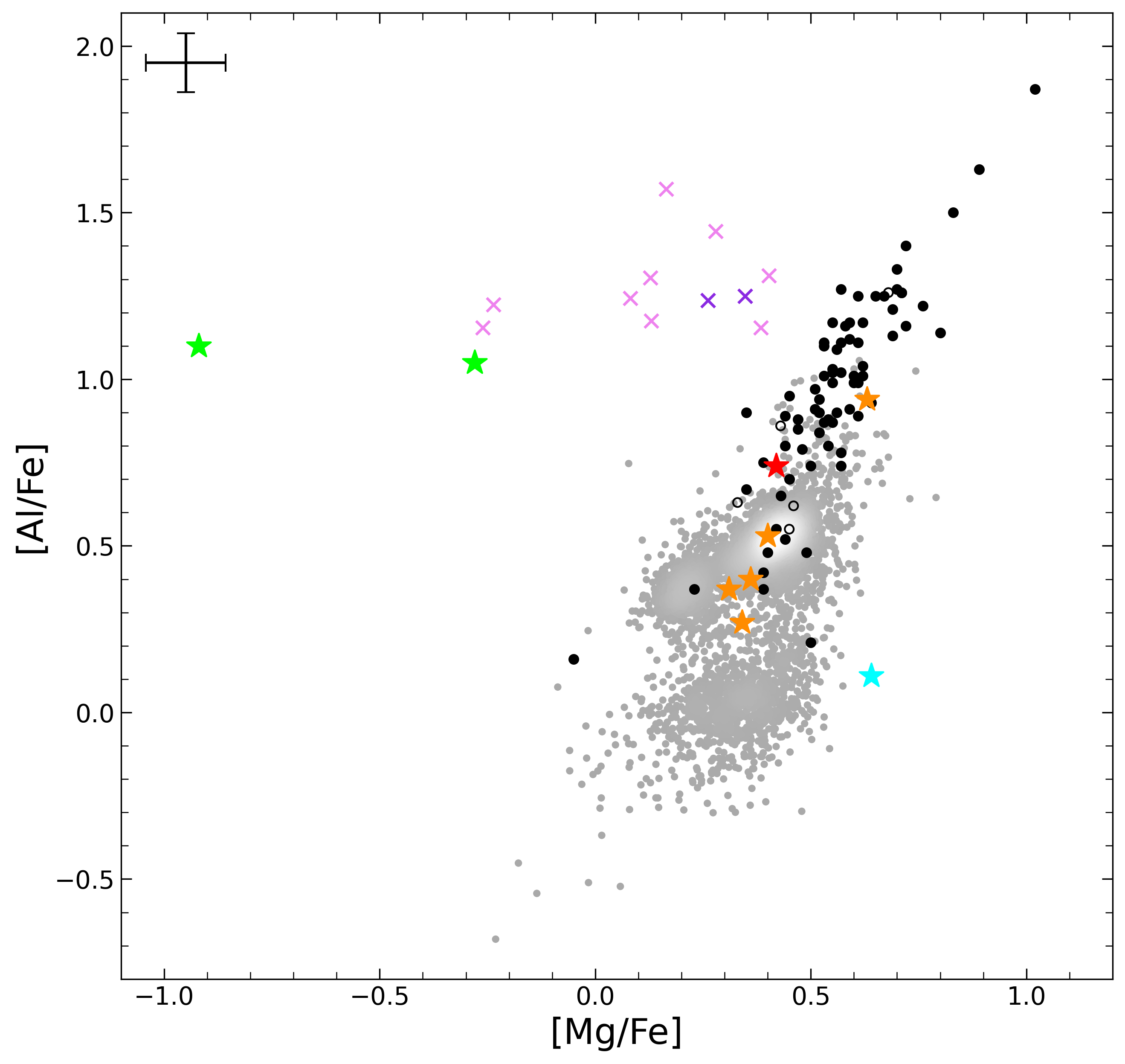}}
    \caption{Abundances in the [Al/Fe]-[Mg/Fe] plane, showing the background stars (gray color-coded density map) and the full P-rich candidates sample (black circles) described in Sect. \ref{Sample}. Empty black circles represent upper limits in [P/Fe]. Outliers identified in this or the plane shown in Fig. \ref{Fig:P_Si_all} are highlighted with orange, green, and blue stars. The red star corresponds to the P-rich GC (M4) member 2M16231729-2624578. Violet crosses indicate outliers in the background sample with GC characteristics, two of which (dark violet crosses; 2M12155306+1431114, 2M17350460-2856477) have a high [O/Fe] abundance ratio. As defined in Sect. \ref{Errors}, the error bar represents the typical standard deviation of the elements listed in Table \ref{tab:stdev+syst}.}
     \label{Fig:Al_Mg_all}
\end{figure}

\begin{figure}
  \resizebox{\hsize}{!}{\includegraphics{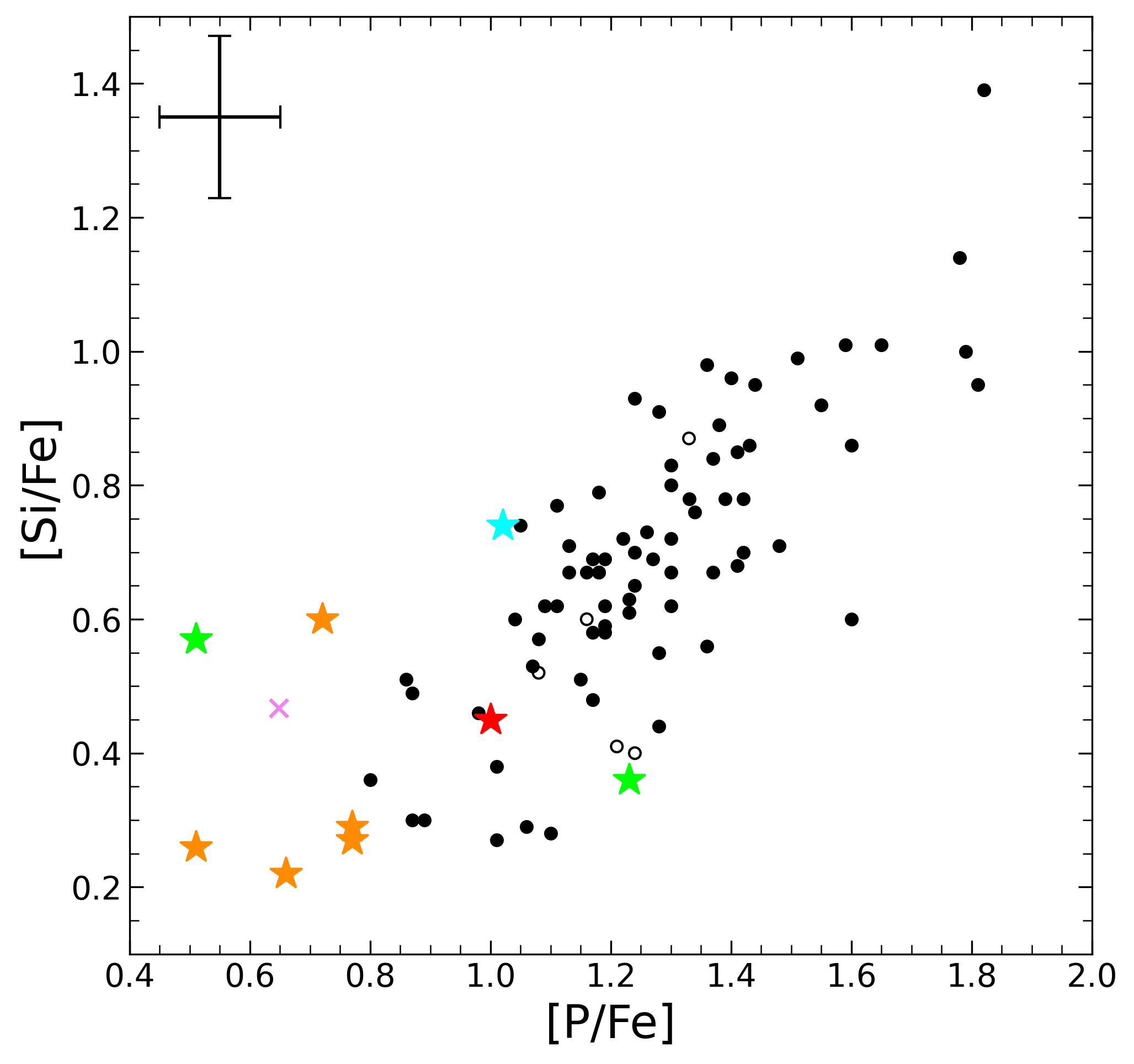}}
  \caption{Abundances in the [Si/Fe] vs. [P/Fe] plane. The symbols are the same as in Fig. \ref{Fig:Al_Mg_all}. The error bar represents the typical standard deviation of the elements listed in Table \ref{tab:stdev+syst}. No background stars are displayed because the reliability of the P abundances obtained by the automatic calculation is limited (see Appendix \ref{LPvalidation} for further justifications).}
  \label{Fig:P_Si_all}
\end{figure}

Stars that show outstanding properties compared to the reference group become visible in various planes of the abundance versus abundance diagrams. In the following, we use the [Al/Fe]-[Mg/Fe] plane to spot outliers that might be related with GCs. Additionally, we examine the [P/Fe]-[Si/Fe] plane with the aim to define a limit on the P abundance that indicates whether a star can be considered as P rich. When outliers were detected, we analyzed their properties and decided whether they should be included in the P-rich sample or be eliminated from the current study. The detected outliers and finally excluded stars are listed in Table \ref{table:Outliers} alongside their abundance values. \\
The [Al/Fe]-[Mg/Fe] plane is shown in Fig. \ref{Fig:Al_Mg_all}. Two stars of the sample can be found in the Mg depleted region of Fig. \ref{Fig:Al_Mg_all}, 2M22375002-1654304 and 2M14513934-0602148 (green stars). For the first star, the abundance values from Table \ref{table:Outliers} are in accordance with the results from \citet{fernandeztrincado20}, showing a chemical pattern consistent with a second-generation GC star or escapee. The low upper limit of the [P/Fe] value of 2M22375002-1654304 caused us to drop this star from the P-rich sample. The chemistry of the second star, 2M14513934-0602148, is also compatible with that of a GC star or escapee. Like for 2M22375002-1654304, it was only possible to give an upper limit instead of an actual value for the P abundance of 2M14513934-0602148. This star was therefore also removed. \\ 
Another discovery from the [Al/Fe]-[Mg/Fe] plane is related to the background sample. Several stars of the background sample in Fig. \ref{Fig:Al_Mg_all} (violet crosses) show characteristics of GC stars, but passed the cross matching of the two GC catalogs (see Appendix \ref{BackgroundSelection}), meaning that these stars are currently not included in these catalogs or are GC escapees. Two of these outliers, 2M12155306+1431114 and 2M17350460-2856477 (dark violet crosses), present high values of [O/Fe] and [N/Fe], indicating that they are not GC escapees. It is possible that they belong to a set of N-rich stars in the inner Galaxy \citep{schiavon17}, although they do not have the same stellar parameters as the N-rich stars. The nature of these outliers still has to be investigated in a different context. In the following, we have eliminated them from the background sample. \\ 
The second plane, [P/Fe]-[Si/Fe] (Fig. \ref{Fig:P_Si_all}), was used to search for a cut in the P abundance that defines P-rich stars. In Fig. \ref{Fig:P_Si_all}, no background sample is plotted because of the bias toward high metallicity in the VAC stars and the low reliability of the automated [P/Fe] determination (see Appendix \ref{LPvalidation}). The plot reveals that five stars, 2M04371861-5916494, 2M06330984-6104151, 2M06394177-6757357, 2M16485945-2647326, and 2M17073023-2717147 (orange star symbols), have relatively low values of P and Si, combined with upper limits or uncertain measurements of P. Furthermore, they do not show similarities with the group regarding the O, Mg, Al, and Ce abundances, as we discuss in the following. Three of these stars, 2M06330984-6104151, 2M06394177-6757357, and 2M16485945-2647326, belong to the subgroup of 18 stars that was filtered out of the VAC. In this case, the P abundances were most likely overestimated by the VAC, and we decided to exclude these stars from the P-rich sample. Finally, given the cut [P/Fe]~$<\unit[0.8]{dex}$ arising from Fig. \ref{Fig:P_Si_all}, stars with [P/Fe]~$<\unit[0.8]{dex}$ are not counted to our P-rich sample. This is a slightly lower value than the empirical cut [P/Fe]~$<\unit[1.0]{dex}$ used by \citet{masseron20a}. \\
The last candidate eliminated from the sample is 2M15170852+4033475 (blue star). This star presents suspiciously high upper limits in N, O, Na, Ce, and Nd, resulting from the poor quality of the spectrum, which also complicates a proper measurement of the P abundance. We therefore decided to reject this star. \\ \\
As mentioned at the beginning of this section, nine stars were eliminated of the P-rich star candidates group. In summary, we excluded stars with [P/Fe]~$<\unit[0.8]{dex}$, GC characteristics (Mg depleted and Al enhanced) and no constraining (i.e., too high) upper limits. The remaining sample of 78 stars represents a fraction of approximately $\unit[0.24]{\%}$ out of all APOGEE-2 DR17 giants with similar stellar parameters ($T_{eff}$, log g, and [M/H]). We emphasize again that this percentage has to be regarded as a lower limit.

\subsubsection{Double-lined spectroscopic binary 2M17171752-2148372\label{binary}}
As pointed out in Sect. \ref{Sample}, while discovering the Si-rich stars, \citet{fernandeztrincado20} also analyzed a binary member with ID 2M17171752-2148372. In DR17, this star is flagged, indicating that 2M17171752-2148372 might be a double-lined spectroscopic binary (SB2)\footnote{see paragraph "Spectroscopic Binary Identification" on \url{https://www.sdss.org/dr17/irspec/radialvelocities/}}. The identification of SBs indicated by the flag is based on the work of \citet{kounkel21}. A visual inspection of the spectrum confirms that 2M17171752-2148372 is part of a binary system. \\
The abundance values corresponding to 2M17171752-2148372 are listed in the penultimate line of Table \ref{table:Outliers}. The strikingly high upper limit of P is caused by the fact that the second \ion{P}{i} line ($\unit[16482.93]{\angstrom}$) is close to the CCD edge, and therefore, the measurement quality is poor. When only the first \ion{P}{i} line ($\unit[15724.00]{\angstrom}$) is considered, the abundance value would be $\text{[P/Fe]}=\unit[1.18]{dex}$, manifesting an enhancement in P. This finding, if confirmed, is interesting because it suggests that the P-rich star progenitor may also affect a binary
system, for example, by polluting the material from which one or more components
were born.
Nevertheless, we decided to exclude this star from the current study because of the spectral quality and because the stellar parameters might be suspect, for example, due to contamination of the spectrum through the companion. We aim to revisit the special case of 2M17171752-2148372 in future works. \\

\begin{figure}
  \resizebox{\hsize}{!}{\includegraphics{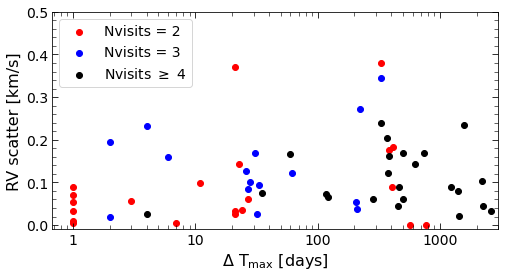}}
  \caption{RV scatter vs. maximum time lag $\Delta T_{max}$ between visits that are color-coded by the number of visits. Stars with fewer than two visits have been excluded.}
  \label{Fig:RV_statistics}
\end{figure}

\subsubsection{Other possible binary stars}
We identified other possible binary stars in the sample by searching for a radial velocity (RV) scatter higher than $\unit[1]{km/s}$ (or higher than ten times the median of the individual visit RV errors)\footnote{\label{RVs}see \url{https://www.sdss4.org/dr17/irspec/use-radial-velocities/}}. Applying this criterion, we found four stars in the sample that are likely binaries: 2M18043255-4819138, 2M19193412-2931210, 2M17033361-2254246, and 2M13303961+2719096 with $\unit[13.14]{km/s}$, $\unit[2.42]{km/s}$, $\unit[1.76]{km/s}$, and $\unit[1.22]{km/s}$, respectively. \\
We recall that although the RV scatter of the remaining 74 stars is lower than $\unit[1]{km/s}$, they might also be binaries. Because the RV scatter is derived from the RV measurements obtained in various visits, a significant scatter can only be detected when the visits are sufficiently separated in time. To evaluate the reliability of the low RV scatter results, we analyzed the RV measurements on the visit level. We applied the recommended quality cuts\footnoteref{RVs} and found that 72 stars have at least one visit that is not flagged. We then extracted the number of valid visits (Nvisits) and the corresponding modified Julian date (MJD) to calculate the maximum time lag $\Delta T_{max}$ between the visits. Out of the 72 stars, 13 have been visited only once, meaning that the RV scatter is set to be zero and no conclusion on binarity can be drawn in these cases. In Fig. \ref{Fig:RV_statistics} we display the RV scatter versus the maximum time lag $\Delta T_{max}$, color-coded by the number of visits. The accumulation of points at $\Delta T_{max}=\unit[1]{day}$ corresponds to stars that have been visited on consecutive days, which means that the reliability of the RV scatter for these stars can be regarded as low. Imposing the conservative condition that at least $\unit[1000]{days}$ should have passed between the visits in order to trust the low RV scatter value, we find that the RV of 7 stars did not vary over a long period of time, meaning that a close binary scenario can be discarded in these cases. For the  less conservative case and when the criterion is applied that $\unit[100]{days}$ should have passed between the visits, 30 stars are likely nonbinaries. Although few binary stars are present in the sample, the previous considerations show that the majority of the sample stars is not part of close binary systems. Instead, if they were part of widely separated binary systems, we would not expect significant interaction to take place between the components. We therefore conclude that the P-richness cannot be attributed to a binary mass-transfer scenario.

\subsubsection{Globular cluster member 2M16231729-2624578\label{GCstar}} 
By cross-matching the GC catalog based on \citet{harris1996} (see Appendix \ref{BackgroundSelection}) and inspecting the information provided by the ASPCAP flag that indicates possible GC members, we are able to report the first detection of a P-rich star in a GC, namely M4. The chemical abundances of this star with ID 2M16231729-2624578 are presented in the last line of Table \ref{table:Outliers}. In Figs. \ref{Fig:Al_Mg_all} and \ref{Fig:P_Si_all}, we highlight 2M16231729-2624578 with a red star. We know from \citet{yong08} that M4 is a peculiar GC, rich in slow-neutron capture (\textit{s}-process) elements (Ce, Ba, and Pb) with a mean [Ce/Fe] abundance ratio of $\unit[0.46\pm 0.07]{dex}$ \citep{contursi22}. The Ce abundance of 2M16231729-2624578 is compatible with this mean. \\
In addition to the slight enhancement in Ce and the C-N anticorrelation typical for GCs, this star also presents an enrichment in P, with $\text{[P/Fe]}=\unit[(1.00\pm 0.14)]{dex}$\footnote{Here the error is the root sum squared of the P standard deviation in this particular star ($\unit[0.04]{dex}$) and the systematic P error averaged over the two stars given in Table \ref{tab:stdev+syst} ($\unit[0.13]{dex}$).}, O and Al, and moderate enhancements in Si and Mg (see Table \ref{table:Outliers}) compared to the P-rich group. Although its [Al/Fe] is higher than $\unit[0.3]{dex}$, the limit used by \citet{meszaros20} to separate first- (unenriched) and second-generation (enriched) GC stars, there is no depletion in O and Mg. This abundance pattern suggests that 2M16231729-2624578 presumably belongs to
the first generation and may have been polluted by the P-rich star progenitor,
implying that the pollution or the progenitor timescale is short
compared to the typical evolution timescale of GCs. A
second-generation status for this star is probably less likely because it would produce a
different chemical pattern, in particular, much higher Al and Si than observed. \\
The presence of a P-rich star in a GC enhanced in \textit{s}-process elements could give rise to the speculation that the P-rich star progenitor must be an AGB star, the stellar site where the \textit{s}-process mainly takes place. However, this hypothesis has already been ruled out by \citet{masseron20b} based on discrepancies in the light element abundance pattern. Our finding that 2M16231729-2624578 belongs to M4, however, supports the assumption from \citet{masseron20b} that the P-rich star progenitor has undergone a neutron-capture process similar to the \textit{s}-process. Overall, we can state that regardless of what the progenitor of the P-rich stars turns out to be, it must have been present in M4 as well.

\FloatBarrier

\subsection{Chemical signature of the final sample\label{EnhanceCorr}}
\subsubsection{Phosphorus}

\begin{figure}
  \resizebox{\hsize}{!}{\includegraphics{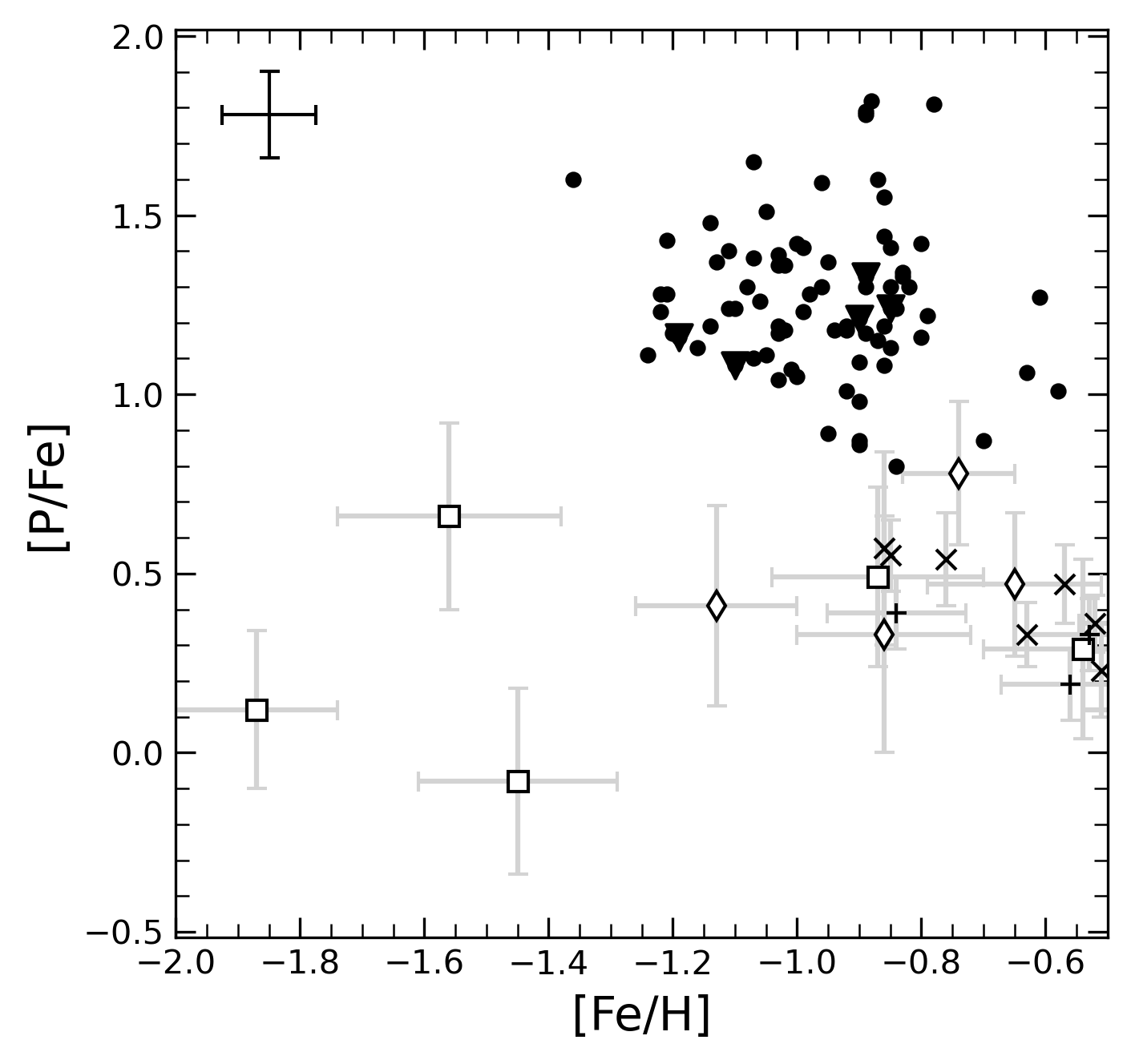}} 
  
  \resizebox{\hsize}{!}{\includegraphics{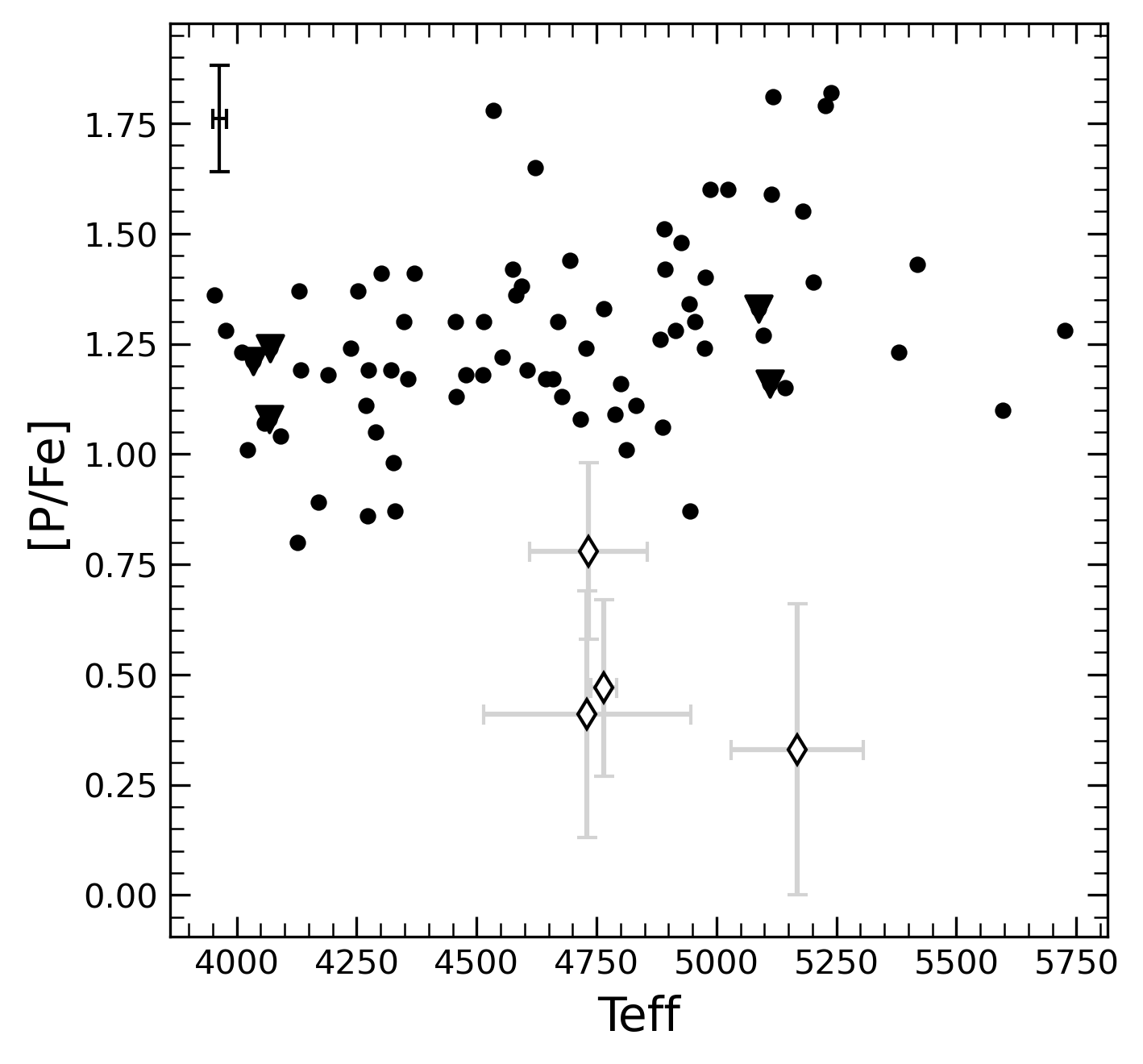}}
  \caption{Abundance ratio [P/Fe] vs. metallicity [Fe/H] (top panel) and vs. effective temperature $T_{eff}$ (bottom panel) of the final sample of 78 P-rich stars (black dots). Downward triangles indicate upper limits in [P/Fe]. Literature values are shown with open black squares \citep{jacobson14}, crosses \citep{maas17,maas19}, diamonds \citep{masseron20a}, and plus signs \citep{caffau11,caffau19}. Typical errors are indicated in the top left corner.}
  \label{Fig:P-richs}
\end{figure}

The behavior of the P abundance in our final sample of 78 P-rich stars compared to literature values is shown in Fig. \ref{Fig:P-richs}. The top panel, an updated version of Fig. 1 from \citet{masseron20a}, shows the abundance ratio [P/Fe] versus metallicity [Fe/H] of the enlarged sample, covering a slightly wider [Fe/H] range than the 16 P-rich stars from \citet{masseron20a,masseron20b}. At this point, we again refer to Fig. \ref{fig:Plines}, which illustrated the spectral difference between a P-rich and a P-normal star with similar stellar parameters. \\
Although some stars near $\text{[Fe/H]}=\unit[-0.8]{dex}$ fall below the empirical line drawn by \citet{masseron20a} at $\text{[P/Fe]}=\unit[1.00]{dex}$, we count them to the P-rich sample because they do not fulfill the exclusion criteria defined in Sect. \ref{Outliers}, and their overall chemical signature, which is discussed in the following subsections, fits the characteristic group of P-rich stars. Nevertheless, this updated version also shows a clear separation between the sample and the reference group. The plot [P/Fe] versus $T_{eff}$ in the bottom panel of Fig. \ref{Fig:P-richs} demonstrates that there is no trend of [P/Fe] with $T_{eff}$, suggesting that the P lines are unaffected by NLTE effects. \\ \\

\begin{figure*}
\centering
   \includegraphics[width=17cm]{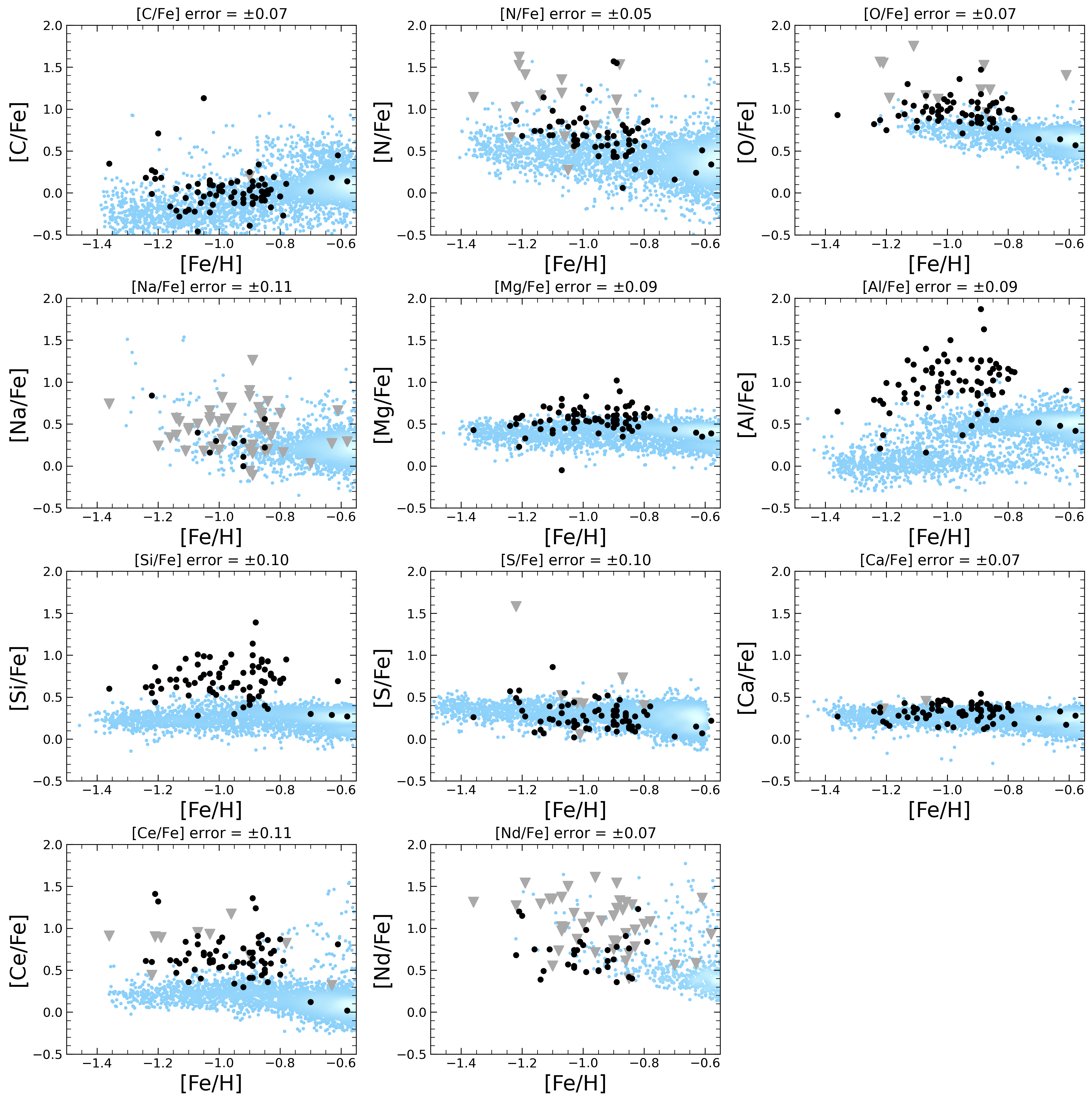}
     \caption{Abundance ratios [X/Fe] vs. metallicity [Fe/H], displaying one element X in each panel. Gray downward triangles indicate upper limits. The background sample was selected and the corresponding abundances were calculated following the procedure described in Sects. \ref{Sample} and \ref{LineSelecProcessing}, respectively. In the case of S, the background stars and their corresponding abundances have been extracted from the VAC from \citet{hayes22}. The color-coding reflects the number density of the stars. For clarity, error bars are suppressed. The errors adopted for each element are given in the title of each panel and correspond to the average standard deviation listed in Table \ref{tab:stdev+syst}. The error of the metallicity [Fe/H] is $\pm 0.08$.}
     \label{fig:MetalMulti}
\end{figure*}

In Fig. \ref{fig:MetalMulti} we plot the abundance ratios [X/Fe] of the remaining elements considered in this work versus [Fe/H]. Figure \ref{fig:TeffMulti} shows the same abundance ratios [X/Fe] versus $T_{eff}$. As in the bottom panel of Fig. \ref{Fig:P-richs}, the [X/Fe] versus $T_{eff}$ in Fig. \ref{fig:TeffMulti} rules out possible trends with $T_{eff}$. In Fig. \ref{fig:MetalMulti} we identify other enriched elements in our P-rich sample: the $\alpha$-elements O and Si, the odd-Z element Al, and the heavy neutron-capture element Ce. In the cases of Mg and Nd, overabundances are present in some stars. We also plotted [(C+N)/Fe] versus [Fe/H] (Fig. \ref{Fig:C+N}) to explore a possible enhancement in C+N, which is discussed in the next section. For the following, more detailed discussion of certain elements, a zoom into the corresponding [X/Fe] versus [Fe/H] panel of Fig. \ref{fig:MetalMulti} is provided in Figs. \ref{Fig:Si} (for Si) and \ref{Fig:O}-\ref{Fig:Nd} (for O, Mg, Al, Ce, and Nd, respectively) alongside a [X/Fe] versus [P/Fe] plot to illustrate their correlated relation with P. A summary of the identified enhancements is given in Fig. \ref{fig:summaryplot}, where we compare the median chemical abundances of the P-rich stars with those of the background sample. In the case of P, we compare the P-rich sample to the median abundance of the literature values shown in Fig. \ref{Fig:P-richs}.

\subsubsection{Carbon and nitrogen}

\begin{figure*}%[ht]
  \resizebox{\hsize}{!}{
  \includegraphics{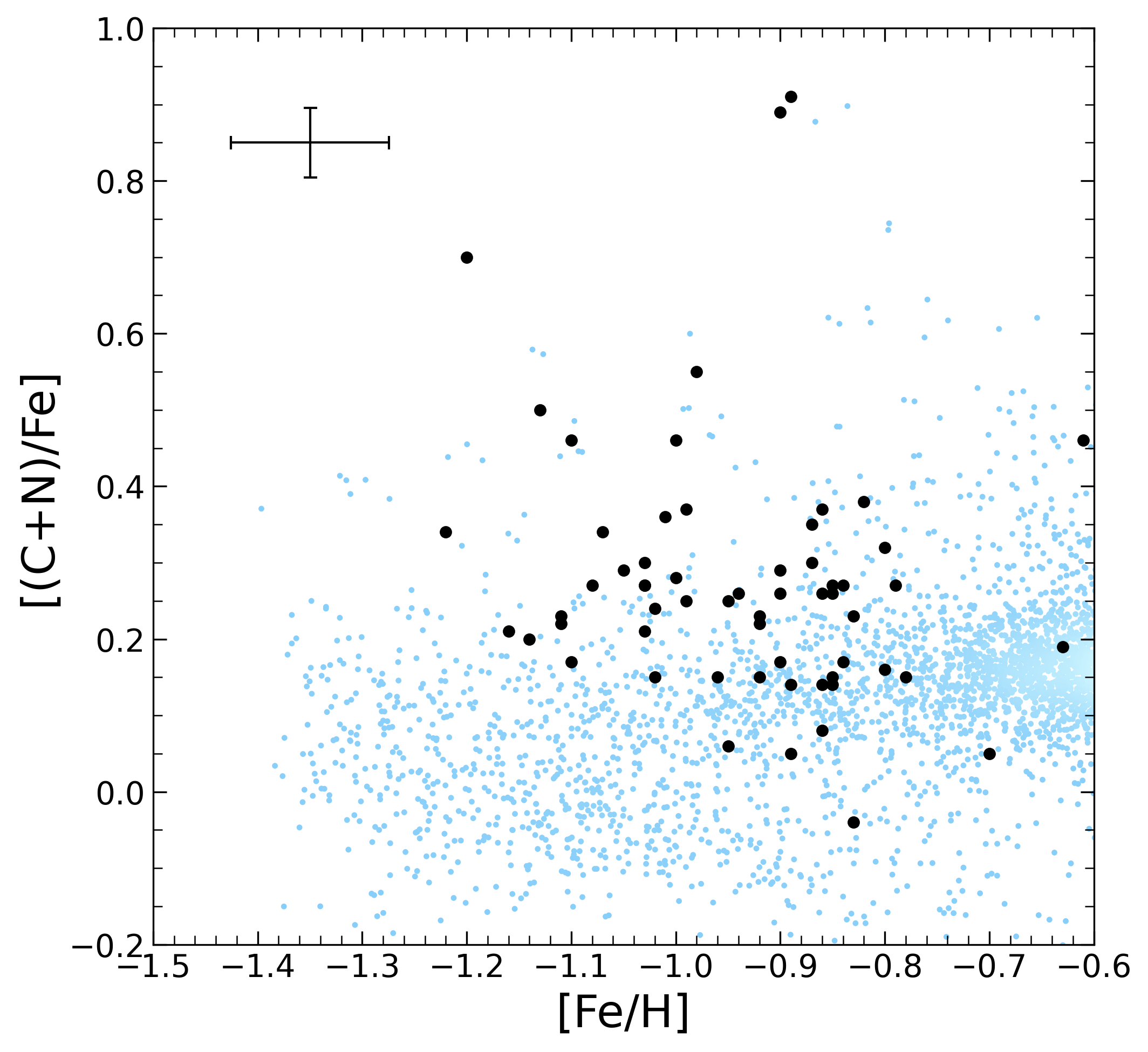}
  \includegraphics{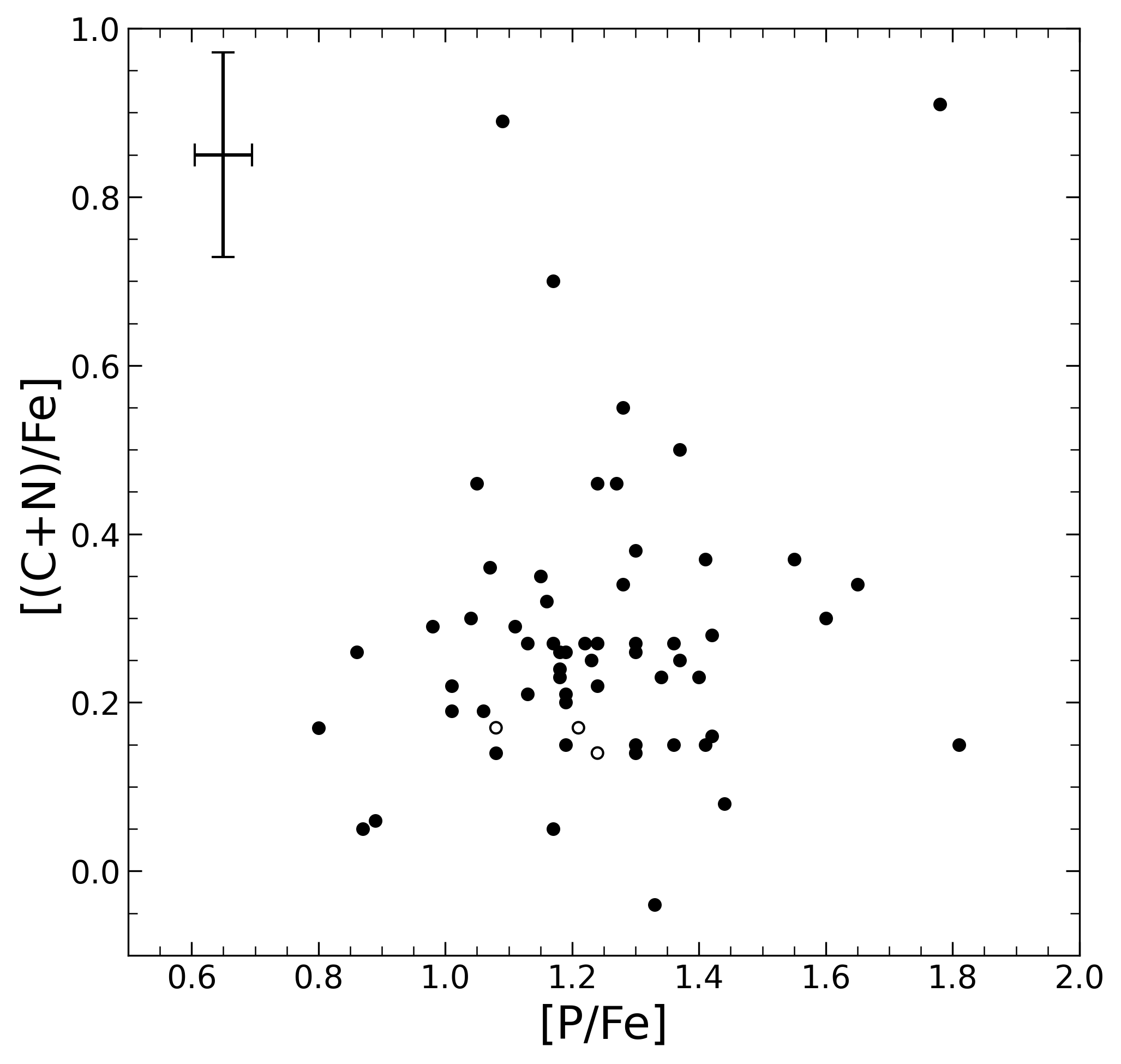}} 
  \caption{Detailed representation of the [(C+N)/Fe] abundance ratio. Left panel: Abundance ratio [(C+N)/Fe] vs. metallicity [Fe/H]. Stars with upper limits in C or N have not been considered. Right panel: [(C+N)/Fe] vs. [P/Fe]. Solid circles correspond to real measurements in
  P, and empty circles denote upper limits in P. In both panels, the error of [(C+N)/Fe] was defined as the quadratic sum of the standard deviation of the C and N measurements.}
  \label{Fig:C+N}
\end{figure*}

The amounts of C and N observed in the P-rich stars may arise from a superposition of different origins. In general, C is made during He-burning in CCSNe and AGB stars \citep{hawkins16,kobayashi:20}. Nitrogen is produced by the CN(O)-cycle in H-burning conditions in nearly all types of stars, including low-mass red giant branch stars such as the P-rich stars. We note a slight enhancement of N in Fig. \ref{fig:MetalMulti}, which can be caused by intrinsic production or pollution from the progenitor. Because the CN-cycle consumes C and produces N, their respective amounts may change over the evolutionary phases, but the sum of C+N remains constant in the P-rich stars \citep[see, e.g.,][]{masseron15,hawkins15}. Thus, it is reasonable to plot the [(C+N)/Fe] abundance ratio versus [Fe/H] in order to explore whether C is produced in the P-rich star progenitors. Figure \ref{Fig:C+N} reveals that the majority of the sample stars is not enhanced in C+N compared to the background stars. \citet{masseron20b} already indicated this lack of an enhancement. This observation is also supported by Fig. \ref{fig:summaryplot}, which does not allow us to state an enhancement of C+N within the current error intervals. The right panel of Fig. \ref{Fig:C+N} also shows that C+N and P are not clearly correlated, and we therefore conclude that the C production is not a direct attribute of the P-rich star progenitors.

\subsubsection{\texorpdfstring{$\alpha$-}-elements\label{alpha-elements}}
\begin{figure*}
  \resizebox{\hsize}{!}{
  \includegraphics{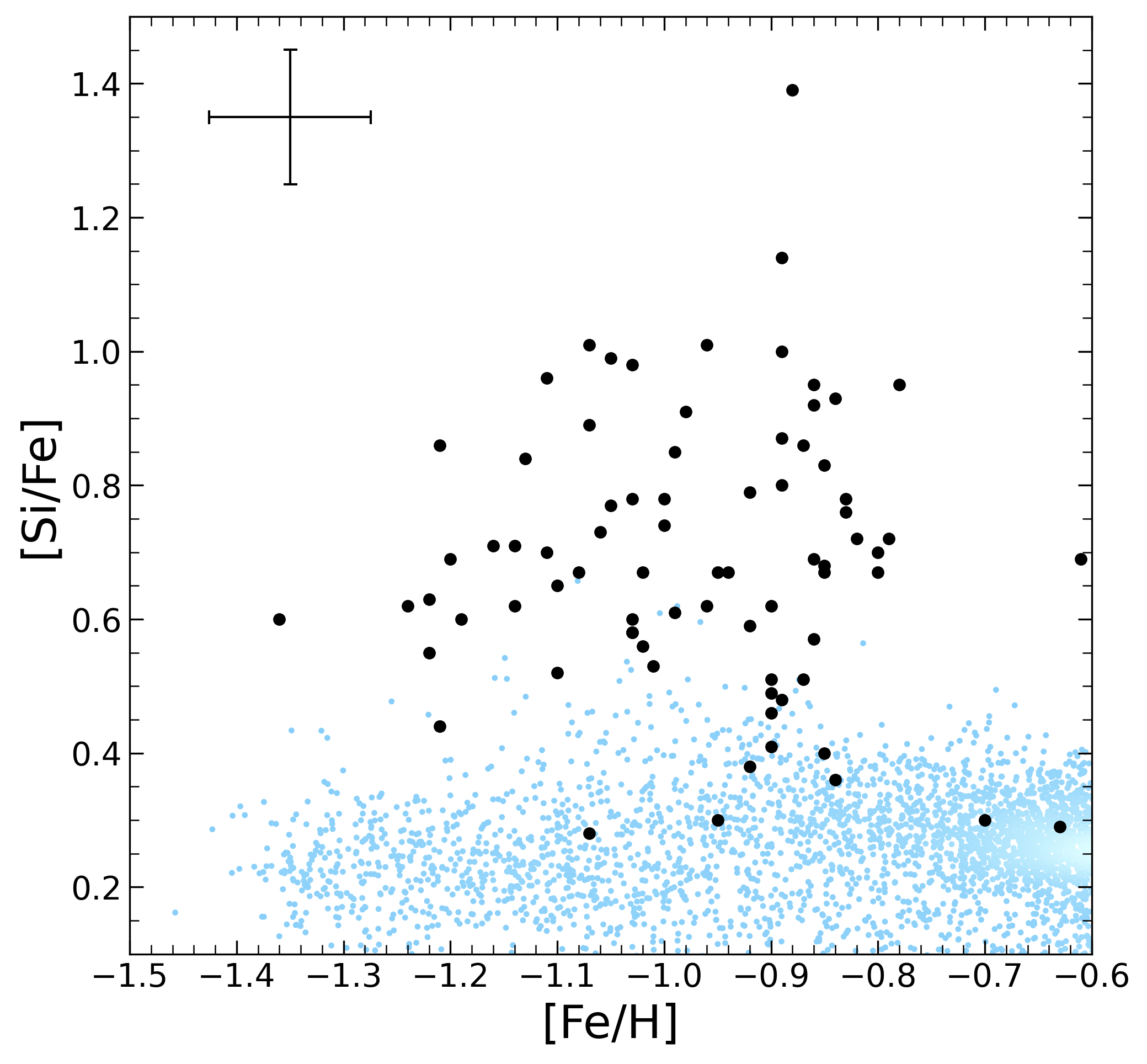}
  \includegraphics{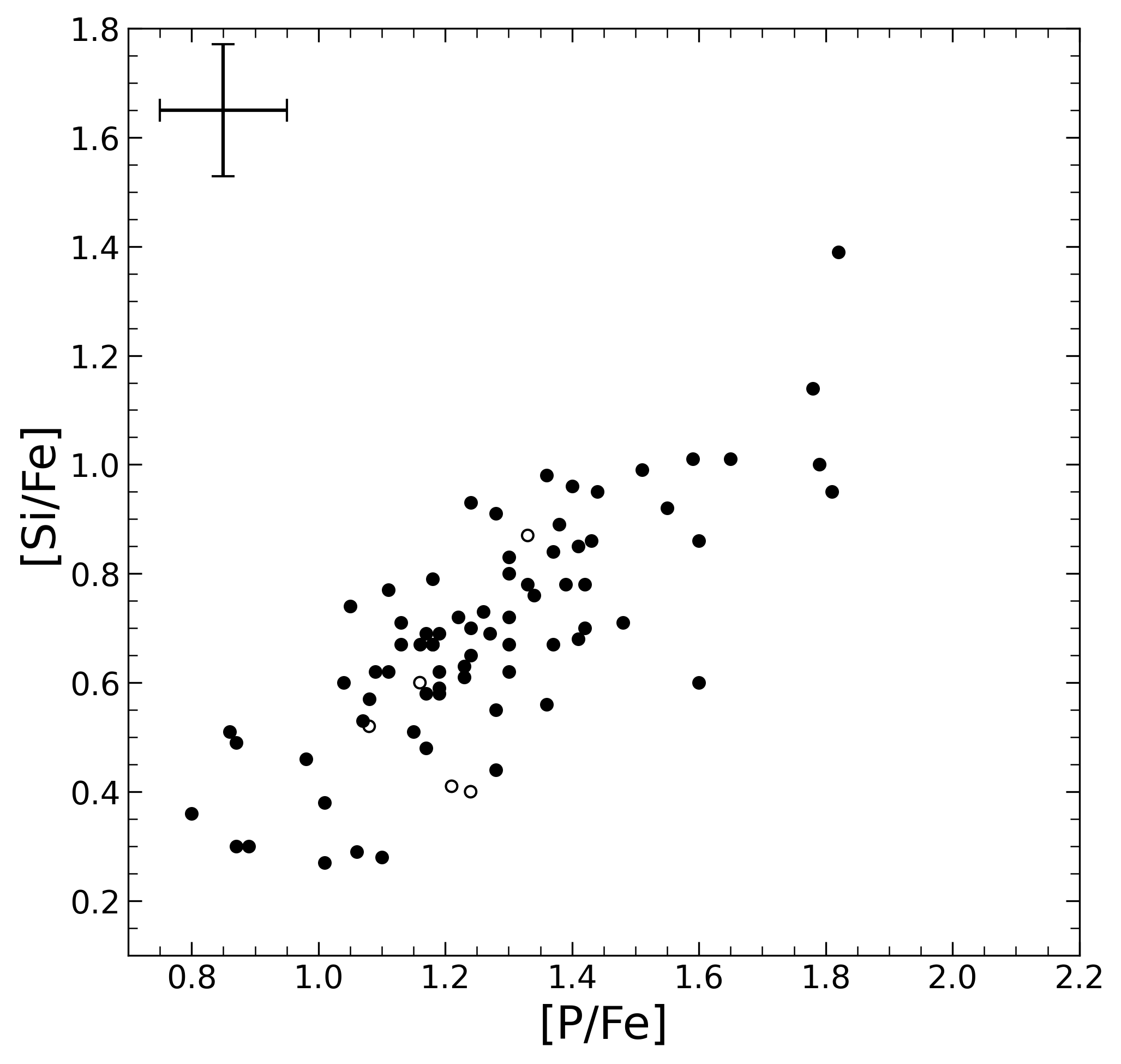}} 
  \caption{Detailed representation of the [Si/Fe] abundance ratio. Left panel: Zoom into the [Si/Fe] vs. [Fe/H] plot of Fig. \ref{fig:MetalMulti}. Right panel: [Si/Fe] vs. [P/Fe]. Solid circles correspond to real measurements in both Si and in P, and empty circles denote upper limits in P.}
  \label{Fig:Si}
\end{figure*}

Figures \ref{fig:MetalMulti}, \ref{Fig:Si}, \ref{fig:summaryplot}, \ref{fig:TeffMulti}, and \ref{Fig:O} show that the $\alpha$-elements O and Si are clearly enhanced in the P-rich stars, confirming the results from \citet{masseron20a,masseron20b}. The right panel of Fig. \ref{Fig:Si} illustrates the correlation between Si and P. This correlation implies that it is possible to detect P-rich stars through the measurement of Si lines, which is convenient because they are much more accessible in the NIR than P lines. The enhancements in O and Si are accompanied by a small enhancement of Mg in only some of the stars. Compared to the background stars, a clear overall Mg-rich signature for the P-rich stars is not observable within the current errors (Fig. \ref{fig:summaryplot}). Nevertheless, Fig. \ref{Fig:Mg} allows us to report a correlation between Mg and P. Figure \ref{Fig:Al_Mg_all} shows that Mg is also correlated with Al. This interconnection leads us to the conclusion that although not clearly enhanced, the behavior of the Mg abundance is coupled with the P-richness, confirming the results from \citet{masseron20a}. The large star-to-star abundance scatter in Fig. \ref{fig:summaryplot} can be explained by the broader P abundance range considered in this work, which naturally leads to a broader Mg abundance range due to the correlation. \\
Similar to \citet{masseron20a,masseron20b}, we find that S and Ca are not enhanced, which is unexpected because these $\alpha$-elements should be correlated with
Si \citep[][]{thielemann:96}. However, it is necessary here to clarify that in the case of S, the background stars in Figs. \ref{fig:MetalMulti}, \ref{fig:TeffMulti} and \ref{fig:summaryplot} are taken from the VAC \citep{hayes22} because S was added to the group of studied elements after the supercomputer run. Furthermore, the results from the VAC were obtained with a careful line selection and are therefore more suitable for the comparison than an automated calculation. Because the S lines in cool metal-poor stars are affected by blends and temperature dependences \citep[see][]{hayes22}, it is difficult to determine a reliable abundance of S, even with a visual inspection as we did for the sample. Hence, the results for S have to be treated with caution. \\
In contrast, there was no difficulty in measuring the Ca abundance based on the five lines that are present. In general, the Ca lines are clear and strong, leading to more robust abundance results. We therefore state that the normal Ca abundance shown in Fig. \ref{fig:MetalMulti} is reliable and that it belongs to the characteristic fingerprint of P-rich stars (Fig. \ref{fig:summaryplot}). We conclude that the process that causes the enhancements of Si does not produce higher S and Ca abundances.

\subsubsection{Sodium and aluminum}
In the spectral range covered by APOGEE-2, the odd-Z element Na exhibits two lines that are both affected by CO blends \citep{hayes22}. We found that the Na abundance could only be measured for a small number of stars. In these cases, the measurements are mostly based on one of the two available lines. For the majority of the sample stars, only upper limits could be provided. Although the measurements for Na given in Figs. \ref{fig:MetalMulti} and \ref{fig:TeffMulti} have to be handled with caution because of the blends, we can conclude from the figures that Na is not enriched in our P-rich stars. \\
Unlike Na, the other odd-Z element Al is easily measurable because three of the four spectral lines are strong and provide consistent abundance values. Thus, the results presented in Figs. \ref{fig:MetalMulti}, \ref{fig:TeffMulti} and \ref{Fig:Al} only consist of real measurements, showing that the detected enrichment and correlation is reliable. It is known that in the NIR H band, the strong Al lines are affected by NLTE effects and that the abundance results obtained in this spectral region differ from the results obtained in the optical \citep{masseron20a}. To account for NLTE effects, it is possible to implement coefficients that correct for departures from LTE \citep[see, e.g.,][]{amarsi20}. As a result, the absolute abundance scale can be improved. In this work, we did not use coefficients like this because we used a differential approach to identify enhancements under the assumption that the background stars are affected by NLTE effects in the same way.

\subsubsection{Cerium and neodymium}
\begin{figure*}
\centering
     \resizebox{\hsize}{!}{
     \includegraphics{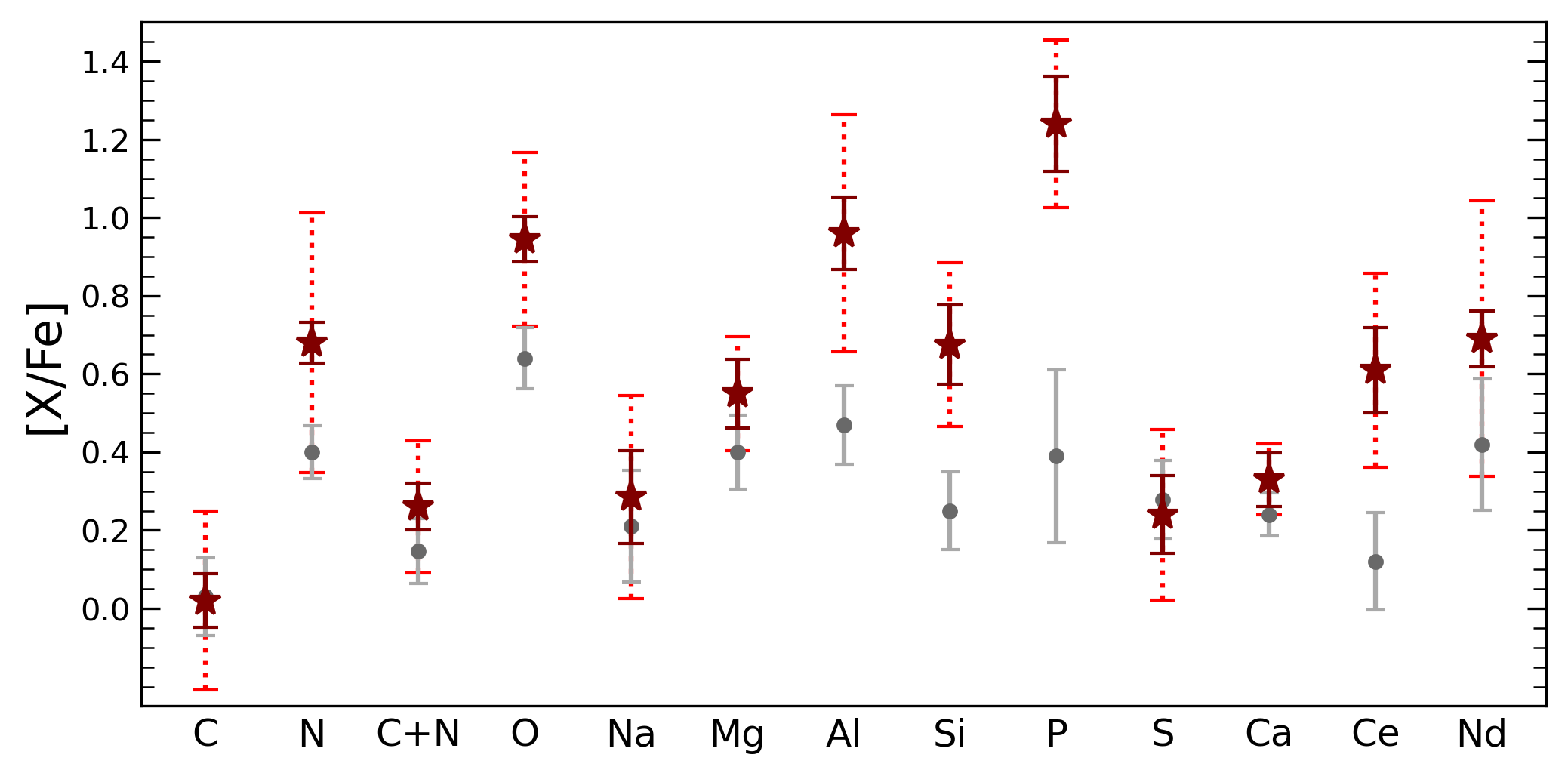}}
     \caption{Chemical abundance pattern of our sample of P-rich stars. Dark red stars correspond to the median chemical abundance of the P-rich sample, and gray dots show the median chemical abundance of the background sample. Stars with upper limits were excluded from the median. The red and gray error bars denote the corresponding average line-to-line abundance scatter (typical standard deviation) from Table \ref{tab:stdev+syst}, which reflects the quality of the measurements. We also provide the star-to-star abundance scatter of the sample as dotted error bars to reflect the spread of abundances. In the case of P, the gray dot indicates the median abundance of the literature values shown in Fig. \ref{Fig:P-richs}, and the gray error bar shows the mean error of the literature values.}
     \label{fig:summaryplot}
\end{figure*}

The first heavy neutron-capture element that we considered is Ce. The measurement of Ce is robust, and its enhancement in the P-rich stars is visible in Figs. \ref{fig:MetalMulti}, \ref{fig:TeffMulti} and \ref{Fig:Ce}. Regarding future investigations, the Ce abundance is of great interest, for example, to explore the relation between P-rich stars and the peculiar GC M4 mentioned in Sect. \ref{GCstar}. \\
Measuring the second heavy neutron-capture element Nd is, similar to Na, a complex issue. The three lines available in the APOGEE-2 DR17 spectral range are in general weak and blended, leading to a small number of real measurements in Figs. \ref{fig:MetalMulti}, \ref{fig:TeffMulti}, and \ref{Fig:Nd}. Instead, there is a high number of upper limits. In the low-metallicity region, there is a lack of reference stars, making it difficult to evaluate a possible enrichment of Nd in the P-rich stars. A correlation with P cannot be observed in Fig. \ref{Fig:Nd} because only a few real measurements are available. Figure \ref{fig:TeffMulti} also reveals the limitation of the Nd measurement with temperature. At higher $T_{eff}$, it is not possible to detect low amounts of Nd, resulting in a biased selection. Although the comparison of the median values in Fig. \ref{fig:summaryplot} suggests a slight enhancement of Nd, we are not able to declare with certainty that P-rich stars are also enhanced in Nd. However, the Nd abundances obtained from optical spectra in the two P-stars from \citet{masseron20b} are consistent with our results. \\ \\

In summary, a comparison of Fig. \ref{fig:summaryplot} with Fig. 4 from \citet{masseron20a} shows that the chemical signature revealed in this work, although containing fewer elements, but a larger number of stars, supports the results reported by \citet{masseron20a}. In addition to P, we confirm the enhancements in O, Mg (for part of the stars), Al, Si, and Ce. Instead, the possible enhancement of Nd remains unclear, and we do not report a relation between the C+N and P abundances. \\
Additionally, we have robust evidence for rather normal abundances of S and Ca. The hypothesis made by \citet{masseron20a} that P-rich stars are part of a larger family of Si- and Al-rich stars is also confirmed by our results. In this context, we highlight the clear correlations between P and the enhanced species, as illustrated in Figs. \ref{Fig:O}-\ref{Fig:Ce}, which naturally leads to a correlation among these species, for example, between Si and Al. We also recall that there is also a correlation between Mg and P, even though Mg is not enhanced in the entire sample. \\
A last general aspect about the chemical fingerprint of the P-rich stars is related to their metallicity. Although our [Fe/H] range of the P-rich stars is slightly broader ($\unit[-1.36]{dex}$ < [Fe/H] < $\unit[-0.58]{dex}$) than the rather narrow range from \citet{masseron20a}, we confirm for now that P-rich stars can only be found at low metallicities, meaning that during this work, we did not detect any P-rich star with [M/H] > $\unit[-0.6]{dex}$. We found that the VAC \citep{hayes22} contains 37 stars with [M/H] > $\unit[-0.6]{dex}$ and [P/Fe] > $\unit[0.8]{dex}$, but a reanalysis of a random sample of 6 of these stars showed that they are not P rich. Extrapolating this result, we tentatively conclude that the low metallicity is indeed characteristic of P-rich stars and not a bias of the searching algorithm used in \citet{masseron20a}.

\section{Search for the P-rich star progenitor}
We begin this section by presenting some detailed GCE models for the Milky Way. We then revisit the possibility of an extragalactic origin of the P-rich stars by performing an orbital analysis.
To conclude the discussion, we compare the outcome of potential nucleosynthetic scenarios with the chemical composition of the P-rich stars, including a new theory that was proposed recently.

\subsection{Galactic chemical evolution of P with theoretical models\label{GCE}}
More recent calculations have provided new insights into possible additional sources of P in the Galaxy that might be related to the formation of P-rich stars. The impact of fast-rotating massive stars for the GCE of P was studied by \cite{kobayashi:11} and more recently by \cite{prantzos:18}. Both studies showed a marginal effect of rotation on the production of P in the Galaxy. \cite{ritter:18} instead studied the impact of C-O shell mergers on the element production in massive stars and on chemical evolution of odd intermediate-mass elements. Their preliminary results show that these events might even boost the production of P in CCSN ejecta by up to an order of magnitude, and if 10-50$\%$ of CCSNe were affected by C-O shell mergers, the GCE evolution of [P/Fe] and its observed dispersion might be reproduced. However, C-O shell mergers are convective-reactive events that require multidimensional hydrodynamics simulations \citep[e.g.,][]{andrassy:20,yadav:20}. While calculations by \cite{ritter:18} provide valuable insights into the possible nucleosynthesis in C-O shell mergers, they are mostly qualitative, without further guidance from stellar hydrodynamics simulations to implement in the next generations of one-dimensional massive star models and CCSN yields for GCE \citep[e.g., ][]{ritter:18,cristini:19,andrassy:20,schatz:22}. In order to place P-rich and P-normal stars in the same astrophysical context and discuss them, we present four GCE models for the Milky Way in this section.\\

The GCE simulations were made using the Python code OMEGA+\footnote{OMEGA+ is available online as part of the JINAPyCEE package \url{https://github.com/becot85/JINAPyCEE}} \citep{Cote2018}. This is a two-zone model comprised of a central, star-forming, cold gas reservoir modeled using the code OMEGA\footnote{OMEGA is available online as part of the NuPyCEE package \url{https://github.com/NuGrid/NUPYCEE}} \citep{Cote2017}, surrounded by a non-star-forming, hot gas reservoir considered as the circumgalactic medium (CGM).\\
At each time step, we followed the evolution of the CGM and the internal star-forming region. The evolution of the mass of the gas in the CGM ($M_{\rm CGM}$) follows
\begin{equation}
    \dot{M}_{\rm CGM}(t) = \dot{M}_{\rm CGM,in}(t)  + \dot{M}_{\rm outflow}(t) - \dot{M}_{\rm inflow}(t) - \dot{M}_{\rm CGM,out}(t),
\end{equation}
where $\dot{M}_{\rm CGM,in}$ is the inflow rate from the external intergalactic medium into the CGM, $\dot{M}_{\rm outflow}$ is the mass removed from the central galaxy and added to the CGM via outflow, $\dot{M}_{\rm inflow}$ is the gas that flows into the galaxy from the CGM, and $\dot{M}_{\rm CGM,out}$ is the outflow rate of gas from the CGM into the intergalactic medium. The evolution of the galactic gas mass is then defined as (\citealt{Tinsley1980}; \citealt{Pagel1997}; \citealt{Matteucci2012}) \begin{equation}\label{eq:gas_evol}
    \dot{M}_{\rm gas}(t) = \dot{M}_{\rm inflow}(t)  + \dot{M}_{\rm ej}(t) - \dot{M}_\star(t) - \dot{M}_{\rm outflow}(t),
\end{equation}
where $\dot{M}_{\rm inflow}$ is the mass added by galactic inflows from the CGM, $\dot{M}_{\rm ej}$ is the mass added by stellar ejecta, $\dot{M}_\star$ is the mass locked away by star formation, and $\dot{M}_{\rm outflow}$ is the mass lost by outflows into the CGM. Gas infalls into the galaxy from the CGM using a dual-infall model based on \citet{Chiappini1997}. This prescription for gas infall combines two exponential gas inflow events and is described as 
\begin{equation} \label{eq:infall}
    \dot{M}_{inflow}(t)=A_{1}{exp}\left(-\frac{t}{\tau_{1}}\right) + A_{2}{exp}\left(\frac{t_{{max}} - t}{\tau_{2}}\right),
\end{equation}
where A$_{1}$, A$_{2}$, $\tau_{1}$, $\tau_{2}$ , and t$_{max}$ are free parameters that are presented in Table \ref{tab:model_properties}.

\begin{table}
    \centering
    \caption{Parameter values of the model.}
    \begin{tabular}{l c} \hline\hline
        Parameter & Value \\
        \hline
        A$_{1}$ [M$_{\odot}$\,yr$^{-1}$] & 46 \\
        A$_{2}$ [M$_{\odot}$\,yr$^{-1}$] & 5.9 \\
        $\tau_{1}$ [Gyr] &  0.8 \\
        $\tau_{2}$ [Gyr] & 7.0 \\
        t$_{\rm{max}}$ [Gyr] & 1.0\\
        $\epsilon_\star$& 0.23\\
        $\tau_\star$ [Gyr] & 1.0\\
        $\eta$ & 0.52\\ \hline
    \end{tabular}
    \tablefoot{A$_{1}$, A$_{2}$, $\tau_{1}$, $\tau_{2}$ , and t$_{\rm{max}}$ are all free parameters of Eq. (\ref{eq:infall}). $\epsilon_\star$ is the star formation efficiency, $\tau_\star$ is the star formation timescale, and $\eta$ is the mass-loading factor. These values are equivalent to the values in the best model of \citet{Cote2019}.}
    \label{tab:model_properties}
\end{table}

The star formation rate is defined as
\begin{equation}
    \dot{M}_\star(t)\,=\,\frac{\epsilon_\star}{\tau_\star}\,M_{\rm gas}(t)\,=f_\star\,M_{\rm gas}(t),
\end{equation}
where $\epsilon_\star$ and $\tau_\star$ are the dimensionless star formation efficiency (sfe) and star formation timescale, respectively. The outflow rate is proportional to the star formation rate and is defined as
\begin{equation}
    \dot{M}_{\rm outflow}(t)\,=\,\eta\,\dot{M}_\star(t),
\end{equation}
where $\eta$, the mass-loading factor, is a free parameter and controls the strength of the outflows. The values for $f_\star$ and $\eta$ are found in Table \ref{tab:model_properties}. To follow the chemical evolution, we calculated the mass of the gas added by stellar ejecta at each time step. To do this, simple stellar populations (SSPs) were generated at each time step by the code stellar yields for galactic modeling applications (SYGMA; \citealp{Ritter2018c}). An SSP is a population of stars with the same age and chemical composition, with the number of stars of each mass given by the initial mass function (IMF) of \citet{Kroupa2001}.

We can include yields from low- and intermediate-mass stars, massive stars, SNIa, and neutron star mergers. Other additional sources can be added manually by the user. Ejecta from SSPs is instantaneously and uniformly mixed into the gas reservoir at each time step. To calculate the mass of the gas added to the galaxy by stellar ejecta, the contribution of every stellar population formed by time $t$ is summed so that \begin{equation}
    \dot{M}_{\rm ej}(t)\,=\,\sum_{j}\,\dot{M}_{\rm ej}^{j}\,(M_{j}, Z_{j}, t-t_{j}),
\end{equation} where $\dot{M}_{\rm ej}^{j}$ is the mass ejected by the $j$th stellar population, $M_{j}$ is the initial mass of the population, $Z_{j}$ is the initial metallicity of the population, and $t-t_{j}$ is the age of the $j$th population at time $t$. The stellar yield sets used by each of these models are summarized in Table \ref{tab:yields}. In summary, all models use the FRUITY asymptotic giant branch (AGB) yields from \citet{Cristallo2015} and the SNIa yields of \citet{Iwamoto1999}. Model CLC uses the rapidly rotating, massive star yields of \citet{limongi:18}, where low-metallicity stars rotate faster than high-metallicity stars according to the mixed rotational velocity prescription of \citet{prantzos:18}. CRit(del) uses the massive star yields of \citet{Ritter2018b} with C-O shell mergers and the delayed explosion prescription, while CRit(rap) uses the same yields, but with the rapid explosion prescription. Finally, CNom uses the massive star yields of \citet{Nomoto2013}. %\marco{[MP: did you define before CRit(del), CRit(rap) and CNom? Check! Not sure that you need acronyms, since you just use them in this section.]}

\begin{table}
    \centering
    \caption{Combinations of yields used for the chemical evolution modeling.}
    \begin{tabular}{c c c c} \hline \hline
        Model Name & AGB & Massive Stars & SN Ia \\
        \hline
        CLC & FRUITY & L\&C V$_{\rm{rot}}$\,=\,mix & Iwamoto \\
        CRit(del) & FRUITY & Ritter delayed & Iwamoto \\
        CRit(rap) & FRUITY & Ritter rapid & Iwamoto \\
        CNom & FRUITY & Nomoto & Iwamoto \\ \hline
    \end{tabular}
    \tablefoot{FRUITY\,=\,\citet{Cristallo2015} and L\&C V$_{\rm{rot}}$\,=\,mix are the rapidly rotating massive stars of \citet{limongi:18}, Ritter delayed and Ritter rapid are from \citet{Ritter2018b}, Nomoto\,=\,\citet{Nomoto2013} and Iwamoto\,=\,\citet{Iwamoto1999}.}
    \label{tab:yields}
\end{table}

\begin{figure}
  \resizebox{\hsize}{!}{\includegraphics{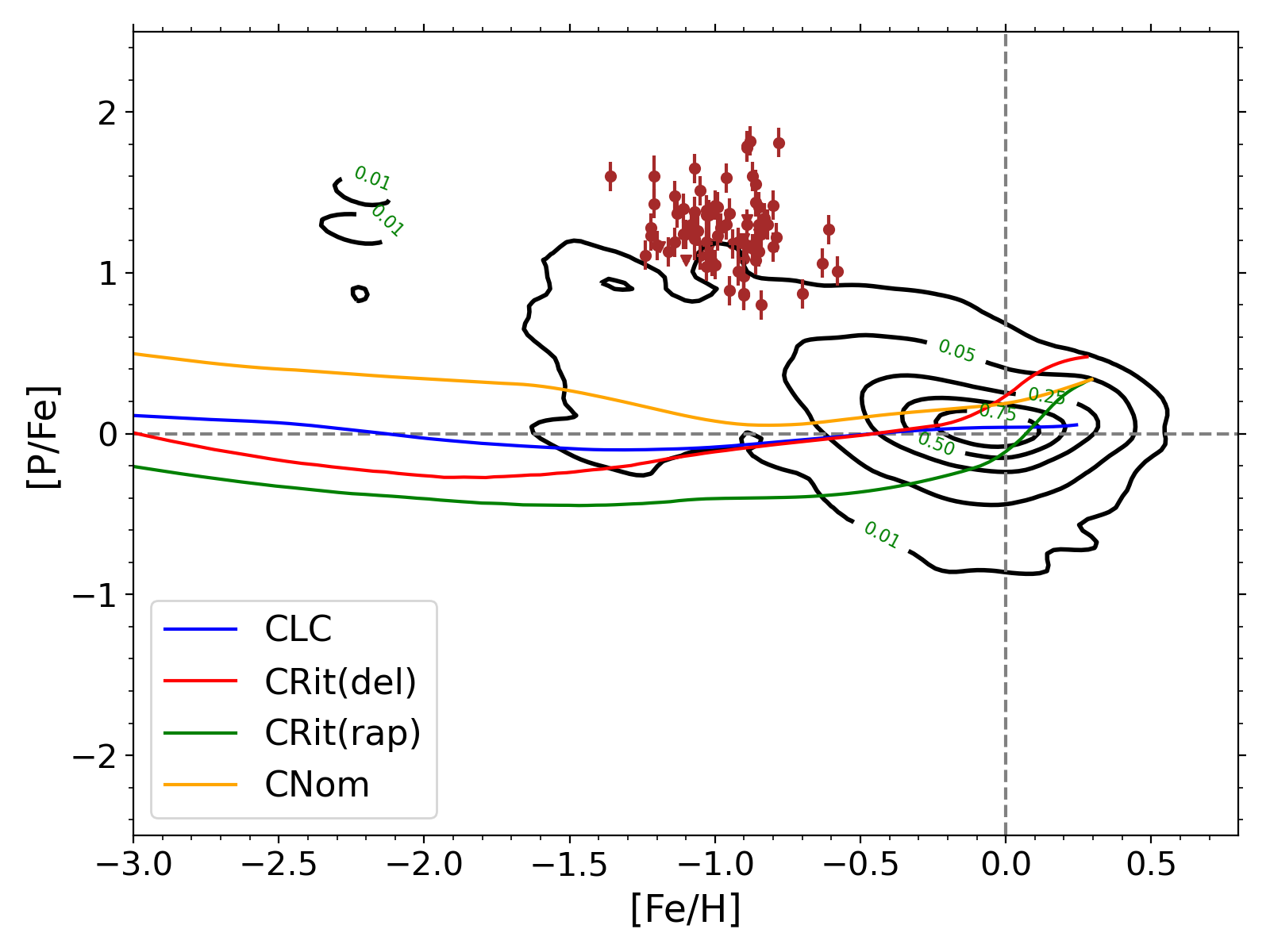}} 
  
  \resizebox{\hsize}{!}{\includegraphics{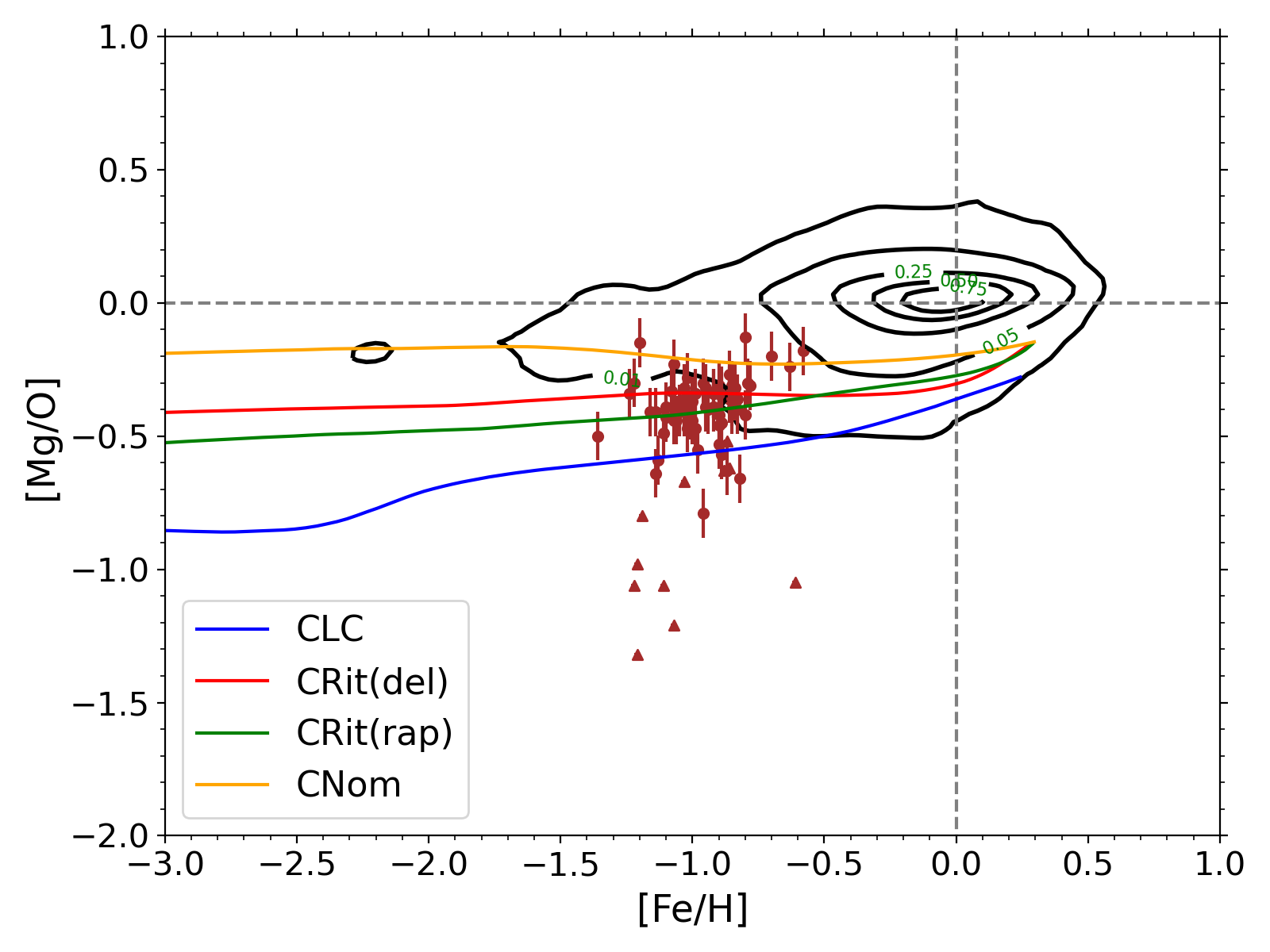}} 
  
  \resizebox{\hsize}{!}{\includegraphics{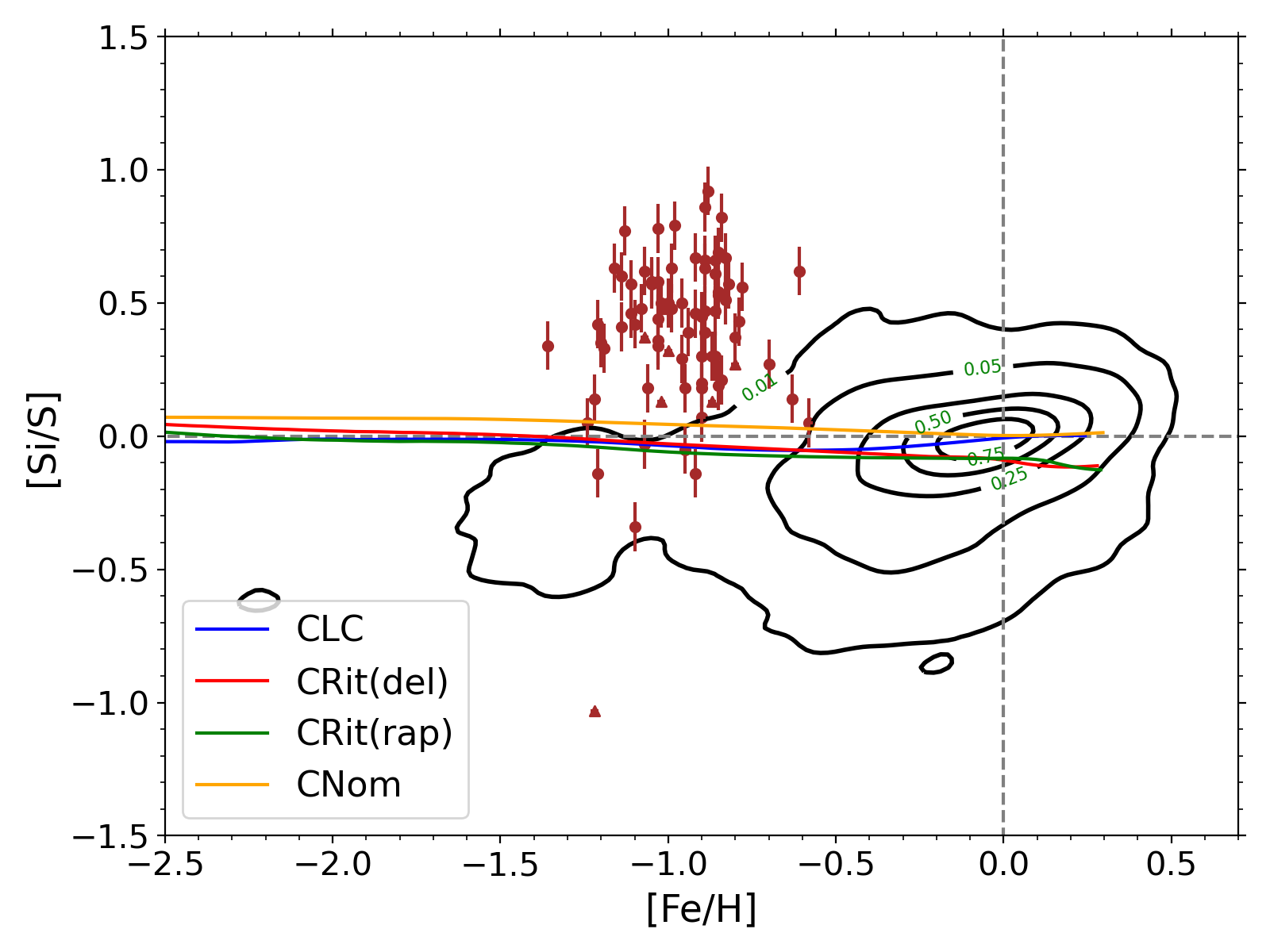}} 
  \caption{Predicted evolution curves. Top panel: [P/Fe] as a function of [Fe/H] in the Galactic disk is shown for four GCE simulations using \texttt{OMEGA+}: the models are CLC (blue line), CRit(del) (red line), CRit(rap) (green line) and CNom (yellow line). For comparison, our new sample of P-rich stars with errors is shown (brown points) together with data from \citet{Hinkel2014} as representative of P-normal stars (black contours). Middle panel: Same as the top panel, but for the [Mg/O] ratio. Bottom panel: Same as the top panel, but for the [Si/S] ratio. For each panel, upper limits in the P-rich sample are represented by downward triangles and lower limits by upward triangles. The black contours allow us to map the distribution density of the stellar abundances.}
  \label{fig:GCE}
\end{figure}

In Fig. \ref{fig:GCE} the results of the GCE simulations are shown compared to the observations of P-normal stars (\citealt{Hinkel2014})
and of P-rich stars. In the top panel, the abundance trend of [P/Fe] confirms what we discussed before for P-normal stars. While the solar [P/Fe] is reproduced from GCE models, theoretical predictions underproduce the [P/Fe] at metallicities lower than solar, consistently with previous results in the literature. In GCE calculations, most of the P is made in CCSN models by local neutron captures starting from Si isotopes, both before the CCSN explosion in the convective C shell and in explosive C-burning \citep[][]{woosley:95,woosley:02,pignatari:16}. Therefore, P is made in different parts of the ejecta compared to Fe, which instead is the main product of explosive Si burning. In these conditions, CCSN yields are the dominant source of uncertainties for GCE simulations \citep[e.g.,][]{romano:10, molla:15,mishenina:17,matteucci:21}. While the production of CCSN yields for GCE calculations are delivered by one-dimensional stellar models, several uncertainties still affect the last evolutionary stages of massive star progenitors \citep[see, e.g.,][]{ritter:18} and multidimensional simulations of CCSN explosions \citep[][and references therein]{burrows:21}. 
Concerning our analysis in this work, although the P-enrichment in P-rich stars is not a GCE product but most likely is a signature of local stellar nucleosynthesis, we can compare these stars with GCE results for qualitative purposes. As shown already by \cite{masseron20a}, the [P/Fe] predicted by GCE is about a factor of ten lower than P-rich stars.

In the middle panel of Fig. \ref{fig:GCE}, we show the evolution of the [Mg/O] ratio with metallicity. As we have discussed in the previous sections, in P-rich stars, both [O/Fe] and [Mg/Fe] may be enhanced with respect to solar. However, the [Mg/O] ratio in P-rich stars is rather consistent with other Milky Way stars considering both errors and lower limits. In particular, [Mg/O] is well reproduced by GCE simulations. The bulk of Mg and O in the Galaxy is made by massive stars \citep[e.g.,][]{woosley:02}. O is mostly made in He-burning conditions in the stellar progenitor by the $^{12}$C($\alpha$,$\gamma$)$^{16}$O reaction. Mg is made as $^{24}$Mg during hydrostatic shell C-burning evolving on the previous He-core ashes, via the $\alpha$-capture sequence $^{16}$O($\alpha$,$\gamma$)$^{20}$Ne($\alpha$,$\gamma$)$^{24}$Mg and directly from the C-fusion products $^{20}$Ne and $^{23}$Na \citep[e.g.,][]{chieffi:98, rauscher:02, pignatari:16, sukhbold:16, curtis:19}. They are both ejected together in the O-rich stellar layers, with only minor modification by CCSN nucleosynthesis \citep[][]{thielemann:96}. Because massive stars are the only relevant galactic source for Mg and O, their relative GCE ratio is therefore simply given by the relative ratios from the CCSN stellar yield sets used for the calculations. The GCE models in Fig. \ref{fig:GCE} show a [Mg/O] variation of about $\unit[0.2]{dex}$ at solar metallicities, which reflects the variation in the CCSN yields we adopted (see Table \ref{tab:yields}). This dispersion tends to increase with decreasing [Fe/H], in particular, for model CLC. This is due to the enhanced production of O in the external layers in massive star models induced by rotation \citep[e.g.,][]{limongi:18}, which are used in model CLC. In summary, while both Mg and O may be enhanced in P-rich stars, their relative ratio is overall consistent with the theoretical ratios from typical CCSN yields used for GCE calculations, and it is consistent with ratios obtained in shell C-burning stellar layers. 

In the bottom panel, we finally show the evolution of the [Si/S] ratio with [Fe/H]. Even though different CCSN yield sets are used, the GCE models show a solar [Si/S], with a variation smaller than 0.1 dex between them. Si and S in the Galaxy are mostly made by CCSNe, with a relevant contribution from SNIa \citep[e.g.,][]{kobayashi:20}. For all the CCSNe models we considered, Si and S are produced in the same CCSN layers and are ejected together, mostly as a direct product of explosive O-burning \citep[e.g.,][]{pignatari:16}. Therefore, GCE models typically carry the same [Si/S] ratios as are generated in these layers from the CCSN yields. Instead, a significant fraction of P-rich stars tend to show a [Si/S] higher than solar due to the Si-enrichment. This decoupling in the nucleosynthesis signature of Si and S observed in P-rich stars is indeed puzzling and difficult to match with typical stellar models. Si and S are also produced in the same parts of the ejecta in SNIa, showing a similar relative production for different progenitor masses
\citep[e.g.,][]{Iwamoto1999, gronow:20, nomoto:18}. We note that P-normal stars at metallicities [Fe/H]$\gtrsim$-0.5 seem to show a significant [Si/S] variation, between $\unit[-0.5]{dex}$ and supersolar [Si/S] up to about $\unit[0.5]{dex}$. Therefore, we cannot exclude that a small number of P-normal stars has anomalous [Si/S] ratios.   

To summarize, P-rich stars represent a really interesting challenge for nuclear astrophysics. For instance, they might be the main fingerprints of a new P stellar nucleosynthesis component that is still missing in GCE calculations. Since the production of P is disentangled from the production of Fe, it is in principle simple to find stellar sources with high [P/Fe] abundance ratios. However, in this section, we discussed the anomalous [Si/S] ratio as an example, which provides strong constraints for defining the correct astrophysics scenario that explains these stars. We defer the detailed analysis of the observed elemental ratios in comparison with theoretical stellar simulations to a future paper. We showed that the observed enhancement in both O and Mg, combined in a rather normal [Mg/O] ratio, is a strong indication that massive stars might be the source of the nucleosynthesis anomalies carried by P-rich stars. However, the details of the astrophysics context still need to be defined.
%______________________________________________________________

\subsection{Orbital analysis\label{Orbits}}

\begin{figure}[htb!]
  \resizebox{\hsize}{!}{
  \includegraphics[width=\hsize]{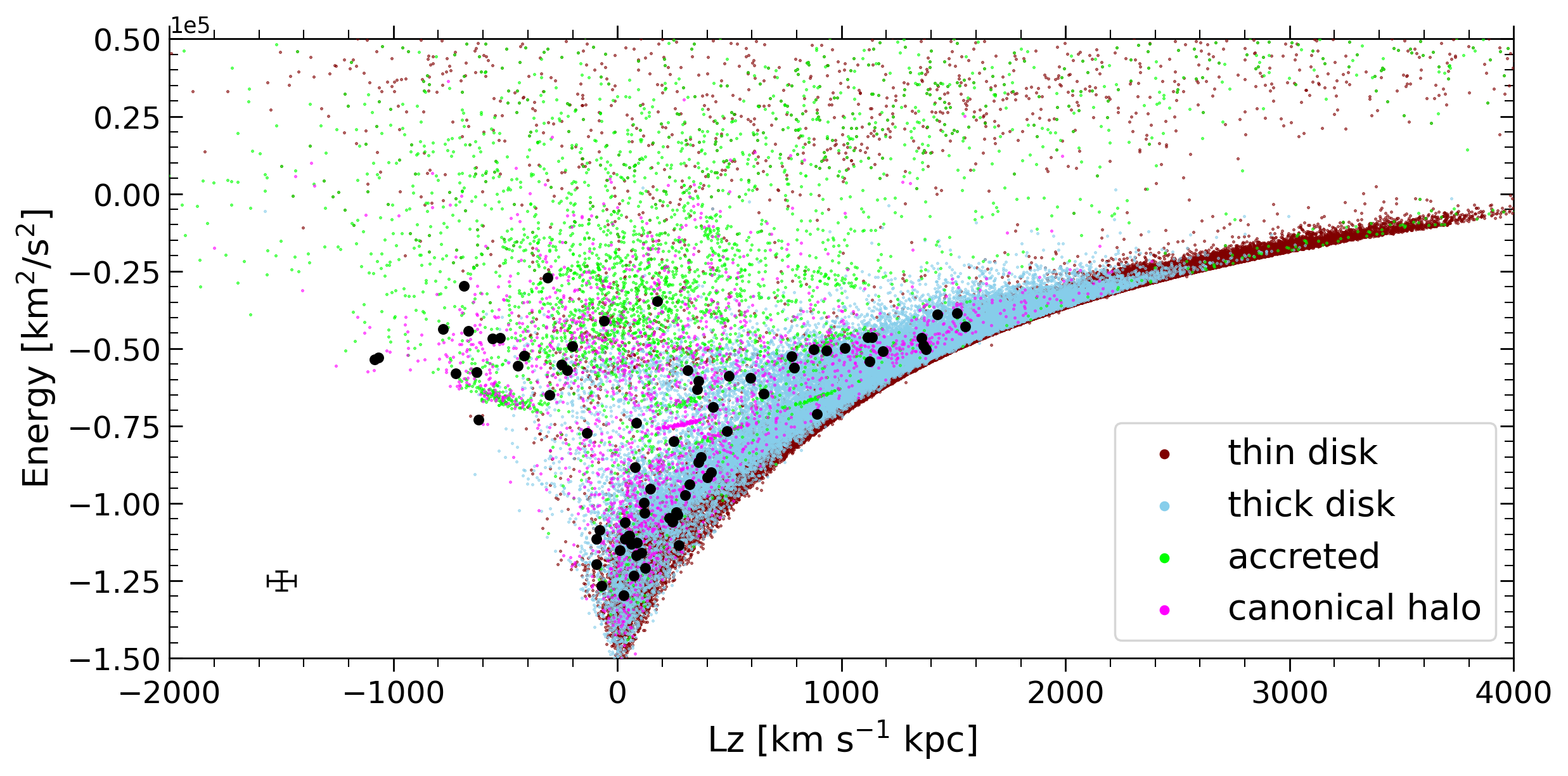}} 
  
  \resizebox{\hsize}{!}{
  \includegraphics[width=\hsize]{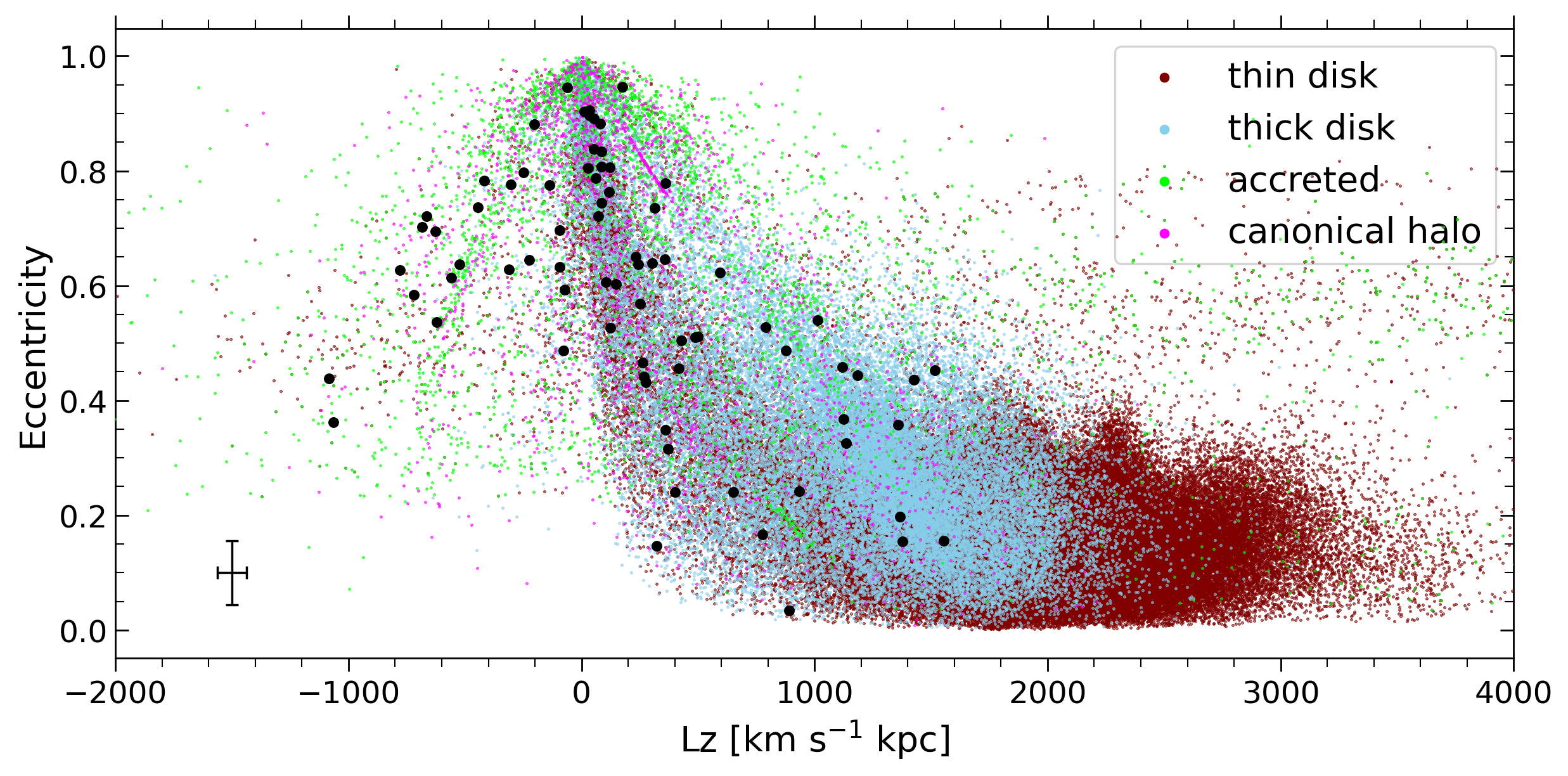}} 
  
  \resizebox{\hsize}{!}{
  \includegraphics[width=\hsize]{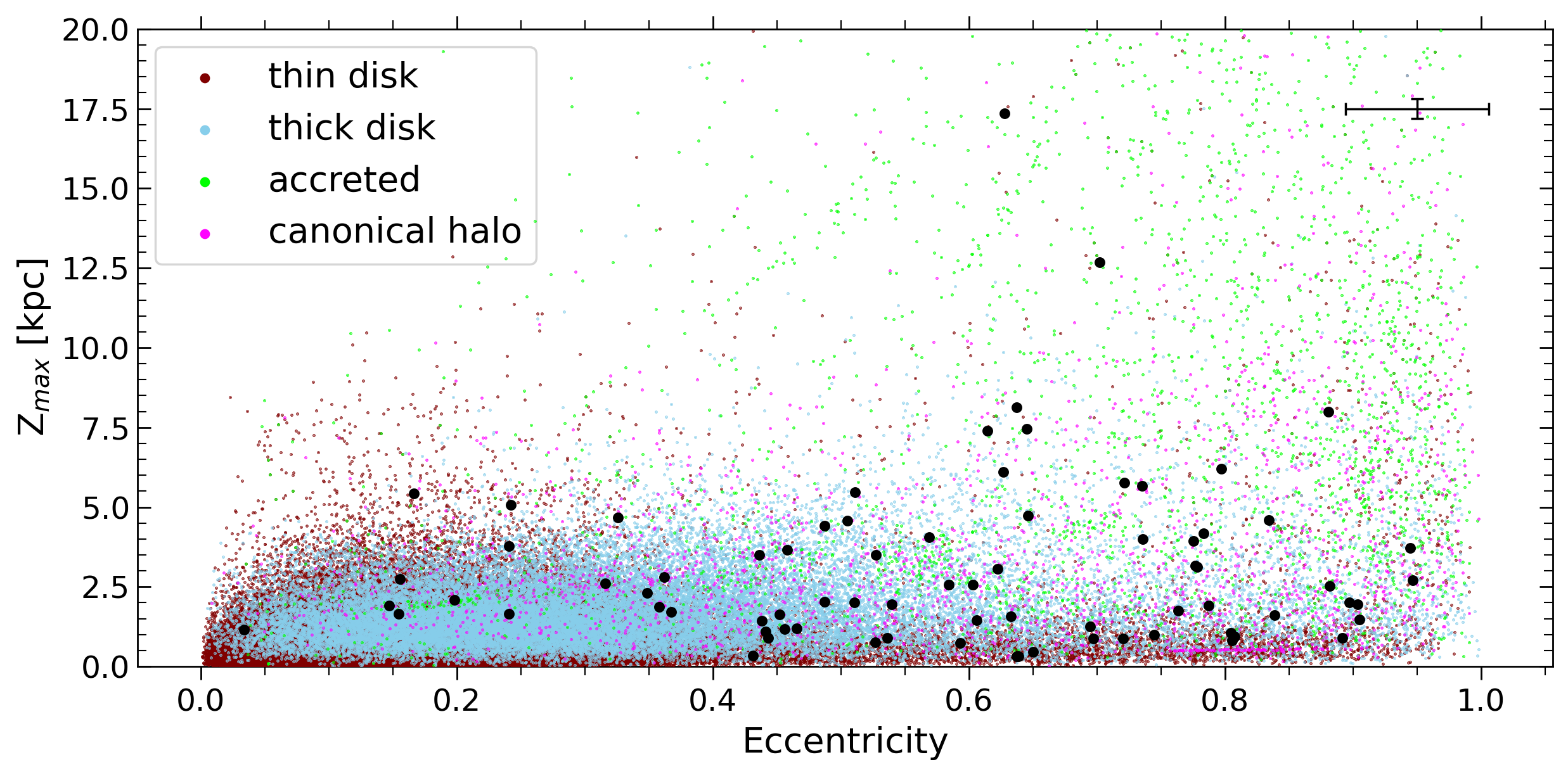}}
  \caption{P-rich stars (black dots) plotted against a background sample in three different dynamical planes: total energy vs. angular momentum $L_z$ (top panel), eccentricity vs. $L_z$ (middle panel), and maximum height above the midplane $Z_{\text{max}}$ vs. eccentricity (bottom panel). To separate the background sample into thin-disk, thick-disk, canonical halo, and accreted halo stars, we used the results from \citet{hawkins15}, \citet{masseron15}, and \citet{hayes18} (see main text for more details). For clarity, we give a typical error bar for the P-rich stars corresponding to the average error of their parameters.}
  \label{Fig:Orbits}
\end{figure}

Taking advantage of the new sample size, we analyzed the dynamics of the P-rich stars using the astroNN\footnote{The VAC can be found at \url{https://github.com/henrysky/astroNN}} VAC described in \citet{abdurrouf22APOGEEDR17}, and references therein. Included in this VAC, among other properties, are the stellar orbits in the Milky Way, determined by applying the deep learning code astroNN to the APOGEE-2 DR17 data. More precisely, the orbital parameters of the VAC were calculated by the fast method from \citet{mackereth18VAC2} using the gravitational potential \verb|MWPotential2014| \citep{bovy15VAC3}. \\
A cross-match between the astroNN VAC and our sample enabled us to extract the orbital parameters and plot them in three dynamical planes shown in Fig. \ref{Fig:Orbits}: total energy versus angular momentum $L_z$, eccentricity versus $L_z$, and maximum height above the midplane $Z_{\text{max}}$ versus eccentricity. The P-rich stars are plotted against a sample of S/N>70 background stars that are color-coded according to the Galactic components. To separate the background sample into thin- and thick-disk stars, we used the chemical information contained in the [$\alpha$/Fe]-[Fe/H] plots presented in Fig. 1 of \citet{hawkins15} and Fig. 4 of \citet{masseron15}. We dropped stars from the intersection region. To separate the canonical halo from the accreted population, we estimated and adopted the cuts made in Fig. 4 of \citet{hayes18}. For clarity, we used the average error of the parameters over the P-rich sample as a typical error bar in Fig. \ref{Fig:Orbits}. Regarding the overall uncertainty, \citet{abdurrouf22APOGEEDR17} stated that the orbital parameters have a precision between 4\% and 8\%. \\     
From the distribution of the P-rich stars in the orbital plots, we conclude that most of the P-rich stars belong to the thick disk and the inner Galactic halo, without any particular motion. This means that the P-rich stars cannot be assigned exclusively to any dynamical subgroup from either an in situ or an accreted origin. However, some P-rich stars are located in regions of accreted material in the planes shown in Fig. \ref{Fig:Orbits}. Our rather rough comparison with the background sample does not allow us to completely discard the possibility that some of the P-rich stars might be the result of a merger event. In a recent work, \citet{dodd22} identified numerous subgroups of stars in the Milky Way based on similarities in their dynamical and chemical properties and interpreted them as remnants of past merger events. For future works, it might be of interest to determine whether some of the P-rich stars belong to the subgroups reported by \citet{dodd22}.  \\
In search of the progenitor, the source of the P-richness, this dynamical diagnostics shows us that the mechanism that causes the overabundance of P is not bound to a specific location or population. Although some P-rich stars might have been accreted from external galaxies, which is in agreement with the low-metallicity range, the P-richness cannot be attributed to ex situ formation alone.

\subsection{Nucleosynthetic scenarios\label{Scenarios}}
With a reliable chemical fingerprint of the P-rich stars, we are able to compare it with other chemical signatures. As the main objective of this work is to find clues on the P-rich star progenitor, we compared our results with nucleosynthetic scenarios that might have caused the strong P pollution of the ISM, out of which the observed low-mass P-rich stars were born. 
As mentioned in Sect. \ref{Introduction}, \citet{masseron20a} already discarded a number of possible P-rich star progenitors, including super-AGB stars and novae, through a comparison with their nucleosynthetic theoretical
predictions. They tentatively suggested that investigating possible
combinations of specific effects, such as rotation \citep[e.g.,][]{prantzos:18} and/or nucleosynthesis
in convective-reactive regions \citep[e.g.,][]{ritter:18} in massive stars, might represent a promising way to explain the peculiar chemical pattern of the P-rich stars. In the following subsections, we continue the discussion from \citet{masseron20a,masseron20b} and focus on three nucleosynthetic scenarios that have not been considered until now
or have been suggested recently.

\subsubsection{Sub-Chandrasekhar supernovae Ia}
It is known that the final yields derived from SNe are bound to numerous parameters, such as the initial mass and the metallicity. A striking example for the strong dependence on the initial conditions are SNeIa, the thermonuclear explosion of carbon-oxygen white dwarfs (WDs) \citep[][and references therein]{sanders21,bravo22}. The properties of these SN explosions depend on the mass of the WD at the time of explosion, which can fall into one of the following two categories: (i) the mass is close to the Chandrasekhar-limit (M $\approx$ 1.4 M$_\odot$) (Ch-m WD), or (ii) the mass is below this limit (M $\lesssim$ 1.4 M$_\odot$) (subCh-m WD). \citet{sanders21} explored the abundance signature of SNeIa by fitting the chemical evolution models from \citet{weinberg17} to the Gaia-Enceladus chemical space. They speculated that subCh-m WD channels constitute a significant fraction of SNeIa in metal-poor systems. This speculation was motivated by their prediction of a subsolar production of Mn and Ni ([Mn/Fe] $\approx \unit[-0.15]{dex}$ and [Ni/Fe] $\approx \unit[-0.3]{dex}$) in these objects. Although P has not been measured in relation with subCh SNe so far, we used the information from \citet{sanders21} to find indications of a possible contribution to the P-rich stars because, as we have shown in the previous section, some P-rich stars might have been part of or may have formed from material of accreted dwarf galaxies where subCh SNe occur. The metallicity of these systems also matches the metallicity range of the P-rich stars. Thus, we compared the predicted Mn and Ni abundances from \citet{sanders21} with the corresponding abundances of the P-rich stars, using the abundances provided by ASPCAP \citep{garciaperez16ASPCAP,abdurrouf22APOGEEDR17}. In the case of Mn, the P-rich stars reproduce the predicted yield with a mean [Mn/Fe] of $\unit[-0.15\pm 0.03]{dex}$. In contrast to Mn, the mean Ni abundance of the P-rich stars is slightly supersolar ($\unit[0.07\pm 0.03]{dex}$) and therefore does not match the result from \citet{sanders21}. We note that the Mn and Ni abundances were determined via an automatic pipeline and thus have to be handled carefully. However, we tentatively conclude that subCh SNe are not the primary polluter.

\subsubsection{Pair-instability supernovae}
Pair-instability SNe (PISNe) are the thermonuclear explosion of very massive stars (M > 100 M$_\odot$) induced by oxygen burning after a gravitational collapse caused by the generation of electron-positron pairs \citep[see, e.g.,][and references therein]{kozyreva14}. PISNe of very massive metal-poor stars are suggested to be relevant in the local Universe \citep{kozyreva14}, even though an observational proof for this type of SN event is still lacking \citep{takahashi18}. One requirement for a PISN to occur is the formation of a high-mass CO core. In metal-rich environments, the formation of this core is hindered by the high mass-loss rates caused by stellar winds \citep{georgy17,takahashi18}. The P-rich stars considered in this work are found in a narrow low-metallicity range, which means that massive stars causing PISNe are a possible progenitor. Another aspect indicating a massive progenitor is the identification of a GC P-rich star, which most likely belongs to the first generation. This discovery, if confirmed, would indicate that the progenitor has evolved on a shorter timescale than the GC, meaning that only a massive progenitor can be considered. In addition, a significant fraction of the P-rich stars are most likely not part of close binary systems due to their low RV scatter. Because PISNe do not leave any remnant (i.e., a possible binary companion of the P-rich stars), it could be speculated that
these very massive stars exploding as a PISN are potential candidates for the P-rich star progenitor. \\
Although PISNe seem to be promising candidates for the P-rich star progenitor based on the arguments above, comparing the abundance pattern of the PISNe with the P-rich stars causes us to doubt this scenario. Stellar evolution simulations performed by \citet{kozyreva14} and \citet{takahashi18} revealed that PISN exhibit a deficiency of odd-Z elements compared to even-Z elements, meaning that the production of P is reduced compared to, for example, Mg, Si, and S. This so-called odd-even effect is not observed in our sample. According to \citet{kozyreva14}, the odd-even effect is weakened with increasing metallicity due to a stronger contribution of CCSNe. Regarding the observed high amounts of P and Al in our sample, weakening the odd-even effect would not be enough to match the abundance patter, making it less plausible that PISNe are the progenitor of the P-rich stars.

\subsubsection{Nonthermal nucleosynthesis}
Since the discovery of the P-rich stars, a new theory beyond standard nucleosynthesis models emerged, aiming to account for the unusual P abundances. \citet{goriely22} explored the irradiation by stellar energetic particles (SEPs) as a nonthermal nucleosynthesis process and found that, globally, the relative abundance pattern of the P-rich stars is reproduced to a great extent. Four different spallation events have been considered by \citet{goriely22}, consisting of either $\alpha$-particles only or $\alpha$-particles in combination with protons as SEPs, and either CNO in solar ratio or pure C as target material. The results show that particles with energies of a few $\unit{MeV}$ are able to produce significant amounts of P. However, there are some main discrepancies between the P-rich stars abundance pattern and the results from \citet{goriely22}. In particular, the [Ba/La] ratio as well as the Al and Si abundances are strongly underpredicted by the considered spallation models. In the case of Al and Si, \citet{goriely22} proposed possible corrections that can be applied to improve the agreement, such as modifying the target material or adding an extra SEP component. Nevertheless, there are only two P-rich stars with measurements of Ba and La \citep{masseron20b}, and therefore, it would be useful to increase this number to have more reliable evidence that this discrepancy indeed exists. \citet{goriely22} also stated that in addition to P, a significant amount of F and Cl is produced in this type of process. However, thus far, we lack observational data for these elements. Furthermore, several questions about the nature of these processes are left unanswered by \citet{goriely22} and will require further studies. The source and the acceleration mechanism of the SEPs, as well as quantitative declarations about the SEPs and the target material that needs to be affected by the SEPs to reproduce the abundances, are not reported. The definition of the site where these spallation processes occur is still lacking. In particular, \citet{goriely22} left open whether the atmosphere of the P-rich stars has been polluted by the outcomes of the spallation processes or if it was directly irradiated by the SEPs. Moreover, the output produced by a superposition of different events and the use of distinct target materials is still unexplored. 

\section{Summary and conclusions}
We presented a follow-up study of P-rich stars that began with the discovery of 16 giants exhibiting unusually high abundances of P, as well as enhancements in O, Al, Si, and Ce \citep{masseron20a}. Because these P-rich stars are low-mass giants, it is most likely that a progenitor caused the observed abundance pattern by polluting the gas out of which the stars were born. To draw statistically reliable conclusions on the progenitor of these chemically peculiar stars, we carried out a chemical abundance analysis on an enlarged sample, mainly consisting of a collection of Si-rich giants \citep{fernandeztrincado19,fernandeztrincado20} that were suspected to be P-rich as well. The final group of 78 giants represents the largest sample of confirmed P-rich stars to date. Using the NIR (H-band) observations from DR17 of the APOGEE-2 survey and the BACCHUS code, we determined the abundances of 13 elements in the stars of the new sample. Finally, we compared our abundance results with detailed GCE models, each using a different prescription for the massive star yields, performed an orbital diagnostic, and discussed possible progenitor candidates. \\
Our main results, interpretations and recommendations for future investigations are summarized in the following bullet points:
\begin{itemize}
    \item The chemical abundance analysis of our enlarged sample revealed the unique fingerprint of the P-rich stars, supporting the abundance pattern that was previously found by \citet{masseron20a}, with high abundances of O, Al, Si, P, and Ce. Consequently, the hypothesis made by \citet{masseron20a} that P-rich stars are part of a larger family of Si- and Al-rich stars is confirmed. We also validated that S and Ca are not enriched, which is at odds with current stellar nucleosynthesis models of massive stars because Si is enhanced instead. The low-metallicity range in which the P-rich stars from \citet{masseron20a} are located is also tentatively confirmed because no P-rich stars with [M/H] > $\unit[-0.6]{dex}$ have been detected during this work.
    \item In addition, we identified several correlations among the enriched elements as well as between Mg and P, converting Mg to a characteristic component of the P-rich stars abundance pattern even though it is not enhanced in the entire sample. In contrast, no correlation between the C+N abundances was detected, although some stars show a C+N enhancement. 
    \item We highlight the correlation between P and Si, which is particularly useful for the search of P-rich stars in large data sets. For example, the performance of machine-learning (ML) techniques could be improved if the algorithm is trained on the more accessible Si lines instead of on P lines. As there was no systematic search for P-rich stars in DR17, it would be interesting to perform this search with improved ML algorithms, bearing in mind that P is correlated with other elements.
    \item \citet{nandakumar22} also discovered two P-rich stars that are enriched in Ce, but not in other elements, however, as reported by \citet{masseron20a,masseron20b}. It still has to be determined whether these two stars and their origin can be related to the P-rich stars from \citet{masseron20a,masseron20b} or if they constitute a new class of P-rich stars. Furthermore, the automated supercomputer calculation of the background sample uncovered 139 stars with [P/Fe] > $\unit[0.8]{dex}$. Although we did not rate the results for P as reliable enough to be used for comparison, it might be reasonable to revisit these stars in order to validate the high abundances and expand the P-rich sample.
    \item From the analysis of the RV scatter, we conclude that the P-rich stars are not exclusively binaries. Although four stars have been found to be likely binaries due to their high RV scatter, the majority of the sample stars shows a low RV scatter. Depending on the time-lag cut between individual visits, the low RV scatter is reliable for a significant fraction, allowing us to conclude that the sample is not composed of close binaries alone, but also of field stars. Hence, the peculiar abundance pattern cannot be assigned to close binary interaction.
    \item We made four GCE models of the MW disk using different stellar yields. We confirm that P-rich stars are not reproduced. In addition to the high [P/Fe], the P-rich stars also show an anomalous high [Si/S] ratio with respect to P-normal stars and GCE simulations, which is difficult to reproduce by standard nucleosynthesis in supernovae.
    \item The orbital analysis implies that the P-rich stars cannot be assigned to one specific population alone. Instead, they can be found in accreted streams as well in the in situ population of the Milky Way. This indicates a universal process that leads to the unusual abundance pattern and that is not locally bound.
    \item By comparison of the abundance pattern with possible progenitor candidates, we discard the scenario of sub-Chandrasekhar SNeIa. We considered this type of SNeIa because they share the same metallicity range as the P-rich stars and are thought to be significant contributors to metal-poor dwarf galaxies, out of which some P-rich stars might have been accreted. Because the Mn and Ni abundances of the P-rich stars do not match the predicted yields from subCh SNeIa, we ruled out subCh SNeIa as possible progenitor, however.
    \item The assignment of one P-rich star to the GC M4 motivated us to study the possibility of a massive progenitor that ends its life through a pair-instability SNe. We suggest that the P-rich star in M4 belongs to the first-generation GC stars, indicating that the corresponding progenitor has evolved faster than the GC itself. This in turn indicates a massive origin. Predictions show that the odd-even effect is typical for PISNe, but this effect is not observed in the P-rich stars, leading us to discard PISNe as a possible progenitor.
    \item As a next observational step, we recommend the expansion of the chemical space to explore heavier elements, for example, \textit{s}-process elements such as Ba and Pb. Observations in the optical range that supplement the APOGEE-2 data, such as those from \citet{masseron20b}, have to be carried out on a larger set of P-rich targets. In particular, further observations of the P-rich star that is part of M4 would be of interest in either the optical or another NIR band, to confirm the results found with the APOGEE-2 data.
    \item A possible alternative to standard nucleosynthesis was presented recently by \citet{goriely22}, who studied nonthermal nucleosynthesis in the form of spallation processes that are able to produce a pattern consistent with the P-rich stars to a great extent. There are still some
    discrepancies, for example, in the [Ba/La] ratio, but [Ba/La] measurements are lacking for an extended sample of P-rich stars. Other indicators could be the F and Cl abundances, which are predicted to be high in P-rich stars if spallation is the mechanism that causes the high P abundances. Unfortunately, F and Cl cannot be measured with the available spectral data. Thus, we emphasize the need for observations to explore the viability of this theory. Furthermore, this process would also need to be tested quantitatively within a GCE context, in particular, if it might affect the ISM at least locally and the abundance of the rare P-rich stars.
\end{itemize}
Following from the conclusions listed above, the next step on the search for the P-rich star progenitor should consist of the exploration of more (heavy) elements to constrain the astrophysical origin. This may be achieved through observations in spectral ranges other than the NIR.

%______________________________________________________________

\begin{acknowledgements}
    The author is grateful for the positive feedback from the anonymous referee that helped to improve this paper. MB acknowledges financial  support  from  the European Union and the State Agency of Investigation of the Spanish Ministry of  Science  and  Innovation  (MICINN)  under  the  grant PRE-2020-095531 of the Severo Ochoa Program for the Training of Pre-Doc Researchers (FPI-SO). 
    MB and TM acknowledge support from the ACIISI, Consejer\'{i}a de Econom\'{i}a, Conocimiento y Empleo del Gobierno de Canarias and the European Regional Development Fund (ERDF) under grant with reference  PROID2021010128. TM also acknowledges financial support from the Spanish Ministry of Science and Innovation (MICINN) through the Spanish State Research Agency, under the Severo Ochoa Program 2020-2023 (CEX2019-000920-S).
    This research made use of computing time available on the high-performance computing systems at the Instituto de Astrofísica de Canarias. The author thankfully acknowledges the technical expertise and assistance provided by the Spanish Supercomputing Network (Red Española de Supercomputacion), as well as the computer resources used: the LaPalma Supercomputer, located at the Instituto de Astrofísica de Canarias. KAW acknowledges the support of the European Union's Horizon 2020 research and innovation programme (ChETEC-INFRA -- Project no. 101008324) and ongoing access to \tt viper\rm, the University of Hull's High Performance Computing Facility. MP and ML thank the "Lendulet-2014" Program of the Hungarian Academy of Sciences (Hungary) and the ERC Consolidator Grant (Hungary) funding scheme (Project RADIOSTAR, G.A. n. 724560).\\
    The data processing of this work relies on the Pandas \citep{reback2020pandas}, NumPy \citep{harris2020array}, Astropy \citep{astropy:2013,astropy:2018,astropy:2022} and Matplotlib \citep{hunter07} Python libraries. \\
    This work has made use of Sloan Digital Sky Survey IV (SDSS-IV) data. Funding for the Sloan Digital Sky Survey IV has been provided by the Alfred P. Sloan Foundation, the U.S. Department of Energy Office of Science, and the Participating Institutions. SDSS-IV acknowledges support and resources from the Center for High Performance Computing  at the University of Utah. The SDSS website is \url{www.sdss4.org}. SDSS-IV is managed by the Astrophysical Research Consortium for the Participating Institutions of the SDSS Collaboration including the Brazilian Participation Group, the Carnegie Institution for Science, Carnegie Mellon University, Center for Astrophysics | Harvard \& Smithsonian, the Chilean Participation Group, the French Participation Group, Instituto de Astrofísica de Canarias, The Johns Hopkins University, Kavli Institute for the Physics and Mathematics of the Universe (IPMU) / University of Tokyo, the Korean Participation Group, Lawrence Berkeley National Laboratory, Leibniz Institut f\"ur Astrophysik Potsdam (AIP),  Max-Planck-Institut f\"ur Astronomie (MPIA Heidelberg), Max-Planck-Institut f\"ur Astrophysik (MPA Garching), Max-Planck-Institut f\"ur Extraterrestrische Physik (MPE), National Astronomical Observatories of China, New Mexico State University, New York University, University of Notre Dame, Observat\'ario Nacional / MCTI, The Ohio State University, Pennsylvania State University, Shanghai Astronomical Observatory, United Kingdom Participation Group, Universidad Nacional Aut\'onoma de M\'exico, University of Arizona, University of Colorado Boulder, University of Oxford, University of Portsmouth, University of Utah, University of Virginia, University of Washington, University of Wisconsin, Vanderbilt University, and Yale University.
\end{acknowledgements}

%______________________________________________________________

\bibliographystyle{aa} % style aa.bst
\bibliography{Refs} % your references Yourfile.bib

\begin{thebibliography}{115}
\expandafter\ifx\csname natexlab\endcsname\relax\def\natexlab#1{#1}\fi

\bibitem[{{Abdurro'uf} {et~al.}(2022){Abdurro'uf}, {Accetta}, {Aerts}, {Silva
  Aguirre}, {Ahumada}, {Ajgaonkar}, {Filiz Ak}, {Alam}, {Allende Prieto},
  {Almeida}, {Anders}, {Anderson}, {Andrews}, {Anguiano}, {Aquino-Ort{\'\i}z},
  {Arag{\'o}n-Salamanca}, {Argudo-Fern{\'a}ndez}, {Ata}, {Aubert},
  {Avila-Reese}, {Badenes}, {Barb{\'a}}, {Barger}, {Barrera-Ballesteros},
  {Beaton}, {Beers}, {Belfiore}, {Bender}, {Bernardi}, {Bershady}, {Beutler},
  {Bidin}, {Bird}, {Bizyaev}, {Blanc}, {Blanton}, {Boardman}, {Bolton},
  {Boquien}, {Borissova}, {Bovy}, {Brandt}, {Brown}, {Brownstein}, {Brusa},
  {Buchner}, {Bundy}, {Burchett}, {Bureau}, {Burgasser}, {Cabang}, {Campbell},
  {Cappellari}, {Carlberg}, {Wanderley}, {Carrera}, {Cash}, {Chen}, {Chen},
  {Cherinka}, {Chiappini}, {Choi}, {Chojnowski}, {Chung}, {Clerc}, {Cohen},
  {Comerford}, {Comparat}, {da Costa}, {Covey}, {Crane}, {Cruz-Gonzalez},
  {Culhane}, {Cunha}, {Dai}, {Damke}, {Darling}, {Davidson}, {Davies},
  {Dawson}, {De Lee}, {Diamond-Stanic}, {Cano-D{\'\i}az}, {S{\'a}nchez},
  {Donor}, {Duckworth}, {Dwelly}, {Eisenstein}, {Elsworth}, {Emsellem},
  {Eracleous}, {Escoffier}, {Fan}, {Farr}, {Feng}, {Fern{\'a}ndez-Trincado},
  {Feuillet}, {Filipp}, {Fillingham}, {Frinchaboy}, {Fromenteau}, {Galbany},
  {Garc{\'\i}a}, {Garc{\'\i}a-Hern{\'a}ndez}, {Ge}, {Geisler}, {Gelfand},
  {G{\'e}ron}, {Gibson}, {Goddy}, {Godoy-Rivera}, {Grabowski}, {Green},
  {Greener}, {Grier}, {Griffith}, {Guo}, {Guy}, {Hadjara}, {Harding},
  {Hasselquist}, {Hayes}, {Hearty}, {Hern{\'a}ndez}, {Hill}, {Hogg},
  {Holtzman}, {Horta}, {Hsieh}, {Hsu}, {Hsu}, {Huber}, {Huertas-Company},
  {Hutchinson}, {Hwang}, {Ibarra-Medel}, {Chitham}, {Ilha}, {Imig}, {Jaekle},
  {Jayasinghe}, {Ji}, {Johnson}, {Jones}, {J{\"o}nsson}, {Katkov}, {Khalatyan},
  {Kinemuchi}, {Kisku}, {Knapen}, {Kneib}, {Kollmeier}, {Kong}, {Kounkel},
  {Kreckel}, {Krishnarao}, {Lacerna}, {Lane}, {Langgin}, {Lavender}, {Law},
  {Lazarz}, {Leung}, {Leung}, {Lewis}, {Li}, {Li}, {Lian}, {Liang}, {Lin},
  {Lin}, {Lin}, {Lintott}, {Long}, {Longa-Pe{\~n}a}, {L{\'o}pez-Cob{\'a}},
  {Lu}, {Lundgren}, {Luo}, {Mackereth}, {de la Macorra}, {Mahadevan},
  {Majewski}, {Manchado}, {Mandeville}, {Maraston}, {Margalef-Bentabol},
  {Masseron}, {Masters}, {Mathur}, {McDermid}, {Mckay}, {Merloni},
  {Merrifield}, {Meszaros}, {Miglio}, {Di Mille}, {Minniti}, {Minsley},
  {Monachesi}, {Moon}, {Mosser}, {Mulchaey}, {Muna}, {Mu{\~n}oz}, {Myers},
  {Myers}, {Nadathur}, {Nair}, {Nandra}, {Neumann}, {Newman}, {Nidever},
  {Nikakhtar}, {Nitschelm}, {O'Connell}, {Garma-Oehmichen}, {Luan Souza de
  Oliveira}, {Olney}, {Oravetz}, {Ortigoza-Urdaneta}, {Osorio}, {Otter},
  {Pace}, {Padilla}, {Pan}, {Pan}, {Parikh}, {Parker}, {Peirani}, {Pe{\~n}a
  Ram{\'\i}rez}, {Penny}, {Percival}, {Perez-Fournon}, {Pinsonneault},
  {Poidevin}, {Poovelil}, {Price-Whelan}, {B{\'a}rbara de Andrade Queiroz},
  {Raddick}, {Ray}, {Rembold}, {Riddle}, {Riffel}, {Riffel}, {Rix}, {Robin},
  {Rodr{\'\i}guez-Puebla}, {Roman-Lopes}, {Rom{\'a}n-Z{\'u}{\~n}iga}, {Rose},
  {Ross}, {Rossi}, {Rubin}, {Salvato}, {S{\'a}nchez}, {S{\'a}nchez-Gallego},
  {Sanderson}, {Santana Rojas}, {Sarceno}, {Sarmiento}, {Sayres}, {Sazonova},
  {Schaefer}, {Schiavon}, {Schlegel}, {Schneider}, {Schultheis}, {Schwope},
  {Serenelli}, {Serna}, {Shao}, {Shapiro}, {Sharma}, {Shen}, {Shetrone}, {Shu},
  {Simon}, {Skrutskie}, {Smethurst}, {Smith}, {Sobeck}, {Spoo}, {Sprague},
  {Stark}, {Stassun}, {Steinmetz}, {Stello}, {Stone-Martinez},
  {Storchi-Bergmann}, {Stringfellow}, {Stutz}, {Su}, {Taghizadeh-Popp},
  {Talbot}, {Tayar}, {Telles}, {Teske}, {Thakar}, {Theissen}, {Tkachenko},
  {Thomas}, {Tojeiro}, {Hernandez Toledo}, {Troup}, {Trump}, {Trussler},
  {Turner}, {Tuttle}, {Unda-Sanzana}, {V{\'a}zquez-Mata}, {Valentini},
  {Valenzuela}, {Vargas-Gonz{\'a}lez}, {Vargas-Maga{\~n}a}, {Alfaro},
  {Villanova}, {Vincenzo}, {Wake}, {Warfield}, {Washington}, {Weaver},
  {Weijmans}, {Weinberg}, {Weiss}, {Westfall}, {Wild}, {Wilde}, {Wilson},
  {Wilson}, {Wilson}, {Wolf}, {Wood-Vasey}, {Yan}, {Zamora}, {Zasowski},
  {Zhang}, {Zhao}, {Zheng}, {Zheng}, \& {Zhu}}]{abdurrouf22APOGEEDR17}
{Abdurro'uf}, {Accetta}, K., {Aerts}, C., {et~al.} 2022, \apjs, 259, 35

\bibitem[{{Abolfathi} {et~al.}(2018){Abolfathi}, {Aguado}, {Aguilar}, {Allende
  Prieto}, {Almeida}, {Ananna}, {Anders}, {Anderson}, {Andrews}, {Anguiano},
  {Arag{\'o}n-Salamanca}, {Argudo-Fern{\'a}ndez}, {Armengaud}, {Ata},
  {Aubourg}, {Avila-Reese}, {Badenes}, {Bailey}, {Balland}, {Barger},
  {Barrera-Ballesteros}, {Bartosz}, {Bastien}, {Bates}, {Baumgarten},
  {Bautista}, {Beaton}, {Beers}, {Belfiore}, {Bender}, {Bernardi}, {Bershady},
  {Beutler}, {Bird}, {Bizyaev}, {Blanc}, {Blanton}, {Blomqvist}, {Bolton},
  {Boquien}, {Borissova}, {Bovy}, {Bradna Diaz}, {Brandt}, {Brinkmann},
  {Brownstein}, {Bundy}, {Burgasser}, {Burtin}, {Busca}, {Ca{\~n}as},
  {Cano-D{\'\i}az}, {Cappellari}, {Carrera}, {Casey}, {Cervantes Sodi}, {Chen},
  {Cherinka}, {Chiappini}, {Choi}, {Chojnowski}, {Chuang}, {Chung}, {Clerc},
  {Cohen}, {Comerford}, {Comparat}, {Correa do Nascimento}, {da Costa},
  {Cousinou}, {Covey}, {Crane}, {Cruz-Gonzalez}, {Cunha}, {da Silva Ilha},
  {Damke}, {Darling}, {Davidson}, {Dawson}, {de Icaza Lizaola}, {de la
  Macorra}, {de la Torre}, {De Lee}, {de Sainte Agathe}, {Deconto Machado},
  {Dell'Agli}, {Delubac}, {Diamond-Stanic}, {Donor}, {Downes}, {Drory}, {du Mas
  des Bourboux}, {Duckworth}, {Dwelly}, {Dyer}, {Ebelke}, {Davis Eigenbrot},
  {Eisenstein}, {Elsworth}, {Emsellem}, {Eracleous}, {Erfanianfar},
  {Escoffier}, {Fan}, {Fern{\'a}ndez Alvar}, {Fernandez-Trincado}, {Fernando
  Cirolini}, {Feuillet}, {Finoguenov}, {Fleming}, {Font-Ribera}, {Freischlad},
  {Frinchaboy}, {Fu}, {G{\'o}mez Maqueo Chew}, {Galbany}, {Garc{\'\i}a
  P{\'e}rez}, {Garcia-Dias}, {Garc{\'\i}a-Hern{\'a}ndez}, {Garma Oehmichen},
  {Gaulme}, {Gelfand}, {Gil-Mar{\'\i}n}, {Gillespie}, {Goddard}, {Gonz{\'a}lez
  Hern{\'a}ndez}, {Gonzalez-Perez}, {Grabowski}, {Green}, {Grier}, {Gueguen},
  {Guo}, {Guy}, {Hagen}, {Hall}, {Harding}, {Hasselquist}, {Hawley}, {Hayes},
  {Hearty}, {Hekker}, {Hernandez}, {Hernandez Toledo}, {Hogg},
  {Holley-Bockelmann}, {Holtzman}, {Hou}, {Hsieh}, {Hunt}, {Hutchinson},
  {Hwang}, {Jimenez Angel}, {Johnson}, {Jones}, {J{\"o}nsson}, {Jullo}, {Khan},
  {Kinemuchi}, {Kirkby}, {Kirkpatrick}, {Kitaura}, {Knapp}, {Kneib},
  {Kollmeier}, {Lacerna}, {Lane}, {Lang}, {Law}, {Le Goff}, {Lee}, {Li}, {Li},
  {Lian}, {Liang}, {Lima}, {Lin}, {Long}, {Lucatello}, {Lundgren}, {Mackereth},
  {MacLeod}, {Mahadevan}, {Maia}, {Majewski}, {Manchado}, {Maraston},
  {Mariappan}, {Marques-Chaves}, {Masseron}, {Masters}, {McDermid}, {McGreer},
  {Melendez}, {Meneses-Goytia}, {Merloni}, {Merrifield}, {Meszaros}, {Meza},
  {Minchev}, {Minniti}, {Mueller}, {Muller-Sanchez}, {Muna}, {Mu{\~n}oz},
  {Myers}, {Nair}, {Nandra}, {Ness}, {Newman}, {Nichol}, {Nidever},
  {Nitschelm}, {Noterdaeme}, {O'Connell}, {Oelkers}, {Oravetz}, {Oravetz},
  {Ort{\'\i}z}, {Osorio}, {Pace}, {Padilla}, {Palanque-Delabrouille},
  {Palicio}, {Pan}, {Pan}, {Parikh}, {P{\^a}ris}, {Park}, {Peirani},
  {Pellejero-Ibanez}, {Penny}, {Percival}, {Perez-Fournon}, {Petitjean},
  {Pieri}, {Pinsonneault}, {Pisani}, {Prada}, {Prakash}, {Queiroz}, {Raddick},
  {Raichoor}, {Barboza Rembold}, {Richstein}, {Riffel}, {Riffel}, {Rix},
  {Robin}, {Rodr{\'\i}guez Torres}, {Rom{\'a}n-Z{\'u}{\~n}iga}, {Ross},
  {Rossi}, {Ruan}, {Ruggeri}, {Ruiz}, {Salvato}, {S{\'a}nchez}, {S{\'a}nchez},
  {Sanchez Almeida}, {S{\'a}nchez-Gallego}, {Santana Rojas}, {Santiago},
  {Schiavon}, {Schimoia}, {Schlafly}, {Schlegel}, {Schneider}, {Schuster},
  {Schwope}, {Seo}, {Serenelli}, {Shen}, {Shen}, {Shetrone}, {Shull}, {Silva
  Aguirre}, {Simon}, {Skrutskie}, {Slosar}, {Smethurst}, {Smith}, {Sobeck},
  {Somers}, {Souter}, {Souto}, {Spindler}, {Stark}, {Stassun}, {Steinmetz},
  {Stello}, {Storchi-Bergmann}, {Streblyanska}, {Stringfellow}, {Su{\'a}rez},
  {Sun}, {Szigeti}, {Taghizadeh-Popp}, {Talbot}, {Tang}, {Tao}, {Tayar},
  {Tembe}, {Teske}, {Thakar}, {Thomas}, {Tissera}, {Tojeiro}, {Tremonti},
  {Troup}, {Urry}, {Valenzuela}, {van den Bosch}, {Vargas-Gonz{\'a}lez},
  {Vargas-Maga{\~n}a}, {Vazquez}, {Villanova}, {Vogt}, {Wake}, {Wang},
  {Weaver}, {Weijmans}, {Weinberg}, {Westfall}, {Whelan}, {Wilcots}, {Wild},
  {Williams}, {Wilson}, {Wood-Vasey}, {Wylezalek}, {Xiao}, {Yan}, {Yang},
  {Ybarra}, {Y{\`e}che}, {Zakamska}, {Zamora}, {Zarrouk}, {Zasowski}, {Zhang},
  {Zhao}, {Zhao}, {Zheng}, {Zheng}, {Zhou}, {Zhu}, {Zinn}, \&
  {Zou}}]{abolfathi18APOGEEDR14}
{Abolfathi}, B., {Aguado}, D.~S., {Aguilar}, G., {et~al.} 2018, \apjs, 235, 42

\bibitem[{{Allende Prieto} {et~al.}(2006){Allende Prieto}, {Beers}, {Wilhelm},
  {Newberg}, {Rockosi}, {Yanny}, \& {Lee}}]{allendeprieto06FERRE}
{Allende Prieto}, C., {Beers}, T.~C., {Wilhelm}, R., {et~al.} 2006, \apj, 636,
  804

\bibitem[{{Amarsi} {et~al.}(2020){Amarsi}, {Grevesse}, {Grumer}, {Asplund},
  {Barklem}, \& {Collet}}]{amarsi20}
{Amarsi}, A.~M., {Grevesse}, N., {Grumer}, J., {et~al.} 2020, \aap, 636, A120

\bibitem[{{Andrassy} {et~al.}(2020){Andrassy}, {Herwig}, {Woodward}, \&
  {Ritter}}]{andrassy:20}
{Andrassy}, R., {Herwig}, F., {Woodward}, P., \& {Ritter}, C. 2020, \mnras,
  491, 972

\bibitem[{{Asplund} {et~al.}(2021){Asplund}, {Amarsi}, \&
  {Grevesse}}]{Asplund21}
{Asplund}, M., {Amarsi}, A.~M., \& {Grevesse}, N. 2021, \aap, 653, A141

\bibitem[{{Astropy Collaboration} {et~al.}(2022){Astropy Collaboration},
  {Price-Whelan}, {Lim}, {Earl}, {Starkman}, {Bradley}, {Shupe}, {Patil},
  {Corrales}, {Brasseur}, {N{"o}the}, {Donath}, {Tollerud}, {Morris},
  {Ginsburg}, {Vaher}, {Weaver}, {Tocknell}, {Jamieson}, {van Kerkwijk},
  {Robitaille}, {Merry}, {Bachetti}, {G{"u}nther}, {Aldcroft},
  {Alvarado-Montes}, {Archibald}, {B{'o}di}, {Bapat}, {Barentsen}, {Baz{'a}n},
  {Biswas}, {Boquien}, {Burke}, {Cara}, {Cara}, {Conroy}, {Conseil}, {Craig},
  {Cross}, {Cruz}, {D'Eugenio}, {Dencheva}, {Devillepoix}, {Dietrich},
  {Eigenbrot}, {Erben}, {Ferreira}, {Foreman-Mackey}, {Fox}, {Freij}, {Garg},
  {Geda}, {Glattly}, {Gondhalekar}, {Gordon}, {Grant}, {Greenfield}, {Groener},
  {Guest}, {Gurovich}, {Handberg}, {Hart}, {Hatfield-Dodds}, {Homeier},
  {Hosseinzadeh}, {Jenness}, {Jones}, {Joseph}, {Kalmbach}, {Karamehmetoglu},
  {Ka{l}uszy{'n}ski}, {Kelley}, {Kern}, {Kerzendorf}, {Koch}, {Kulumani},
  {Lee}, {Ly}, {Ma}, {MacBride}, {Maljaars}, {Muna}, {Murphy}, {Norman},
  {O'Steen}, {Oman}, {Pacifici}, {Pascual}, {Pascual-Granado}, {Patil},
  {Perren}, {Pickering}, {Rastogi}, {Roulston}, {Ryan}, {Rykoff}, {Sabater},
  {Sakurikar}, {Salgado}, {Sanghi}, {Saunders}, {Savchenko}, {Schwardt},
  {Seifert-Eckert}, {Shih}, {Jain}, {Shukla}, {Sick}, {Simpson},
  {Singanamalla}, {Singer}, {Singhal}, {Sinha}, {Sip{H{o}}cz}, {Spitler},
  {Stansby}, {Streicher}, {{{S}}umak}, {Swinbank}, {Taranu}, {Tewary},
  {Tremblay}, {Val-Borro}, {Van Kooten}, {Vasovi{'c}}, {Verma}, {de Miranda
  Cardoso}, {Williams}, {Wilson}, {Winkel}, {Wood-Vasey}, {Xue}, {Yoachim},
  {Zhang}, {Zonca}, \& {Astropy Project Contributors}}]{astropy:2022}
{Astropy Collaboration}, {Price-Whelan}, A.~M., {Lim}, P.~L., {et~al.} 2022,
  apj, 935, 167

\bibitem[{{Astropy Collaboration} {et~al.}(2018){Astropy Collaboration},
  {Price-Whelan}, {Sip{\H{o}}cz}, {G{\"u}nther}, {Lim}, {Crawford}, {Conseil},
  {Shupe}, {Craig}, {Dencheva}, {Ginsburg}, {Vand erPlas}, {Bradley},
  {P{\'e}rez-Su{\'a}rez}, {de Val-Borro}, {Aldcroft}, {Cruz}, {Robitaille},
  {Tollerud}, {Ardelean}, {Babej}, {Bach}, {Bachetti}, {Bakanov}, {Bamford},
  {Barentsen}, {Barmby}, {Baumbach}, {Berry}, {Biscani}, {Boquien}, {Bostroem},
  {Bouma}, {Brammer}, {Bray}, {Breytenbach}, {Buddelmeijer}, {Burke},
  {Calderone}, {Cano Rodr{\'\i}guez}, {Cara}, {Cardoso}, {Cheedella}, {Copin},
  {Corrales}, {Crichton}, {D'Avella}, {Deil}, {Depagne}, {Dietrich}, {Donath},
  {Droettboom}, {Earl}, {Erben}, {Fabbro}, {Ferreira}, {Finethy}, {Fox},
  {Garrison}, {Gibbons}, {Goldstein}, {Gommers}, {Greco}, {Greenfield},
  {Groener}, {Grollier}, {Hagen}, {Hirst}, {Homeier}, {Horton}, {Hosseinzadeh},
  {Hu}, {Hunkeler}, {Ivezi{\'c}}, {Jain}, {Jenness}, {Kanarek}, {Kendrew},
  {Kern}, {Kerzendorf}, {Khvalko}, {King}, {Kirkby}, {Kulkarni}, {Kumar},
  {Lee}, {Lenz}, {Littlefair}, {Ma}, {Macleod}, {Mastropietro}, {McCully},
  {Montagnac}, {Morris}, {Mueller}, {Mumford}, {Muna}, {Murphy}, {Nelson},
  {Nguyen}, {Ninan}, {N{\"o}the}, {Ogaz}, {Oh}, {Parejko}, {Parley}, {Pascual},
  {Patil}, {Patil}, {Plunkett}, {Prochaska}, {Rastogi}, {Reddy Janga},
  {Sabater}, {Sakurikar}, {Seifert}, {Sherbert}, {Sherwood-Taylor}, {Shih},
  {Sick}, {Silbiger}, {Singanamalla}, {Singer}, {Sladen}, {Sooley},
  {Sornarajah}, {Streicher}, {Teuben}, {Thomas}, {Tremblay}, {Turner},
  {Terr{\'o}n}, {van Kerkwijk}, {de la Vega}, {Watkins}, {Weaver}, {Whitmore},
  {Woillez}, {Zabalza}, \& {Astropy Contributors}}]{astropy:2018}
{Astropy Collaboration}, {Price-Whelan}, A.~M., {Sip{\H{o}}cz}, B.~M., {et~al.}
  2018, \aj, 156, 123

\bibitem[{{Astropy Collaboration} {et~al.}(2013){Astropy Collaboration},
  {Robitaille}, {Tollerud}, {Greenfield}, {Droettboom}, {Bray}, {Aldcroft},
  {Davis}, {Ginsburg}, {Price-Whelan}, {Kerzendorf}, {Conley}, {Crighton},
  {Barbary}, {Muna}, {Ferguson}, {Grollier}, {Parikh}, {Nair}, {Unther},
  {Deil}, {Woillez}, {Conseil}, {Kramer}, {Turner}, {Singer}, {Fox}, {Weaver},
  {Zabalza}, {Edwards}, {Azalee Bostroem}, {Burke}, {Casey}, {Crawford},
  {Dencheva}, {Ely}, {Jenness}, {Labrie}, {Lim}, {Pierfederici}, {Pontzen},
  {Ptak}, {Refsdal}, {Servillat}, \& {Streicher}}]{astropy:2013}
{Astropy Collaboration}, {Robitaille}, T.~P., {Tollerud}, E.~J., {et~al.} 2013,
  \aap, 558, A33

\bibitem[{{Blanton} {et~al.}(2017){Blanton}, {Bershady}, {Abolfathi},
  {Albareti}, {Allende Prieto}, {Almeida}, {Alonso-Garc{\'\i}a}, {Anders},
  {Anderson}, {Andrews}, {Aquino-Ort{\'\i}z}, {Arag{\'o}n-Salamanca},
  {Argudo-Fern{\'a}ndez}, {Armengaud}, {Aubourg}, {Avila-Reese}, {Badenes},
  {Bailey}, {Barger}, {Barrera-Ballesteros}, {Bartosz}, {Bates}, {Baumgarten},
  {Bautista}, {Beaton}, {Beers}, {Belfiore}, {Bender}, {Berlind}, {Bernardi},
  {Beutler}, {Bird}, {Bizyaev}, {Blanc}, {Blomqvist}, {Bolton}, {Boquien},
  {Borissova}, {van den Bosch}, {Bovy}, {Brandt}, {Brinkmann}, {Brownstein},
  {Bundy}, {Burgasser}, {Burtin}, {Busca}, {Cappellari}, {Delgado Carigi},
  {Carlberg}, {Carnero Rosell}, {Carrera}, {Chanover}, {Cherinka}, {Cheung},
  {G{\'o}mez Maqueo Chew}, {Chiappini}, {Choi}, {Chojnowski}, {Chuang},
  {Chung}, {Cirolini}, {Clerc}, {Cohen}, {Comparat}, {da Costa}, {Cousinou},
  {Covey}, {Crane}, {Croft}, {Cruz-Gonzalez}, {Garrido Cuadra}, {Cunha},
  {Damke}, {Darling}, {Davies}, {Dawson}, {de la Macorra}, {Dell'Agli}, {De
  Lee}, {Delubac}, {Di Mille}, {Diamond-Stanic}, {Cano-D{\'\i}az}, {Donor},
  {Downes}, {Drory}, {du Mas des Bourboux}, {Duckworth}, {Dwelly}, {Dyer},
  {Ebelke}, {Eigenbrot}, {Eisenstein}, {Emsellem}, {Eracleous}, {Escoffier},
  {Evans}, {Fan}, {Fern{\'a}ndez-Alvar}, {Fernandez-Trincado}, {Feuillet},
  {Finoguenov}, {Fleming}, {Font-Ribera}, {Fredrickson}, {Freischlad},
  {Frinchaboy}, {Fuentes}, {Galbany}, {Garcia-Dias},
  {Garc{\'\i}a-Hern{\'a}ndez}, {Gaulme}, {Geisler}, {Gelfand},
  {Gil-Mar{\'\i}n}, {Gillespie}, {Goddard}, {Gonzalez-Perez}, {Grabowski},
  {Green}, {Grier}, {Gunn}, {Guo}, {Guy}, {Hagen}, {Hahn}, {Hall}, {Harding},
  {Hasselquist}, {Hawley}, {Hearty}, {Gonzalez Hern{\'a}ndez}, {Ho}, {Hogg},
  {Holley-Bockelmann}, {Holtzman}, {Holzer}, {Huehnerhoff}, {Hutchinson},
  {Hwang}, {Ibarra-Medel}, {da Silva Ilha}, {Ivans}, {Ivory}, {Jackson},
  {Jensen}, {Johnson}, {Jones}, {J{\"o}nsson}, {Jullo}, {Kamble}, {Kinemuchi},
  {Kirkby}, {Kitaura}, {Klaene}, {Knapp}, {Kneib}, {Kollmeier}, {Lacerna},
  {Lane}, {Lang}, {Law}, {Lazarz}, {Lee}, {Le Goff}, {Liang}, {Li}, {Li},
  {Lian}, {Lima}, {Lin}, {Lin}, {Bertran de Lis}, {Liu}, {de Icaza Lizaola},
  {Long}, {Lucatello}, {Lundgren}, {MacDonald}, {Deconto Machado}, {MacLeod},
  {Mahadevan}, {Geimba Maia}, {Maiolino}, {Majewski}, {Malanushenko},
  {Malanushenko}, {Manchado}, {Mao}, {Maraston}, {Marques-Chaves}, {Masseron},
  {Masters}, {McBride}, {McDermid}, {McGrath}, {McGreer}, {Medina Pe{\~n}a},
  {Melendez}, {Merloni}, {Merrifield}, {Meszaros}, {Meza}, {Minchev},
  {Minniti}, {Miyaji}, {More}, {Mulchaey}, {M{\"u}ller-S{\'a}nchez}, {Muna},
  {Munoz}, {Myers}, {Nair}, {Nandra}, {Correa do Nascimento}, {Negrete},
  {Ness}, {Newman}, {Nichol}, {Nidever}, {Nitschelm}, {Ntelis}, {O'Connell},
  {Oelkers}, {Oravetz}, {Oravetz}, {Pace}, {Padilla}, {Palanque-Delabrouille},
  {Alonso Palicio}, {Pan}, {Parejko}, {Parikh}, {P{\^a}ris}, {Park}, {Patten},
  {Peirani}, {Pellejero-Ibanez}, {Penny}, {Percival}, {Perez-Fournon},
  {Petitjean}, {Pieri}, {Pinsonneault}, {Pisani}, {Poleski}, {Prada},
  {Prakash}, {Queiroz}, {Raddick}, {Raichoor}, {Barboza Rembold}, {Richstein},
  {Riffel}, {Riffel}, {Rix}, {Robin}, {Rockosi}, {Rodr{\'\i}guez-Torres},
  {Roman-Lopes}, {Rom{\'a}n-Z{\'u}{\~n}iga}, {Rosado}, {Ross}, {Rossi}, {Ruan},
  {Ruggeri}, {Rykoff}, {Salazar-Albornoz}, {Salvato}, {S{\'a}nchez}, {Aguado},
  {S{\'a}nchez-Gallego}, {Santana}, {Santiago}, {Sayres}, {Schiavon}, {da Silva
  Schimoia}, {Schlafly}, {Schlegel}, {Schneider}, {Schultheis}, {Schuster},
  {Schwope}, {Seo}, {Shao}, {Shen}, {Shetrone}, {Shull}, {Simon}, {Skinner},
  {Skrutskie}, {Slosar}, {Smith}, {Sobeck}, {Sobreira}, {Somers}, {Souto},
  {Stark}, {Stassun}, {Stauffer}, {Steinmetz}, {Storchi-Bergmann},
  {Streblyanska}, {Stringfellow}, {Su{\'a}rez}, {Sun}, {Suzuki}, {Szigeti},
  {Taghizadeh-Popp}, {Tang}, {Tao}, {Tayar}, {Tembe}, {Teske}, {Thakar},
  {Thomas}, {Thompson}, {Tinker}, {Tissera}, {Tojeiro}, {Hernandez Toledo}, {de
  la Torre}, {Tremonti}, {Troup}, {Valenzuela}, {Martinez Valpuesta},
  {Vargas-Gonz{\'a}lez}, {Vargas-Maga{\~n}a}, {Vazquez}, {Villanova}, {Vivek},
  {Vogt}, {Wake}, {Walterbos}, {Wang}, {Weaver}, {Weijmans}, {Weinberg},
  {Westfall}, {Whelan}, {Wild}, {Wilson}, {Wood-Vasey}, {Wylezalek}, {Xiao},
  {Yan}, {Yang}, {Ybarra}, {Y{\`e}che}, {Zakamska}, {Zamora}, {Zarrouk},
  {Zasowski}, {Zhang}, {Zhao}, {Zheng}, {Zheng}, {Zhou}, {Zhou}, {Zhu},
  {Zoccali}, \& {Zou}}]{blanton17SDSS}
{Blanton}, M.~R., {Bershady}, M.~A., {Abolfathi}, B., {et~al.} 2017, \aj, 154,
  28

\bibitem[{{Bovy}(2015)}]{bovy15VAC3}
{Bovy}, J. 2015, \apjs, 216, 29

\bibitem[{{Bowen} \& {Vaughan}(1973)}]{bowen73Telescope2}
{Bowen}, I.~S. \& {Vaughan}, A.~H., J. 1973, \ao, 12, 1430

\bibitem[{{Bravo} {et~al.}(2022){Bravo}, {Piersanti}, {Blondin},
  {Dom{\'\i}nguez}, {Straniero}, \& {Cristallo}}]{bravo22}
{Bravo}, E., {Piersanti}, L., {Blondin}, S., {et~al.} 2022, \mnras, 517, L31

\bibitem[{{Brooke} {et~al.}(2016){Brooke}, {Bernath}, {Western}, {Sneden},
  {Af{\c{s}}ar}, {Li}, \& {Gordon}}]{brooke16OH}
{Brooke}, J. S.~A., {Bernath}, P.~F., {Western}, C.~M., {et~al.} 2016, \jqsrt,
  168, 142

\bibitem[{{Burrows} \& {Vartanyan}(2021)}]{burrows:21}
{Burrows}, A. \& {Vartanyan}, D. 2021, \nat, 589, 29

\bibitem[{Butusov \& Jernelov(2013)}]{butusov13}
Butusov, M. \& Jernelov, A. 2013, Phosphorus. An Element that could have been
  called Lucifer, Vol.~9 (Springer New York, NY)

\bibitem[{{Caffau} {et~al.}(2016){Caffau}, {Andrievsky}, {Korotin}, {Origlia},
  {Oliva}, {Sanna}, {Ludwig}, \& {Bonifacio}}]{caffau16}
{Caffau}, E., {Andrievsky}, S., {Korotin}, S., {et~al.} 2016, \aap, 585, A16

\bibitem[{{Caffau} {et~al.}(2011){Caffau}, {Bonifacio}, {Faraggiana}, \&
  {Steffen}}]{caffau11}
{Caffau}, E., {Bonifacio}, P., {Faraggiana}, R., \& {Steffen}, M. 2011, \aap,
  532, A98

\bibitem[{{Caffau} {et~al.}(2019){Caffau}, {Bonifacio}, {Oliva}, {Korotin},
  {Capitanio}, {Andrievsky}, {Collet}, {Sbordone}, {Duffau}, {Sanna}, {Tozzi},
  {Origlia}, {Ryde}, \& {Ludwig}}]{caffau19}
{Caffau}, E., {Bonifacio}, P., {Oliva}, E., {et~al.} 2019, \aap, 622, A68

\bibitem[{{Cescutti} {et~al.}(2012){Cescutti}, {Matteucci}, {Caffau}, \&
  {Fran{\c{c}}ois}}]{cescutti12}
{Cescutti}, G., {Matteucci}, F., {Caffau}, E., \& {Fran{\c{c}}ois}, P. 2012,
  \aap, 540, A33

\bibitem[{{Chiappini} {et~al.}(1997){Chiappini}, {Matteucci}, \&
  {Gratton}}]{Chiappini1997}
{Chiappini}, C., {Matteucci}, F., \& {Gratton}, R. 1997, \apj, 477, 765

\bibitem[{{Chieffi} {et~al.}(1998){Chieffi}, {Limongi}, \&
  {Straniero}}]{chieffi:98}
{Chieffi}, A., {Limongi}, M., \& {Straniero}, O. 1998, \apj, 502, 737

\bibitem[{{Clayton}(2003)}]{clayton03}
{Clayton}, D. 2003, {Handbook of Isotopes in the Cosmos} (Cambridge University
  Press)

\bibitem[{{Contursi} {et~al.}(2022){Contursi}, {de Laverny}, {Recio-Blanco},
  {Spitoni}, {Palicio}, {Poggio}, {Grisoni}, {Cescutti}, {Matteucci}, {Spina},
  {Alvarez}, {Kordopatis}, {Ordenovic}, {Oreshina-Slezak}, \&
  {Zhao}}]{contursi22}
{Contursi}, G., {de Laverny}, P., {Recio-Blanco}, A., {et~al.} 2022, arXiv
  e-prints, arXiv:2207.05368

\bibitem[{{C{\^o}t{\'e}} {et~al.}(2019){C{\^o}t{\'e}}, {Lugaro}, {Reifarth},
  {Pignatari}, {Vil{\'a}gos}, {Yag{\"u}e}, \& {Gibson}}]{Cote2019}
{C{\^o}t{\'e}}, B., {Lugaro}, M., {Reifarth}, R., {et~al.} 2019, \apj, 878, 156

\bibitem[{{C{\^o}t{\'e}} {et~al.}(2017){C{\^o}t{\'e}}, {O'Shea}, {Ritter},
  {Herwig}, \& {Venn}}]{Cote2017}
{C{\^o}t{\'e}}, B., {O'Shea}, B.~W., {Ritter}, C., {Herwig}, F., \& {Venn},
  K.~A. 2017, \apj, 835, 128

\bibitem[{{C{\^o}t{\'e}} {et~al.}(2018){C{\^o}t{\'e}}, {Silvia}, {O'Shea},
  {Smith}, \& {Wise}}]{Cote2018}
{C{\^o}t{\'e}}, B., {Silvia}, D.~W., {O'Shea}, B.~W., {Smith}, B., \& {Wise},
  J.~H. 2018, \apj, 859, 67

\bibitem[{{Cristallo} {et~al.}(2015){Cristallo}, {Straniero}, {Piersanti}, \&
  {Gobrecht}}]{Cristallo2015}
{Cristallo}, S., {Straniero}, O., {Piersanti}, L., \& {Gobrecht}, D. 2015,
  \apjs, 219, 40

\bibitem[{{Cristini} {et~al.}(2019){Cristini}, {Hirschi}, {Meakin}, {Arnett},
  {Georgy}, \& {Walkington}}]{cristini:19}
{Cristini}, A., {Hirschi}, R., {Meakin}, C., {et~al.} 2019, \mnras, 484, 4645

\bibitem[{{Curtis} {et~al.}(2019){Curtis}, {Ebinger}, {Fr{\"o}hlich}, {Hempel},
  {Perego}, {Liebend{\"o}rfer}, \& {Thielemann}}]{curtis:19}
{Curtis}, S., {Ebinger}, K., {Fr{\"o}hlich}, C., {et~al.} 2019, \apj, 870, 2

\bibitem[{{Dodd} {et~al.}(2022){Dodd}, {Callingham}, {Helmi}, {Matsuno},
  {Ruiz-Lara}, {Balbinot}, \& {L{\"o}vdal}}]{dodd22}
{Dodd}, E., {Callingham}, T.~M., {Helmi}, A., {et~al.} 2022, arXiv e-prints,
  arXiv:2206.11248

\bibitem[{{Fern{\'a}ndez-Trincado} {et~al.}(2020){Fern{\'a}ndez-Trincado},
  {Beers}, \& {Minniti}}]{fernandeztrincado20}
{Fern{\'a}ndez-Trincado}, J.~G., {Beers}, T.~C., \& {Minniti}, D. 2020, \aap,
  644, A83

\bibitem[{{Fern{\'a}ndez-Trincado} {et~al.}(2019){Fern{\'a}ndez-Trincado},
  {Beers}, {Placco}, {Moreno}, {Alves-Brito}, {Minniti}, {Tang},
  {P{\'e}rez-Villegas}, {Reyl{\'e}}, {Robin}, \&
  {Villanova}}]{fernandeztrincado19}
{Fern{\'a}ndez-Trincado}, J.~G., {Beers}, T.~C., {Placco}, V.~M., {et~al.}
  2019, \apjl, 886, L8

\bibitem[{{Gaia Collaboration} {et~al.}(2021){Gaia Collaboration}, {Brown},
  {Vallenari}, {Prusti}, {de Bruijne}, {Babusiaux}, {Biermann}, {Creevey},
  {Evans}, {Eyer}, {Hutton}, {Jansen}, {Jordi}, {Klioner}, {Lammers},
  {Lindegren}, {Luri}, {Mignard}, {Panem}, {Pourbaix}, {Randich}, {Sartoretti},
  {Soubiran}, {Walton}, {Arenou}, {Bailer-Jones}, {Bastian}, {Cropper},
  {Drimmel}, {Katz}, {Lattanzi}, {van Leeuwen}, {Bakker}, {Cacciari},
  {Casta{\~n}eda}, {De Angeli}, {Ducourant}, {Fabricius}, {Fouesneau},
  {Fr{\'e}mat}, {Guerra}, {Guerrier}, {Guiraud}, {Jean-Antoine Piccolo},
  {Masana}, {Messineo}, {Mowlavi}, {Nicolas}, {Nienartowicz}, {Pailler},
  {Panuzzo}, {Riclet}, {Roux}, {Seabroke}, {Sordo}, {Tanga}, {Th{\'e}venin},
  {Gracia-Abril}, {Portell}, {Teyssier}, {Altmann}, {Andrae}, {Bellas-Velidis},
  {Benson}, {Berthier}, {Blomme}, {Brugaletta}, {Burgess}, {Busso}, {Carry},
  {Cellino}, {Cheek}, {Clementini}, {Damerdji}, {Davidson}, {Delchambre},
  {Dell'Oro}, {Fern{\'a}ndez-Hern{\'a}ndez}, {Galluccio}, {Garc{\'\i}a-Lario},
  {Garcia-Reinaldos}, {Gonz{\'a}lez-N{\'u}{\~n}ez}, {Gosset}, {Haigron},
  {Halbwachs}, {Hambly}, {Harrison}, {Hatzidimitriou}, {Heiter},
  {Hern{\'a}ndez}, {Hestroffer}, {Hodgkin}, {Holl}, {Jan{\ss}en}, {Jevardat de
  Fombelle}, {Jordan}, {Krone-Martins}, {Lanzafame}, {L{\"o}ffler}, {Lorca},
  {Manteiga}, {Marchal}, {Marrese}, {Moitinho}, {Mora}, {Muinonen}, {Osborne},
  {Pancino}, {Pauwels}, {Petit}, {Recio-Blanco}, {Richards}, {Riello},
  {Rimoldini}, {Robin}, {Roegiers}, {Rybizki}, {Sarro}, {Siopis}, {Smith},
  {Sozzetti}, {Ulla}, {Utrilla}, {van Leeuwen}, {van Reeven}, {Abbas}, {Abreu
  Aramburu}, {Accart}, {Aerts}, {Aguado}, {Ajaj}, {Altavilla}, {{\'A}lvarez},
  {{\'A}lvarez Cid-Fuentes}, {Alves}, {Anderson}, {Anglada Varela}, {Antoja},
  {Audard}, {Baines}, {Baker}, {Balaguer-N{\'u}{\~n}ez}, {Balbinot}, {Balog},
  {Barache}, {Barbato}, {Barros}, {Barstow}, {Bartolom{\'e}}, {Bassilana},
  {Bauchet}, {Baudesson-Stella}, {Becciani}, {Bellazzini}, {Bernet}, {Bertone},
  {Bianchi}, {Blanco-Cuaresma}, {Boch}, {Bombrun}, {Bossini}, {Bouquillon},
  {Bragaglia}, {Bramante}, {Breedt}, {Bressan}, {Brouillet}, {Bucciarelli},
  {Burlacu}, {Busonero}, {Butkevich}, {Buzzi}, {Caffau}, {Cancelliere},
  {C{\'a}novas}, {Cantat-Gaudin}, {Carballo}, {Carlucci}, {Carnerero},
  {Carrasco}, {Casamiquela}, {Castellani}, {Castro-Ginard}, {Castro Sampol},
  {Chaoul}, {Charlot}, {Chemin}, {Chiavassa}, {Cioni}, {Comoretto}, {Cooper},
  {Cornez}, {Cowell}, {Crifo}, {Crosta}, {Crowley}, {Dafonte}, {Dapergolas},
  {David}, {David}, {de Laverny}, {De Luise}, {De March}, {De Ridder}, {de
  Souza}, {de Teodoro}, {de Torres}, {del Peloso}, {del Pozo}, {Delbo},
  {Delgado}, {Delgado}, {Delisle}, {Di Matteo}, {Diakite}, {Diener},
  {Distefano}, {Dolding}, {Eappachen}, {Edvardsson}, {Enke}, {Esquej}, {Fabre},
  {Fabrizio}, {Faigler}, {Fedorets}, {Fernique}, {Fienga}, {Figueras},
  {Fouron}, {Fragkoudi}, {Fraile}, {Franke}, {Gai}, {Garabato},
  {Garcia-Gutierrez}, {Garc{\'\i}a-Torres}, {Garofalo}, {Gavras}, {Gerlach},
  {Geyer}, {Giacobbe}, {Gilmore}, {Girona}, {Giuffrida}, {Gomel}, {Gomez},
  {Gonzalez-Santamaria}, {Gonz{\'a}lez-Vidal}, {Granvik},
  {Guti{\'e}rrez-S{\'a}nchez}, {Guy}, {Hauser}, {Haywood}, {Helmi}, {Hidalgo},
  {Hilger}, {H{\l}adczuk}, {Hobbs}, {Holland}, {Huckle}, {Jasniewicz},
  {Jonker}, {Juaristi Campillo}, {Julbe}, {Karbevska}, {Kervella}, {Khanna},
  {Kochoska}, {Kontizas}, {Kordopatis}, {Korn}, {Kostrzewa-Rutkowska},
  {Kruszy{\'n}ska}, {Lambert}, {Lanza}, {Lasne}, {Le Campion}, {Le Fustec},
  {Lebreton}, {Lebzelter}, {Leccia}, {Leclerc}, {Lecoeur-Taibi}, {Liao},
  {Licata}, {Lindstr{\o}m}, {Lister}, {Livanou}, {Lobel}, {Madrero Pardo},
  {Managau}, {Mann}, {Marchant}, {Marconi}, {Marcos Santos}, {Marinoni},
  {Marocco}, {Marshall}, {Martin Polo}, {Mart{\'\i}n-Fleitas}, {Masip},
  {Massari}, {Mastrobuono-Battisti}, {Mazeh}, {McMillan}, {Messina},
  {Michalik}, {Millar}, {Mints}, {Molina}, {Molinaro}, {Moln{\'a}r},
  {Montegriffo}, {Mor}, {Morbidelli}, {Morel}, {Morris}, {Mulone}, {Munoz},
  {Muraveva}, {Murphy}, {Musella}, {Noval}, {Ord{\'e}novic}, {Orr{\`u}},
  {Osinde}, {Pagani}, {Pagano}, {Palaversa}, {Palicio}, {Panahi}, {Pawlak},
  {Pe{\~n}alosa Esteller}, {Penttil{\"a}}, {Piersimoni}, {Pineau}, {Plachy},
  {Plum}, {Poggio}, {Poretti}, {Poujoulet}, {Pr{\v{s}}a}, {Pulone}, {Racero},
  {Ragaini}, {Rainer}, {Raiteri}, {Rambaux}, {Ramos}, {Ramos-Lerate}, {Re
  Fiorentin}, {Regibo}, {Reyl{\'e}}, {Ripepi}, {Riva}, {Rixon}, {Robichon},
  {Robin}, {Roelens}, {Rohrbasser}, {Romero-G{\'o}mez}, {Rowell}, {Royer},
  {Rybicki}, {Sadowski}, {Sagrist{\`a} Sell{\'e}s}, {Sahlmann}, {Salgado},
  {Salguero}, {Samaras}, {Sanchez Gimenez}, {Sanna}, {Santove{\~n}a},
  {Sarasso}, {Schultheis}, {Sciacca}, {Segol}, {Segovia}, {S{\'e}gransan},
  {Semeux}, {Shahaf}, {Siddiqui}, {Siebert}, {Siltala}, {Slezak}, {Smart},
  {Solano}, {Solitro}, {Souami}, {Souchay}, {Spagna}, {Spoto}, {Steele},
  {Steidelm{\"u}ller}, {Stephenson}, {S{\"u}veges}, {Szabados}, {Szegedi-Elek},
  {Taris}, {Tauran}, {Taylor}, {Teixeira}, {Thuillot}, {Tonello}, {Torra},
  {Torra}, {Turon}, {Unger}, {Vaillant}, {van Dillen}, {Vanel}, {Vecchiato},
  {Viala}, {Vicente}, {Voutsinas}, {Weiler}, {Wevers}, {Wyrzykowski}, {Yoldas},
  {Yvard}, {Zhao}, {Zorec}, {Zucker}, {Zurbach}, \& {Zwitter}}]{gaiaedr3}
{Gaia Collaboration}, {Brown}, A.~G.~A., {Vallenari}, A., {et~al.} 2021, \aap,
  649, A1

\bibitem[{{Garc{\'\i}a P{\'e}rez} {et~al.}(2016){Garc{\'\i}a P{\'e}rez},
  {Allende Prieto}, {Holtzman}, {Shetrone}, {M{\'e}sz{\'a}ros}, {Bizyaev},
  {Carrera}, {Cunha}, {Garc{\'\i}a-Hern{\'a}ndez}, {Johnson}, {Majewski},
  {Nidever}, {Schiavon}, {Shane}, {Smith}, {Sobeck}, {Troup}, {Zamora},
  {Weinberg}, {Bovy}, {Eisenstein}, {Feuillet}, {Frinchaboy}, {Hayden},
  {Hearty}, {Nguyen}, {O'Connell}, {Pinsonneault}, {Wilson}, \&
  {Zasowski}}]{garciaperez16ASPCAP}
{Garc{\'\i}a P{\'e}rez}, A.~E., {Allende Prieto}, C., {Holtzman}, J.~A.,
  {et~al.} 2016, \aj, 151, 144

\bibitem[{{Georgy} {et~al.}(2017){Georgy}, {Meynet}, {Ekstr{\"o}m}, {Wade},
  {Petit}, {Keszthelyi}, \& {Hirschi}}]{georgy17}
{Georgy}, C., {Meynet}, G., {Ekstr{\"o}m}, S., {et~al.} 2017, \aap, 599, L5

\bibitem[{{Goriely}(2022)}]{goriely22}
{Goriely}, S. 2022, \aap, 658, A197

\bibitem[{{Goswami} \& {Prantzos}(2000)}]{goswami:00}
{Goswami}, A. \& {Prantzos}, N. 2000, \aap, 359, 191

\bibitem[{{Grevesse} {et~al.}(2007){Grevesse}, {Asplund}, \&
  {Sauval}}]{Grevesse2007}
{Grevesse}, N., {Asplund}, M., \& {Sauval}, A.~J. 2007, \ssr, 130, 105

\bibitem[{Gronow {et~al.}(2020)Gronow, Collins, Ohlmann, Pakmor, Kromer,
  Seitenzahl, Sim, \& R{\"o}pke}]{gronow:20}
Gronow, S., Collins, C., Ohlmann, S.~T., {et~al.} 2020, Astronomy \&
  Astrophysics, 635, A169

\bibitem[{Gulick(1955)}]{gulick1955}
Gulick, A. 1955, American Scientist, 43, 479

\bibitem[{{Gunn} {et~al.}(2006){Gunn}, {Siegmund}, {Mannery}, {Owen}, {Hull},
  {Leger}, {Carey}, {Knapp}, {York}, {Boroski}, {Kent}, {Lupton}, {Rockosi},
  {Evans}, {Waddell}, {Anderson}, {Annis}, {Barentine}, {Bartoszek}, {Bastian},
  {Bracker}, {Brewington}, {Briegel}, {Brinkmann}, {Brown}, {Carr},
  {Czarapata}, {Drennan}, {Dombeck}, {Federwitz}, {Gillespie}, {Gonzales},
  {Hansen}, {Harvanek}, {Hayes}, {Jordan}, {Kinney}, {Klaene}, {Kleinman},
  {Kron}, {Kresinski}, {Lee}, {Limmongkol}, {Lindenmeyer}, {Long}, {Loomis},
  {McGehee}, {Mantsch}, {Neilsen}, {Neswold}, {Newman}, {Nitta}, {Peoples},
  {Pier}, {Prieto}, {Prosapio}, {Rivetta}, {Schneider}, {Snedden}, \&
  {Wang}}]{gunn06Telescope1}
{Gunn}, J.~E., {Siegmund}, W.~A., {Mannery}, E.~J., {et~al.} 2006, \aj, 131,
  2332

\bibitem[{{Gustafsson} {et~al.}(2008){Gustafsson}, {Edvardsson}, {Eriksson},
  {J{\o}rgensen}, {Nordlund}, \& {Plez}}]{gustafsson08}
{Gustafsson}, B., {Edvardsson}, B., {Eriksson}, K., {et~al.} 2008, \aap, 486,
  951

\bibitem[{Harris {et~al.}(2020)Harris, Millman, van~der Walt, Gommers,
  Virtanen, Cournapeau, Wieser, Taylor, Berg, Smith, Kern, Picus, Hoyer, van
  Kerkwijk, Brett, Haldane, del R{\'{i}}o, Wiebe, Peterson,
  G{\'{e}}rard-Marchant, Sheppard, Reddy, Weckesser, Abbasi, Gohlke, \&
  Oliphant}]{harris2020array}
Harris, C.~R., Millman, K.~J., van~der Walt, S.~J., {et~al.} 2020, Nature, 585,
  357

\bibitem[{{Harris}(1996)}]{harris1996}
{Harris}, W.~E. 1996, \aj, 112, 1487

\bibitem[{{Hawkins} {et~al.}(2015){Hawkins}, {Jofr{\'e}}, {Masseron}, \&
  {Gilmore}}]{hawkins15}
{Hawkins}, K., {Jofr{\'e}}, P., {Masseron}, T., \& {Gilmore}, G. 2015, \mnras,
  453, 758

\bibitem[{{Hawkins} {et~al.}(2016){Hawkins}, {Masseron}, {Jofr{\'e}},
  {Gilmore}, {Elsworth}, \& {Hekker}}]{hawkins16}
{Hawkins}, K., {Masseron}, T., {Jofr{\'e}}, P., {et~al.} 2016, \aap, 594, A43

\bibitem[{{Hayes} {et~al.}(2018){Hayes}, {Majewski}, {Shetrone},
  {Fern{\'a}ndez-Alvar}, {Allende Prieto}, {Schuster}, {Carigi}, {Cunha},
  {Smith}, {Sobeck}, {Almeida}, {Beers}, {Carrera}, {Fern{\'a}ndez-Trincado},
  {Garc{\'\i}a-Hern{\'a}ndez}, {Geisler}, {Lane}, {Lucatello}, {Matthews},
  {Minniti}, {Nitschelm}, {Tang}, {Tissera}, \& {Zamora}}]{hayes18}
{Hayes}, C.~R., {Majewski}, S.~R., {Shetrone}, M., {et~al.} 2018, \apj, 852, 49

\bibitem[{{Hayes} {et~al.}(2022){Hayes}, {Masseron}, {Sobeck},
  {Garcia-Hernandez}, {Allende Prieto}, {Beaton}, {Cunha}, {Hasselquist},
  {Holtzman}, {Jonsson}, {Majewski}, {Shetrone}, {Smith}, \&
  {Almeida}}]{hayes22}
{Hayes}, C.~R., {Masseron}, T., {Sobeck}, J., {et~al.} 2022, arXiv e-prints,
  arXiv:2208.00071

\bibitem[{{Hinkel} {et~al.}(2014){Hinkel}, {Timmes}, {Young}, {Pagano}, \&
  {Turnbull}}]{Hinkel2014}
{Hinkel}, N.~R., {Timmes}, F.~X., {Young}, P.~A., {Pagano}, M.~D., \&
  {Turnbull}, M.~C. 2014, \aj, 148, 54

\bibitem[{{Holtzman} {et~al.}(2018){Holtzman}, {Hasselquist}, {Shetrone},
  {Cunha}, {Allende Prieto}, {Anguiano}, {Bizyaev}, {Bovy}, {Casey},
  {Edvardsson}, {Johnson}, {J{\"o}nsson}, {Meszaros}, {Smith}, {Sobeck},
  {Zamora}, {Chojnowski}, {Fernandez-Trincado}, {Garcia-Hernandez}, {Majewski},
  {Pinsonneault}, {Souto}, {Stringfellow}, {Tayar}, {Troup}, \&
  {Zasowski}}]{holtzman18}
{Holtzman}, J.~A., {Hasselquist}, S., {Shetrone}, M., {et~al.} 2018, \aj, 156,
  125

\bibitem[{{Holtzman} {et~al.}(2015){Holtzman}, {Shetrone}, {Johnson}, {Allende
  Prieto}, {Anders}, {Andrews}, {Beers}, {Bizyaev}, {Blanton}, {Bovy},
  {Carrera}, {Chojnowski}, {Cunha}, {Eisenstein}, {Feuillet}, {Frinchaboy},
  {Galbraith-Frew}, {Garc{\'\i}a P{\'e}rez}, {Garc{\'\i}a-Hern{\'a}ndez},
  {Hasselquist}, {Hayden}, {Hearty}, {Ivans}, {Majewski}, {Martell},
  {Meszaros}, {Muna}, {Nidever}, {Nguyen}, {O'Connell}, {Pan}, {Pinsonneault},
  {Robin}, {Schiavon}, {Shane}, {Sobeck}, {Smith}, {Troup}, {Weinberg},
  {Wilson}, {Wood-Vasey}, {Zamora}, \& {Zasowski}}]{holtzman15}
{Holtzman}, J.~A., {Shetrone}, M., {Johnson}, J.~A., {et~al.} 2015, \aj, 150,
  148

\bibitem[{{Hubeny} \& {Lanz}(2011)}]{hubeny11}
{Hubeny}, I. \& {Lanz}, T. 2011, {Synspec: General Spectrum Synthesis Program},
  Astrophysics Source Code Library, record ascl:1109.022

\bibitem[{Hunter(2007)}]{hunter07}
Hunter, J.~D. 2007, Computing in Science \& Engineering, 9, 90

\bibitem[{{Iwamoto} {et~al.}(1999){Iwamoto}, {Brachwitz}, {Nomoto},
  {Kishimoto}, {Umeda}, {Hix}, \& {Thielemann}}]{Iwamoto1999}
{Iwamoto}, K., {Brachwitz}, F., {Nomoto}, K., {et~al.} 1999, \apjs, 125, 439

\bibitem[{{Jacobson} {et~al.}(2014){Jacobson}, {Thanathibodee}, {Frebel},
  {Roederer}, {Cescutti}, \& {Matteucci}}]{jacobson14}
{Jacobson}, H.~R., {Thanathibodee}, T., {Frebel}, A., {et~al.} 2014, \apjl,
  796, L24

\bibitem[{{J{\"o}nsson} {et~al.}(2020){J{\"o}nsson}, {Holtzman}, {Allende
  Prieto}, {Cunha}, {Garc{\'\i}a-Hern{\'a}ndez}, {Hasselquist}, {Masseron},
  {Osorio}, {Shetrone}, {Smith}, {Stringfellow}, {Bizyaev}, {Edvardsson},
  {Majewski}, {M{\'e}sz{\'a}ros}, {Souto}, {Zamora}, {Beaton}, {Bovy}, {Donor},
  {Pinsonneault}, {Poovelil}, \& {Sobeck}}]{jonsson20}
{J{\"o}nsson}, H., {Holtzman}, J.~A., {Allende Prieto}, C., {et~al.} 2020, \aj,
  160, 120

\bibitem[{{Jorissen} {et~al.}(2016){Jorissen}, {Van Eck}, {Van Winckel},
  {Merle}, {Boffin}, {Andersen}, {Nordstr{\"o}m}, {Udry}, {Masseron},
  {Lenaerts}, \& {Waelkens}}]{jorissen16}
{Jorissen}, A., {Van Eck}, S., {Van Winckel}, H., {et~al.} 2016, \aap, 586,
  A158

\bibitem[{{Karakas} \& {Lugaro}(2016)}]{Karakas16}
{Karakas}, A.~I. \& {Lugaro}, M. 2016, \apj, 825, 26

\bibitem[{{Kobayashi} {et~al.}(2020){Kobayashi}, {Karakas}, \&
  {Lugaro}}]{kobayashi:20}
{Kobayashi}, C., {Karakas}, A.~I., \& {Lugaro}, M. 2020, \apj, 900, 179

\bibitem[{{Kobayashi} {et~al.}(2011){Kobayashi}, {Karakas}, \&
  {Umeda}}]{kobayashi:11}
{Kobayashi}, C., {Karakas}, A.~I., \& {Umeda}, H. 2011, \mnras, 414, 3231

\bibitem[{{Kobayashi} {et~al.}(2006){Kobayashi}, {Umeda}, {Nomoto}, {Tominaga},
  \& {Ohkubo}}]{kobayashi06}
{Kobayashi}, C., {Umeda}, H., {Nomoto}, K., {Tominaga}, N., \& {Ohkubo}, T.
  2006, \apj, 653, 1145

\bibitem[{{Kounkel} {et~al.}(2021){Kounkel}, {Covey}, {Stassun},
  {Price-Whelan}, {Holtzman}, {Chojnowski}, {Longa-Pe{\~n}a},
  {Rom{\'a}n-Z{\'u}{\~n}iga}, {Hernandez}, {Serna}, {Badenes}, {De Lee},
  {Majewski}, {Stringfellow}, {Kratter}, {Moe}, {Frinchaboy}, {Beaton},
  {Fern{\'a}ndez-Trincado}, {Mahadevan}, {Minniti}, {Beers}, {Schneider},
  {Barba}, {Brownstein}, {Garc{\'\i}a-Hern{\'a}ndez}, {Pan}, \&
  {Bizyaev}}]{kounkel21}
{Kounkel}, M., {Covey}, K.~R., {Stassun}, K.~G., {et~al.} 2021, \aj, 162, 184

\bibitem[{{Kozyreva} {et~al.}(2014){Kozyreva}, {Yoon}, \&
  {Langer}}]{kozyreva14}
{Kozyreva}, A., {Yoon}, S.~C., \& {Langer}, N. 2014, \aap, 566, A146

\bibitem[{{Kroupa}(2001)}]{Kroupa2001}
{Kroupa}, P. 2001, \mnras, 322, 231

\bibitem[{{Leung} \& {Nomoto}(2018)}]{leung18}
{Leung}, S.-C. \& {Nomoto}, K. 2018, \apj, 861, 143

\bibitem[{{Li} {et~al.}(2015){Li}, {Gordon}, {Rothman}, {Tan}, {Hu}, {Kassi},
  {Campargue}, \& {Medvedev}}]{li15CO}
{Li}, G., {Gordon}, I.~E., {Rothman}, L.~S., {et~al.} 2015, \apjs, 216, 15

\bibitem[{{Limongi} \& {Chieffi}(2018)}]{limongi:18}
{Limongi}, M. \& {Chieffi}, A. 2018, \apjs, 237, 13

\bibitem[{{Maas} {et~al.}(2019){Maas}, {Cescutti}, \& {Pilachowski}}]{maas19}
{Maas}, Z.~G., {Cescutti}, G., \& {Pilachowski}, C.~A. 2019, \aj, 158, 219

\bibitem[{{Maas} {et~al.}(2017){Maas}, {Pilachowski}, \& {Cescutti}}]{maas17}
{Maas}, Z.~G., {Pilachowski}, C.~A., \& {Cescutti}, G. 2017, \apj, 841, 108

\bibitem[{{Mackereth} \& {Bovy}(2018)}]{mackereth18VAC2}
{Mackereth}, J.~T. \& {Bovy}, J. 2018, \pasp, 130, 114501

\bibitem[{{Majewski} {et~al.}(2017){Majewski}, {Schiavon}, {Frinchaboy},
  {Allende Prieto}, {Barkhouser}, {Bizyaev}, {Blank}, {Brunner}, {Burton},
  {Carrera}, {Chojnowski}, {Cunha}, {Epstein}, {Fitzgerald}, {Garc{\'\i}a
  P{\'e}rez}, {Hearty}, {Henderson}, {Holtzman}, {Johnson}, {Lam}, {Lawler},
  {Maseman}, {M{\'e}sz{\'a}ros}, {Nelson}, {Nguyen}, {Nidever}, {Pinsonneault},
  {Shetrone}, {Smee}, {Smith}, {Stolberg}, {Skrutskie}, {Walker}, {Wilson},
  {Zasowski}, {Anders}, {Basu}, {Beland}, {Blanton}, {Bovy}, {Brownstein},
  {Carlberg}, {Chaplin}, {Chiappini}, {Eisenstein}, {Elsworth}, {Feuillet},
  {Fleming}, {Galbraith-Frew}, {Garc{\'\i}a}, {Garc{\'\i}a-Hern{\'a}ndez},
  {Gillespie}, {Girardi}, {Gunn}, {Hasselquist}, {Hayden}, {Hekker}, {Ivans},
  {Kinemuchi}, {Klaene}, {Mahadevan}, {Mathur}, {Mosser}, {Muna}, {Munn},
  {Nichol}, {O'Connell}, {Parejko}, {Robin}, {Rocha-Pinto}, {Schultheis},
  {Serenelli}, {Shane}, {Silva Aguirre}, {Sobeck}, {Thompson}, {Troup},
  {Weinberg}, \& {Zamora}}]{majewski17APOGEE1}
{Majewski}, S.~R., {Schiavon}, R.~P., {Frinchaboy}, P.~M., {et~al.} 2017, \aj,
  154, 94

\bibitem[{{Masseron} {et~al.}(2019){Masseron}, {Garc{\'\i}a-Hern{\'a}ndez},
  {M{\'e}sz{\'a}ros}, {Zamora}, {Dell'Agli}, {Allende Prieto}, {Edvardsson},
  {Shetrone}, {Plez}, {Fern{\'a}ndez-Trincado}, {Cunha}, {J{\"o}nsson},
  {Geisler}, {Beers}, \& {Cohen}}]{masseron19}
{Masseron}, T., {Garc{\'\i}a-Hern{\'a}ndez}, D.~A., {M{\'e}sz{\'a}ros}, S.,
  {et~al.} 2019, \aap, 622, A191

\bibitem[{{Masseron} {et~al.}(2020{\natexlab{a}}){Masseron},
  {Garc{\'\i}a-Hern{\'a}ndez}, {Santove{\~n}a}, {Manchado}, {Zamora},
  {Manteiga}, \& {Dafonte}}]{masseron20a}
{Masseron}, T., {Garc{\'\i}a-Hern{\'a}ndez}, D.~A., {Santove{\~n}a}, R.,
  {et~al.} 2020{\natexlab{a}}, Nature Communications, 11, 3759

\bibitem[{{Masseron} {et~al.}(2020{\natexlab{b}}){Masseron},
  {Garc{\'\i}a-Hern{\'a}ndez}, {Zamora}, \& {Manchado}}]{masseron20b}
{Masseron}, T., {Garc{\'\i}a-Hern{\'a}ndez}, D.~A., {Zamora}, O., \&
  {Manchado}, A. 2020{\natexlab{b}}, \apjl, 904, L1

\bibitem[{{Masseron} \& {Gilmore}(2015)}]{masseron15}
{Masseron}, T. \& {Gilmore}, G. 2015, \mnras, 453, 1855

\bibitem[{{Masseron} {et~al.}(2016){Masseron}, {Merle}, \&
  {Hawkins}}]{masseron2016}
{Masseron}, T., {Merle}, T., \& {Hawkins}, K. 2016, {BACCHUS: Brussels
  Automatic Code for Characterizing High accUracy Spectra}, Astrophysics Source
  Code Library [\href{http://ascl.net/1605.004}{record ascl:1605.004}]

\bibitem[{Matteucci(2012)}]{Matteucci2012}
Matteucci, F. 2012, Chemical evolution of the Galaxy (Springer)

\bibitem[{{Matteucci}(2021)}]{matteucci:21}
{Matteucci}, F. 2021, \aapr, 29, 5

\bibitem[{{M{\'e}sz{\'a}ros} {et~al.}(2020){M{\'e}sz{\'a}ros}, {Masseron},
  {Garc{\'\i}a-Hern{\'a}ndez}, {Allende Prieto}, {Beers}, {Bizyaev},
  {Chojnowski}, {Cohen}, {Cunha}, {Dell'Agli}, {Ebelke},
  {Fern{\'a}ndez-Trincado}, {Frinchaboy}, {Geisler}, {Hasselquist}, {Hearty},
  {Holtzman}, {Johnson}, {Lane}, {Lacerna}, {Longa-Pe{\~n}a}, {Majewski},
  {Martell}, {Minniti}, {Nataf}, {Nidever}, {Pan}, {Schiavon}, {Shetrone},
  {Smith}, {Sobeck}, {Stringfellow}, {Szigeti}, {Tang}, {Wilson}, \&
  {Zamora}}]{meszaros20}
{M{\'e}sz{\'a}ros}, S., {Masseron}, T., {Garc{\'\i}a-Hern{\'a}ndez}, D.~A.,
  {et~al.} 2020, \mnras, 492, 1641

\bibitem[{{Mishenina} {et~al.}(2017){Mishenina}, {Pignatari}, {C{\^o}t{\'e}},
  {Thielemann}, {Soubiran}, {Basak}, {Gorbaneva}, {Korotin}, {Kovtyukh},
  {Wehmeyer}, {Bisterzo}, {Travaglio}, {Gibson}, {Jordan}, {Paul}, {Ritter},
  {Herwig}, \& {NuGrid Collaboration}}]{mishenina:17}
{Mishenina}, T., {Pignatari}, M., {C{\^o}t{\'e}}, B., {et~al.} 2017, \mnras,
  469, 4378

\bibitem[{{Moll{\'a}} {et~al.}(2015){Moll{\'a}}, {Cavichia}, {Gavil{\'a}n}, \&
  {Gibson}}]{molla:15}
{Moll{\'a}}, M., {Cavichia}, O., {Gavil{\'a}n}, M., \& {Gibson}, B.~K. 2015,
  \mnras, 451, 3693

\bibitem[{{Nandakumar} {et~al.}(2022){Nandakumar}, {Ryde}, {Montelius},
  {Thorsbro}, {J{\"o}nsson}, {Mace}, {Observatory}, {Astronomy}, {Physics},
  {University}, {43}, {Lund}, {Sweden}, {Astronomical Institute}, {Groningen},
  {12}, {Groningen}, {Netherlands}, {Astronomy}, {Science}, {Tokyo}, {Hongo},
  {Bunkyo-ku}, {113-0033}, {Japan}, {Science}, {Mathematics}, {University},
  {Malm{\"o}}, {Sweden}, {Astronomy}, {Observatory}, {Texas}, {Austin},
  {78712}, \& {USA}}]{nandakumar22}
{Nandakumar}, G., {Ryde}, N., {Montelius}, M., {et~al.} 2022, arXiv e-prints,
  arXiv:2210.04940

\bibitem[{{Nidever} {et~al.}(2015){Nidever}, {Holtzman}, {Allende Prieto},
  {Beland}, {Bender}, {Bizyaev}, {Burton}, {Desphande}, {Fleming}, {Garc{\'\i}a
  P{\'e}rez}, {Hearty}, {Majewski}, {M{\'e}sz{\'a}ros}, {Muna}, {Nguyen},
  {Schiavon}, {Shetrone}, {Skrutskie}, {Sobeck}, \& {Wilson}}]{nidever15}
{Nidever}, D.~L., {Holtzman}, J.~A., {Allende Prieto}, C., {et~al.} 2015, \aj,
  150, 173

\bibitem[{{Nomoto} {et~al.}(2013){Nomoto}, {Kobayashi}, \&
  {Tominaga}}]{Nomoto2013}
{Nomoto}, K., {Kobayashi}, C., \& {Tominaga}, N. 2013, \araa, 51, 457

\bibitem[{Nomoto \& Leung(2018)}]{nomoto:18}
Nomoto, K. \& Leung, S.-C. 2018, Space Science Reviews, 214, 67

\bibitem[{{Osorio} {et~al.}(2020){Osorio}, {Allende Prieto}, {Hubeny},
  {M{\'e}sz{\'a}ros}, \& {Shetrone}}]{osorio20}
{Osorio}, Y., {Allende Prieto}, C., {Hubeny}, I., {M{\'e}sz{\'a}ros}, S., \&
  {Shetrone}, M. 2020, \aap, 637, A80

\bibitem[{Pagel(1997)}]{Pagel1997}
Pagel, B. E.~J. 1997, Nucleosynthesis and Chemical Evolution of the Galaxies
  (Cambridge University Press)

\bibitem[{{Pignatari} {et~al.}(2016){Pignatari}, {Herwig}, {Hirschi},
  {Bennett}, {Rockefeller}, {Fryer}, {Timmes}, {Ritter}, {Heger}, {Jones},
  {Battino}, {Dotter}, {Trappitsch}, {Diehl}, {Frischknecht}, {Hungerford},
  {Magkotsios}, {Travaglio}, \& {Young}}]{pignatari:16}
{Pignatari}, M., {Herwig}, F., {Hirschi}, R., {et~al.} 2016, \apjs, 225, 24

\bibitem[{{Plez}(2012)}]{plez12}
{Plez}, B. 2012, {Turbospectrum: Code for spectral synthesis}, Astrophysics
  Source Code Library [\href{http://ascl.net/1205.004}{record ascl:1205.004}]

\bibitem[{{Prantzos} {et~al.}(2018){Prantzos}, {Abia}, {Limongi}, {Chieffi}, \&
  {Cristallo}}]{prantzos:18}
{Prantzos}, N., {Abia}, C., {Limongi}, M., {Chieffi}, A., \& {Cristallo}, S.
  2018, \mnras, 476, 3432

\bibitem[{{Rauscher} {et~al.}(2002){Rauscher}, {Heger}, {Hoffman}, \&
  {Woosley}}]{rauscher:02}
{Rauscher}, T., {Heger}, A., {Hoffman}, R.~D., \& {Woosley}, S.~E. 2002, \apj,
  576, 323

\bibitem[{Reback {et~al.}(2022)Reback, jbrockmendel, McKinney, den Bossche,
  Augspurger, Cloud, Hawkins, Roeschke, gfyoung, Sinhrks, Klein, Hoefler,
  Petersen, Tratner, She, Ayd, Naveh, Garcia, Darbyshire, Schendel, Shadrach,
  Hayden, Saxton, Gorelli, Li, Zeitlin, Jancauskas, McMaster, Battiston, \&
  Seabold}]{reback2020pandas}
Reback, J., jbrockmendel, McKinney, W., {et~al.} 2022

\bibitem[{{Ritter} {et~al.}(2018{\natexlab{a}}){Ritter}, {Andrassy},
  {C{\^o}t{\'e}}, {Herwig}, {Woodward}, {Pignatari}, \& {Jones}}]{ritter:18}
{Ritter}, C., {Andrassy}, R., {C{\^o}t{\'e}}, B., {et~al.} 2018{\natexlab{a}},
  \mnras, 474, L1

\bibitem[{{Ritter} {et~al.}(2018{\natexlab{b}}){Ritter}, {C{\^o}t{\'e}},
  {Herwig}, {Navarro}, \& {Fryer}}]{Ritter2018c}
{Ritter}, C., {C{\^o}t{\'e}}, B., {Herwig}, F., {Navarro}, J.~F., \& {Fryer},
  C.~L. 2018{\natexlab{b}}, \apjs, 237, 42

\bibitem[{{Ritter} {et~al.}(2018{\natexlab{c}}){Ritter}, {Herwig}, {Jones},
  {Pignatari}, {Fryer}, \& {Hirschi}}]{Ritter2018b}
{Ritter}, C., {Herwig}, F., {Jones}, S., {et~al.} 2018{\natexlab{c}}, \mnras,
  480, 538

\bibitem[{{Roederer} {et~al.}(2014){Roederer}, {Jacobson}, {Thanathibodee},
  {Frebel}, \& {Toller}}]{roederer14}
{Roederer}, I.~U., {Jacobson}, H.~R., {Thanathibodee}, T., {Frebel}, A., \&
  {Toller}, E. 2014, \apj, 797, 69

\bibitem[{{Romano} {et~al.}(2010){Romano}, {Karakas}, {Tosi}, \&
  {Matteucci}}]{romano:10}
{Romano}, D., {Karakas}, A.~I., {Tosi}, M., \& {Matteucci}, F. 2010, \aap, 522,
  A32

\bibitem[{{Sanders} {et~al.}(2021){Sanders}, {Belokurov}, \& {Man}}]{sanders21}
{Sanders}, J.~L., {Belokurov}, V., \& {Man}, K. T.~F. 2021, \mnras, 506, 4321

\bibitem[{{Schatz} {et~al.}(2022){Schatz}, {Becerril Reyes}, {Best}, {Brown},
  {Chatziioannou}, {Chipps}, {Deibel}, {Ezzeddine}, {Galloway}, {Hansen},
  {Herwig}, {Ji}, {Lugaro}, {Meisel}, {Norman}, {Read}, {Roberts}, {Spyrou},
  {Tews}, {Timmes}, {Travaglio}, {Vassh}, {Abia}, {Adsley}, {Agarwal},
  {Aliotta}, {Aoki}, {Arcones}, {Aryan}, {Bandyopadhyay}, {Banu}, {Bardayan},
  {Barnes}, {Bauswein}, {Beers}, {Bishop}, {Boztepe}, {C{\^o}t{\'e}}, {Caplan},
  {Champagne}, {Clark}, {Couder}, {Couture}, {de Mink}, {Debnath}, {deBoer},
  {den Hartogh}, {Denissenkov}, {Dexheimer}, {Dillmann}, {Escher}, {Famiano},
  {Farmer}, {Fisher}, {Fr{\"o}hlich}, {Frebel}, {Fryer}, {Fuller}, {Ganguly},
  {Ghosh}, {Gibson}, {Gorda}, {Gourgouliatos}, {Graber}, {Gupta}, {Haxton},
  {Heger}, {Hix}, {Ho}, {Holmbeck}, {Hood}, {Huth}, {Imbriani}, {Izzard},
  {Jain}, {Jayatissa}, {Johnston}, {Kajino}, {Kankainen}, {Kiss},
  {Kwiatkowski}, {La Cognata}, {Laird}, {Lamia}, {Landry}, {Laplace}, {Launey},
  {Leahy}, {Leckenby}, {Lennarz}, {Longfellow}, {Lovell}, {Lynch}, {Lyons},
  {Maeda}, {Masha}, {Matei}, {Merc}, {Messer}, {Montes}, {Mukherjee},
  {Mumpower}, {Neto}, {Nevins}, {Newton}, {Nguyen}, {Nishikawa}, {Nishimura},
  {Nunes}, {O'Connor}, {O'Shea}, {Ong}, {Pain}, {Pajkos}, {Pignatari},
  {Pizzone}, {Placco}, {Plewa}, {Pritychenko}, {Psaltis}, {Puentes}, {Qian},
  {Radice}, {Rapagnani}, {Rebeiro}, {Reifarth}, {Richard}, {Rijal}, {Roederer},
  {Rojo}, {K}, {Saito}, {Schwenk}, {Sergi}, {Sidhu}, {Simon}, {Sivarani},
  {Sk{\'u}lad{\'o}ttir}, {Smith}, {Spiridon}, {Sprouse}, {Starrfield},
  {Steiner}, {Strieder}, {Sultana}, {Surman}, {Sz{\"u}cs}, {Tawfik},
  {Thielemann}, {Trache}, {Trappitsch}, {Tsang}, {Tumino}, {Upadhyayula},
  {Valle Mart{\'\i}nez}, {Van der Swaelmen}, {Viscasillas V{\'a}zquez},
  {Watts}, {Wehmeyer}, {Wiescher}, {Wrede}, {Yoon}, {Zegers}, {Zermane}, \&
  {Zingale}}]{schatz:22}
{Schatz}, H., {Becerril Reyes}, A.~D., {Best}, A., {et~al.} 2022, arXiv
  e-prints, arXiv:2205.07996

\bibitem[{{Schiavon} {et~al.}(2017){Schiavon}, {Zamora}, {Carrera},
  {Lucatello}, {Robin}, {Ness}, {Martell}, {Smith},
  {Garc{\'\i}a-Hern{\'a}ndez}, {Manchado}, {Sch{\"o}nrich}, {Bastian},
  {Chiappini}, {Shetrone}, {Mackereth}, {Williams}, {M{\'e}sz{\'a}ros},
  {Allende Prieto}, {Anders}, {Bizyaev}, {Beers}, {Chojnowski}, {Cunha},
  {Epstein}, {Frinchaboy}, {Garc{\'\i}a P{\'e}rez}, {Hearty}, {Holtzman},
  {Johnson}, {Kinemuchi}, {Majewski}, {Muna}, {Nidever}, {Nguyen}, {O'Connell},
  {Oravetz}, {Pan}, {Pinsonneault}, {Schneider}, {Schultheis}, {Simmons},
  {Skrutskie}, {Sobeck}, {Wilson}, \& {Zasowski}}]{schiavon17}
{Schiavon}, R.~P., {Zamora}, O., {Carrera}, R., {et~al.} 2017, \mnras, 465, 501

\bibitem[{{Shetrone} {et~al.}(2015){Shetrone}, {Bizyaev}, {Lawler}, {Allende
  Prieto}, {Johnson}, {Smith}, {Cunha}, {Holtzman}, {Garc{\'\i}a P{\'e}rez},
  {M{\'e}sz{\'a}ros}, {Sobeck}, {Zamora}, {Garc{\'\i}a-Hern{\'a}ndez}, {Souto},
  {Chojnowski}, {Koesterke}, {Majewski}, \& {Zasowski}}]{shetrone15}
{Shetrone}, M., {Bizyaev}, D., {Lawler}, J.~E., {et~al.} 2015, \apjs, 221, 24

\bibitem[{{Smith} {et~al.}(2021){Smith}, {Bizyaev}, {Cunha}, {Shetrone},
  {Souto}, {Allende Prieto}, {Masseron}, {M{\'e}sz{\'a}ros}, {J{\"o}nsson},
  {Hasselquist}, {Osorio}, {Garc{\'\i}a-Hern{\'a}ndez}, {Plez}, {Beaton},
  {Holtzman}, {Majewski}, {Stringfellow}, \& {Sobeck}}]{smith21}
{Smith}, V.~V., {Bizyaev}, D., {Cunha}, K., {et~al.} 2021, \aj, 161, 254

\bibitem[{{Sneden} {et~al.}(2014){Sneden}, {Lucatello}, {Ram}, {Brooke}, \&
  {Bernath}}]{sneden14CN}
{Sneden}, C., {Lucatello}, S., {Ram}, R.~S., {Brooke}, J. S.~A., \& {Bernath},
  P. 2014, \apjs, 214, 26

\bibitem[{{Sukhbold} {et~al.}(2016){Sukhbold}, {Ertl}, {Woosley}, {Brown}, \&
  {Janka}}]{sukhbold:16}
{Sukhbold}, T., {Ertl}, T., {Woosley}, S.~E., {Brown}, J.~M., \& {Janka}, H.~T.
  2016, \apj, 821, 38

\bibitem[{{Takahashi} {et~al.}(2018){Takahashi}, {Yoshida}, \&
  {Umeda}}]{takahashi18}
{Takahashi}, K., {Yoshida}, T., \& {Umeda}, H. 2018, \apj, 857, 111

\bibitem[{{Thielemann} {et~al.}(1996){Thielemann}, {Nomoto}, \&
  {Hashimoto}}]{thielemann:96}
{Thielemann}, F.-K., {Nomoto}, K., \& {Hashimoto}, M.-A. 1996, \apj, 460, 408

\bibitem[{Tinsley(1980)}]{Tinsley1980}
Tinsley, B.~M. 1980, Fundamentals of Cosmic Physics, 5, 287

\bibitem[{{Weinberg} {et~al.}(2017){Weinberg}, {Andrews}, \&
  {Freudenburg}}]{weinberg17}
{Weinberg}, D.~H., {Andrews}, B.~H., \& {Freudenburg}, J. 2017, \apj, 837, 183

\bibitem[{{Wilson} {et~al.}(2019){Wilson}, {Hearty}, {Skrutskie}, {Majewski},
  {Holtzman}, {Eisenstein}, {Gunn}, {Blank}, {Henderson}, {Smee}, {Nelson},
  {Nidever}, {Arns}, {Barkhouser}, {Barr}, {Beland}, {Bershady}, {Blanton},
  {Brunner}, {Burton}, {Carey}, {Carr}, {Colque}, {Crane}, {Damke}, {Davidson},
  {Dean}, {Di Mille}, {Don}, {Ebelke}, {Evans}, {Fitzgerald}, {Gillespie},
  {Hall}, {Harding}, {Harding}, {Hammond}, {Hancock}, {Harrison}, {Hope},
  {Horne}, {Karakla}, {Lam}, {Leger}, {MacDonald}, {Maseman}, {Matsunari},
  {Melton}, {Mitcheltree}, {O'Brien}, {O'Connell}, {Patten}, {Richardson},
  {Rieke}, {Rieke}, {Roman-Lopes}, {Schiavon}, {Sobeck}, {Stolberg}, {Stoll},
  {Tembe}, {Trujillo}, {Uomoto}, {Vernieri}, {Walker}, {Weinberg}, {Young},
  {Anthony-Brumfield}, {Bizyaev}, {Breslauer}, {De Lee}, {Downey}, {Halverson},
  {Huehnerhoff}, {Klaene}, {Leon}, {Long}, {Mahadevan}, {Malanushenko},
  {Nguyen}, {Owen}, {S{\'a}nchez-Gallego}, {Sayres}, {Shane}, {Shectman},
  {Shetrone}, {Skinner}, {Stauffer}, \& {Zhao}}]{wilson19APOSpec}
{Wilson}, J.~C., {Hearty}, F.~R., {Skrutskie}, M.~F., {et~al.} 2019, \pasp,
  131, 055001

\bibitem[{{Woosley} {et~al.}(2002){Woosley}, {Heger}, \& {Weaver}}]{woosley:02}
{Woosley}, S.~E., {Heger}, A., \& {Weaver}, T.~A. 2002, Reviews of Modern
  Physics, 74, 1015

\bibitem[{{Woosley} \& {Weaver}(1995)}]{woosley:95}
{Woosley}, S.~E. \& {Weaver}, T.~A. 1995, \apjs, 101, 181

\bibitem[{{Yadav} {et~al.}(2020){Yadav}, {M{\"u}ller}, {Janka}, {Melson}, \&
  {Heger}}]{yadav:20}
{Yadav}, N., {M{\"u}ller}, B., {Janka}, H.~T., {Melson}, T., \& {Heger}, A.
  2020, \apj, 890, 94

\bibitem[{{Yong} {et~al.}(2008){Yong}, {Karakas}, {Lambert}, {Chieffi}, \&
  {Limongi}}]{yong08}
{Yong}, D., {Karakas}, A.~I., {Lambert}, D.~L., {Chieffi}, A., \& {Limongi}, M.
  2008, \apj, 689, 1031

\bibitem[{{Zamora} {et~al.}(2015){Zamora}, {Garc{\'\i}a-Hern{\'a}ndez},
  {Allende Prieto}, {Carrera}, {Koesterke}, {Edvardsson}, {Castelli}, {Plez},
  {Bizyaev}, {Cunha}, {Garc{\'\i}a P{\'e}rez}, {Gustafsson}, {Holtzman},
  {Lawler}, {Majewski}, {Manchado}, {M{\'e}sz{\'a}ros}, {Shane}, {Shetrone},
  {Smith}, \& {Zasowski}}]{zamora15}
{Zamora}, O., {Garc{\'\i}a-Hern{\'a}ndez}, D.~A., {Allende Prieto}, C.,
  {et~al.} 2015, \aj, 149, 181

\end{thebibliography}

\clearpage
\onecolumn % looks better but maybe not allowed. check when all appendices are finished
%______________________________________________________________

\begin{appendix}

\section{Selection of the background sample\label{BackgroundSelection}}
To define the background sample, we applied the following criteria to the full APOGEE-2 DR17 data contained in the ASPCAP table\footnote{\url{https://data.sdss.org/sas/dr17/apogee/spectro/aspcap/dr17/synspec/}}:
\begin{enumerate}
    \item $\unit[4000]{K} < T_{eff} < \unit[5000]{K}$
    \item log g $< 3.5$
    \item RV scatter $< \unit[1]{km~s^{-1}}$
    \item $\unit[-1.5]{dex} <$ [M/H] $ < \unit[-0.6]{dex}$
    \item $\unit[56]{px^{-1}}\lesssim$ SNR $\lesssim \unit[640]{px^{-1}}$
    \item ASPCAPFLAG = 0; STARFLAG = 0; ANDFLAG = 0.
\end{enumerate}
These conditions mainly serve to filter out reference stars with similar stellar parameters as the P-star candidates, but also to reject possible binary members\footnote{We note that binary systems are known to show abundance anomalies \citep[e.g.,][and references therein]{jorissen16} that are not representative of standard field stars and the global chemical enrichment of the Galaxy.} by restricting the radial velocity (RV) scatter (criterion 3). We also rejected targets with flagged entries (criterion 6). These quality flags are given in the form of bitmasks, which may indicate several issues in the ASPCAP results. In short, ASPCAPFLAG takes into account any potential problem with the derived stellar parameters, while STARFLAG indicates issues in the spectra, for example, a high number of bad pixels \citep{holtzman15}. The ANDFLAG bitmask has the same meaning as STARFLAG, indicating whether all the spectra used for the final combination present the same issue \citep{holtzman18}. In addition, both ASPCAPFLAG and STARFLAG may also identify possible spectroscopic binaries \citep{abdurrouf22APOGEEDR17}. \\
Similar to the case of binary systems mentioned above, giant stars in Galactic globular clusters (GCs) are also well known to display chemical abundance anomalies, especially the so-called second-generation stars. They neither represent the standard chemical composition of field stars nor the average Galactic chemical enrichment \citep[see, e.g.,][and references therein]{meszaros20}. Thus, we aimed to exclude members of Galactic GCs from our reference group by cross-matching the reference stars with the GC stars contained in Table 2 of \citet{meszaros20}. In addition, we cross-matched with another GC catalog that has been created by selecting stars in APOGEE-2 DR17 within the tidal radius of GCs in the \citet[][2010 edition]{harris1996} GC catalog, which are selected as GC members according to an iterative sigma clipping on RVs and {\it Gaia} EDR3 proper motions \citep{gaiaedr3}.

\section{Line list\label{LineList}}

\begin{ThreePartTable}

\begin{TableNotes}
\item \textbf{Notes.} Measurements of C, N, and O are predominantly based on molecular lines of CO, CN, and OH, respectively. For the values of excitation potential and log(gf), we directly refer to the cited lists because several lines may contribute to the flux at the given wavelength. The values corresponding to the five C I lines as well as the other atomic lines were taken from the APOGEE DR17 line list \citep{smith21,shetrone15}. Wavelengths in parentheses were removed from the line list of the automated abundance calculation of the background sample as they have been identified to provide problematic results during the visual examination of the sample fits.
\end{TableNotes}

\begin{longtable}{c c c c }%[ht]
\caption{\label{tab:LL} Spectral lines alongside the corresponding excitation potential and log(gf) used for the abundance determination of each elemental species considered in this work. } \\        
\hline\hline                 
Element & Wavelength [$\angstrom$] & Excitation potential [eV] & log(gf) \\    
\hline
\endfirsthead
\caption{continued.} \\
\hline\hline
Species & Wavelength [$\angstrom$] & Excitation potential [eV] & log(gf) \\ 
\hline
\endhead
   \multirow{12}{*}{C} & (15578.0) &  \multicolumn{2}{c}{\multirow{4}{*}{see \citet{li15CO}}} \\ 
   & (15775.5) &  &  \\
   & (15783.9) &  &  \\
   & 15978.7 &  &  \\
   & (C I: 16004.9) & 9.631 & 0.434 \\
   & (C I: 16021.7) & 9.631 & 0.163 \\
   & 16185.5 & \multicolumn{2}{c}{\multirow{3}{*}{see \citet{li15CO}}}  \\
   & 16397.2 & &  \\
   & 16481.5 & & \\
   & C I: 16613.9 & 10.427 & -1.239 \\
   & C I: 16836.0 & 10.396 & -7.294 \\
   & C I: 16890.4 & 9.002 & 0.503 \\ \hline
   \multirow{14}{*}{N} & 15210.2 & \multicolumn{2}{c}{\multirow{14}{*}{see \citet{sneden14CN}}} \\
   & 15222.0 & & \\
   & (15228.8) & & \\
   & (15242.5) & & \\
   & (15251.8) & & \\
   & 15309.0 & & \\
   & 15317.6 & & \\
   & 15363.5 & &  \\
   & (15410.5) & &  \\
   & 15447.0 & &  \\
   & (15462.4) & &  \\
   & (15466.2) & & \\
   & (15495.0) & & \\
   & 15514.0 & & \\
   \multirow{4}{*}{N} & 15581.0 & \multicolumn{2}{c}{\multirow{4}{*}{see \citet{sneden14CN}}} \\
   & (15636.5) & & \\
   & (15659.0) & &  \\
   & (15706.9) & &  \\
   & (15708.5) & &  \\ \hline
   \multirow{12}{*}{O} & (15373.5) & \multicolumn{2}{c}{\multirow{12}{*}{see \citet{brooke16OH}}}   \\
   & 15391.0 & &  \\
   & 15569.0 & &  \\
   & 15719.7 & &  \\
   & (15778.5) &  &  \\
   & 16052.8 & &  \\
   & 16055.5 & &  \\
   & 16650.0 & &  \\
   & 16704.8 & &  \\
   & 16714.5 & & \\
   & 16872.0 & & \\
   & 16909.4 &  &  \\ \hline
   \multirow{2}{*}{Na I} & 16373.9 & 3.753 & -1.853 \\
   & 16388.9  & 3.753 & -1.406 \\ \hline
   \multirow{11}{*}{Mg I} & (15231.8) & 6.719 & -1.158 \\
   & (15366.9) & 6.779 & -1.801 \\
   & (15693.6) & 6.719 & -1.650 \\
   & 15740.7 & 5.931 & -0.323 \\
   & 15749.0 & 5.932 &  0.049 \\
   & 15765.8 & 5.933 & 0.320 \\
   & (15879.6) & 5.946 & -1.354 \\
   & (15954.5) & 6.588 & -0.828 \\
   & 16365.0 & 6.719 & -1.469 \\
   & 16624.7 & 6.726 & -1.468 \\
   & 16632.0 & 6.726 & -1.193 \\ \hline
   \multirow{4}{*}{Al I} & (15956.6) & 4.827 & -1.498 \\
   & 16719.0 & 4.085 & -0.186 \\
   & 16750.5 & 4.087 & -0.021 \\
   & 16763.4 & 4.087 & -1.091 \\ \hline
   \multirow{14}{*}{Si I} & (15361.2) & 5.954 & -2.033 \\
   & 15376.8 & 6.223 & -0.701 \\
   & (15557.8) & 5.964 & -0.820 \\
   & (15884.5) & 5.954 & -0.945 \\
   & 15960.1 & 5.984 & 0.130 \\
   & 16060.0 & 5.954 & -0.452 \\
   & (16094.8) & 5.964 & -0.088 \\
   & (16129.0) & 7.139 & -0.878 \\
   & 16163.7 & 5.954 & -0.858 \\
   & (16170.2) & 6.734 & -1.465 \\
   & 16215.7 & 5.954 & -0.565 \\
   & 16241.8 & 5.964 & -0.762 \\
   & 16680.8 & 5.984 & -0.138 \\
   & 16828.2 & 5.984 & -1.058 \\ \hline
   \multirow{2}{*}{P I} & 15711.5 & 7.176 & -0.404 \\
   & 16482.9 & 7.213 & -0.273 \\ \hline
   \multirow{5}{*}{S I} & 15403.8 & 8.699 & 0.348 \\
   & 15422.3 & 8.700 & 0.775 \\
   & 15469.8 & 8.045 & -0.310 \\
   & 15475.6 & 8.045 & -0.744 \\
   & 15478.5 & 8.045 & -0.040 \\ \hline
   \multirow{5}{*}{Ca I} & 16136.8 & 4.531 & -0.585 \\
   & 16150.8 & 4.532 & -0.237 \\
   & 16155.2 & 4.532 & -0.758 \\
   & 16157.4 & 4.554 & -0.219 \\
   & 16197.1 & 4.535 & 0.089 \\ \hline 
   & (15219.6) & 5.587 & -0.665 \\
   \multirow{18}{*}{Fe I} & 15501.3 & 6.286 & 0.115 \\
   & 15648.5 & 5.426 & -0.668 \\
   & 15677.5 & 6.246 &  0.184 \\
   & 15774.1 & 6.299 & 0.352 \\
   & (15892.4) & 6.306 & 0.245 \\
   & (15895.2) & 6.258 & 0.269 \\
   & 15898.0 & 6.306 & 0.343 \\
   & 15906.0 & 5.621 & -0.307 \\
   & 15920.6 & 6.258 & -0.011 \\
   & 16006.8 & 6.347 &  0.481 \\
   & (16102.4) & 5.874 & 0.150 \\
   & 16175.0 & 6.380 & 0.208 \\
   & (16179.6) & 6.319 & 0.118 \\
   & (16185.8) & 6.393 & 0.185 \\
   & (16382.3) & 6.280 & 0.163 \\
   & 16541.4 & 5.946 & -0.416 \\
   & 16561.8 & 5.979 & 0.028 \\
   & 16753.1 & 6.380 & 0.184 \\ \hline
   \multirow{4}{*}{Ce II} & 15784.8 & 0.318 & -1.540 \\
   & 16376.5 & 0.122 & -1.965 \\
   & 16595.2 & 0.122 & -2.190 \\
   & 16722.6 & 0.470 & -1.830 \\ \hline
   \multirow{3}{*}{Nd II} & (15368.1) & 1.264 & -1.550 \\
   & 16053.6 & 0.745 & -2.200 \\
   & 16262.0 & 0.986 & -1.990 \\ \hline
\hline
\insertTableNotes
\end{longtable}

\end{ThreePartTable}

\FloatBarrier
\clearpage

\section{Background star abundance validation\label{LPvalidation}}

\begin{figure*}[ht!]
\sidecaption
  \includegraphics[width=12cm]{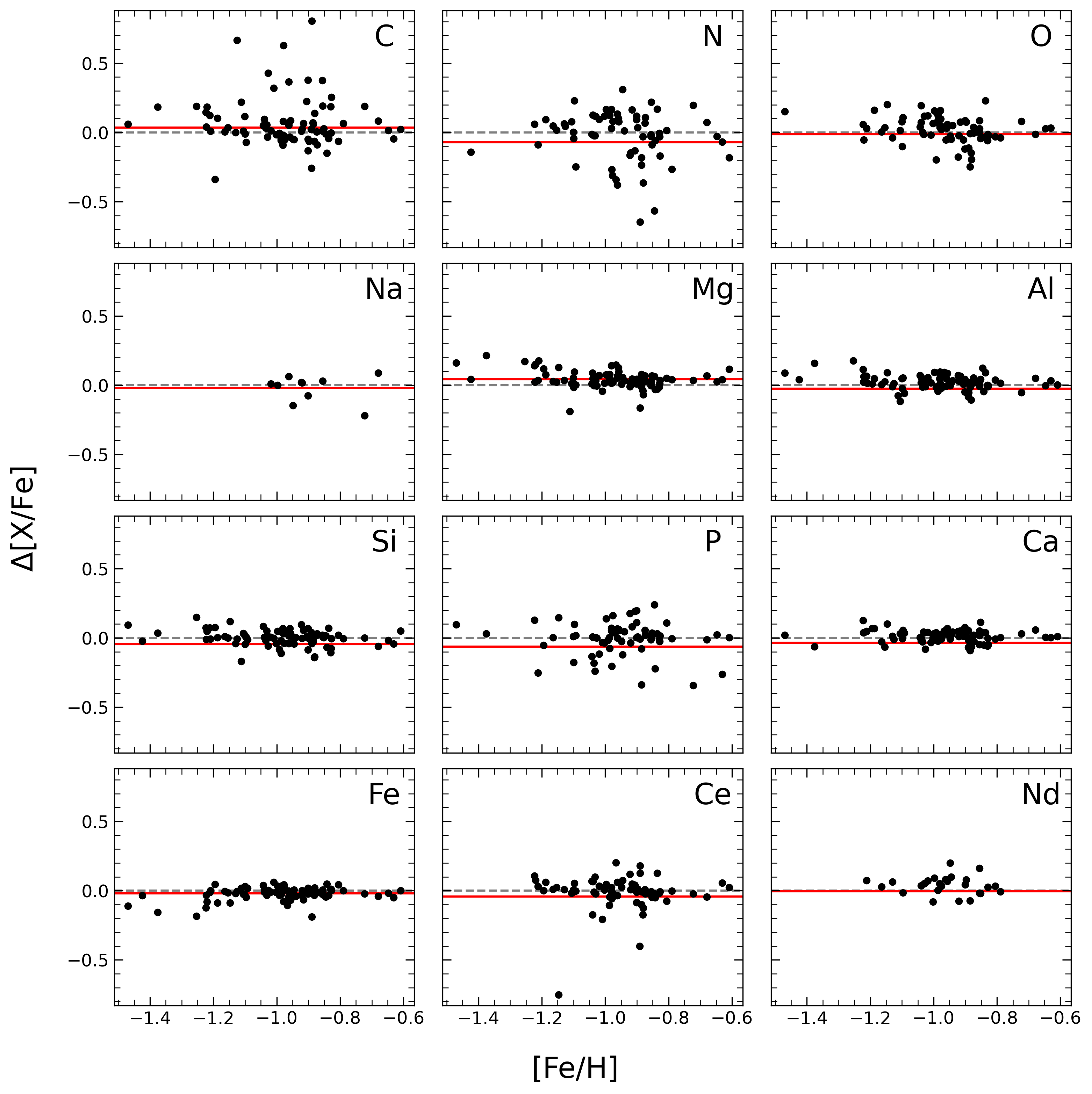}
     \caption{Difference in the abundance ratio $\Delta$[X/Fe] between manual and automated calculation vs. metallicity [Fe/H] for all the elements studied in this work, except for S, because this element was added to our analysis after the calculation with the supercomputer was concluded. The average difference (red line) can be extrapolated to the background sample to validate the performance of the automatic calculation.}
     \label{fig:LP_validation}
\end{figure*}

To probe the quality of the supercomputer calculation of the abundances corresponding to the background stars, we added the initial 87 sample stars to the automated run described in Sect. \ref{LineSelecProcessing}. We then compared their abundances obtained from the automatic run and the manual inspection. As mentioned before, the manual sanity check of the lines was used to refine the line list used for the automatic run (see \ref{tab:LL}) in order to minimize the impact of poorly fitted lines. \\
To illustrate the performance of the automatic abundance calculation, we plot the difference in the abundance ratio $\Delta$[X/Fe] between manual and automated calculation versus metallicity [Fe/H] for all the elements studied in this work (Fig. \ref{fig:LP_validation}), except for S, because this element was included after the calculation with the supercomputer was concluded. The average difference (red line) can be extrapolated to the background sample, giving an impression of the uncertainty related to the automatic calculation mode. Ideally, the average difference should be close to zero, as is the case for mostly all the elements. Nevertheless, because the focus of this work lies on P and its two intrinsically weak lines included in the APOGEE-2 wavelength range, we compared the P abundances of the sample with literature values instead of the background sample, because extracting reliable P abundances require a careful (manual) inspection. In this context, we also point out that the stars contained in the BAWLAS VAC \citep{hayes22} present a metallicity bias, meaning that the number of stars with [P/Fe] values decreases with metallicity. As a result, only a few stars present [P/Fe] values in the metallicity range in which the sample is located. Therefore, the VAC is likewise not suitable for constituting a reference group, which supports our choice of using literature values for comparison.

\FloatBarrier
\clearpage

\section{Uncertainty tables\label{UncertaintyTables}}

\begin{table*}[ht]
    \centering
    \tiny
    \caption{Details of the systematic error calculation.}
    \setlength{\tabcolsep}{3pt}
    \begin{tabular}{c  c  c c c c c c c c c c c c c c} \hline\hline
    & & [C/Fe] & [N/Fe] & [O/Fe] & [Na/Fe] & [Mg/Fe] & [Al/Fe] & [Si/Fe] & [P/Fe] & [S/Fe] & [Ca/Fe] & [Fe/H] & [Ce/Fe] & [Nd/Fe] \\
    \hline 
    \multirow{11}{*}{2M16441013-1850478} & original & -0.01 & 0.74 & 1.08 & 0.77 & 0.76 & 1.22 & 0.93 & 1.24 & 0.11 & 0.38 & -0.84 & 0.86 & 1.24 \\ \cline{2-15} & T$_{eff}+$ & -0.11 & 0.77 & 1.12 & 0.81 & 0.76 & 1.26 & 0.9 & 1.14 & 0.02 & 0.38 & -0.78 & 0.87 & 1.23 \\
   & T$_{eff}-$ & 0.09 & 0.70 & 1.01 & 0.72 & 0.73 & 1.16 & 0.96 & 1.31 & 0.15 & 0.37 & -0.89 & 0.85 & 1.31 \\ \cline{2-15} & $log g+$ & 0.11 & 0.78 & 1.04 & 0.78 & 0.68 & 1.13 & 0.90 & 1.31 & 0.18 & 0.34 & -0.83 & 0.99 & 1.38 \\ & $log g-$ & -0.10 & 0.70 & 1.05 & 0.78 & 0.81 & 1.30 & 0.94 & 1.14 & 0.06 & 0.40 & -0.84 & 0.73 & 1.19 \\ \cline{2-15} & $\mu_{t}+$ & 0.01 & 0.74 & 1.04 & 0.77 & 0.75 & 1.21 & 0.92 & 1.23 & 0.11 & 0.37 & -0.84 & 0.85 & 1.31 \\ & $\mu_{t}-$ & 0.01 & 0.73 & 1.04 & 0.76 & 0.75 & 1.23 & 0.93 & 1.23 & 0.10 & 0.37 & -0.83 & 0.86 & 1.32 \\ \cline{2-15} & $CONVOL+$ & 0.10 & 0.74 & 1.04 & 0.78 & 0.74 & 1.24 & 0.93 & 1.22 & 0.11 & 0.38 & -0.8 & 0.86 & 1.25 \\ & $CONVOL-$ & -0.01 & 0.75 & 1.06 & 0.76 & 0.75 & 1.19 & 0.91 & 1.22 & 0.10 & 0.36 & -0.87 & 0.85 & 1.29 \\ \cline{2-15} & $[M/H]+$ & 0.00 & 0.77 & 1.03 & 0.79 & 0.72 & 1.19 & 0.92 & 1.23 & 0.07 & 0.36 & -0.82 & 0.91 & 1.30 \\ & $[M/H]-$ & -0.01 & 0.71 & 1.05 & 0.77 & 0.78 & 1.25 & 0.93 & 1.22 & 0.11 & 0.38 & -0.84 & 0.81 & 1.24 \\ \cline{2-15} & $\sigma^2_{[X/Fe]}$ & 0.037 & 0.004 & 0.012 & 0.003 & 0.009 & 0.014 & 0.002 & 0.021 & 0.015 & 0.003 & 0.006 & 0.020 & 0.017 \\
    \\ \toprule \hline    
   \multirow{11}{*}{2M18070782-1517393} & original & 0.10 & 0.69 & 0.97 & 0.54 & 0.61 & 0.89 & 0.60 & 1.04 & 0.02 & 0.46 & -1.03 & 0.62 & 0.69 \\ \cline{2-15} & T$_{eff}+$ & 0.12 & 0.65 & 1.11 & 0.61 & 0.61 & 0.99 & 0.56 & 0.95 & -0.16 & 0.51 & -0.98 & 0.65 & 0.7 \\
   & T$_{eff}-$ & 0.10 & 0.74 & 0.78 & 0.50 & 0.58 & 0.74 & 0.68 & 1.14 & 0.13 & 0.39 & -1.07 & 0.58 & 0.69 \\ \cline{2-15} & $log g+$ & 0.16 & 0.75 & 0.91 & 0.50 & 0.54 & 0.86 & 0.65 & 1.09 & -0.17 & 0.43 & -0.97 & 0.75 & 0.77 \\ & $log g-$ & 0.06 & 0.66 & 1.01 & 0.57 & 0.62 & 0.88 & 0.55 & 0.98 & 0.15 & 0.47 & -1.09 & 0.54 & 0.6 \\ \cline{2-15} & $\mu_{t}+$ & 0.11 & 0.70 & 0.96 & 0.54 & 0.58 & 0.87 & 0.58 & 1.04 & -0.01 & 0.45 & -1.03 & 0.61 & 0.69 \\ & $\mu_{t}-$ & 0.10 & 0.69 & 0.96 & 0.54 & 0.60 & 0.91 & 0.61 & 1.03 & -0.02 & 0.45 & -1.02 & 0.62 & 0.68 \\ \cline{2-15} & $CONVOL+$ & 0.10 & 0.69 & 0.97 & 0.53 & 0.61 & 0.94 & 0.66 & 1.04 & 0.06 & 0.47 & -1.0 & 0.64 & 0.71 \\ & $CONVOL-$ & 0.15 & 0.71 & 0.96 & 0.65 & 0.60 & 0.87 & 0.58 & 1.04 & -0.01 & 0.45 & -1.08 & 0.62 & 0.70 \\ \cline{2-15} & $[M/H]+$ & 0.14 & 0.73 & 0.92 & 0.53 & 0.56 & 0.83 & 0.61 & 1.03 & 0.06 & 0.44 & -1.00 & 0.66 & 0.73 \\ & $[M/H]-$ & 0.08 & 0.63 & 0.99 & 0.53 & 0.62 & 0.92 & 0.59 & 1.04 & 0.01 & 0.45 & -1.05 & 0.57 & 0.64 \\ \cline{2-15} & $\sigma^2_{[X/Fe]}$ & 0.008 & 0.010 & 0.042 & 0.019 & 0.009 & 0.030 & 0.013 & 0.014 & 0.073 & 0.006 & 0.010 & 0.022 & 0.011 \\ \hline
    \end{tabular}
    \label{tab:PlusMinus}
    \tablefoot{For each of the two stars considered, the original abundances are given, alongside the abundances obtained from the  + and $-$ run. The last line corresponds to the total systematic error according to Eq. \ref{eq:totalsysterr}. To estimate the upper limits of Na and Nd of 2M16441013-1850478, we used the Na I $\unit[16388.8]{\angstrom}$ and Nd I $\unit[16262.0]{\angstrom}$ lines, and for the upper limits of Na and S of 2M18070782-1517393, we used the Na I $\unit[16388.8]{\angstrom}$ and S I $\unit[15478.5]{\angstrom}$ lines.}
\end{table*}

\begin{table*}[ht]
    \centering
    \caption{Typical standard deviation $\sigma_{stdev,[X/Fe]}$ and total systematic $\sigma_{[X/Fe]}$ error associated with each element.}
    \begin{tabular}{l c c c} \hline\hline
         & $\sigma_{stdev,[X/Fe]}$ & \multicolumn{2}{c}{$\sigma_{[X/Fe]}$} \\
         & & 2M16441013-1850478 & 2M18070782-1517393 \\
        \hline 
        C & 0.07 & 0.19 & 0.09 \\
        N & 0.05 & 0.06 & 0.10 \\ 
        O & 0.07 & 0.11 & 0.20 \\
        Na & 0.11 & 0.05 & 0.14 \\
        Mg & 0.09 &  0.09 & 0.09 \\ 
        Al & 0.09 & 0.12 & 0.17 \\
        Si & 0.10 & 0.04 & 0.11 \\
        P & 0.12 & 0.14 & 0.12 \\
        S & 0.10 & 0.12 & 0.27 \\
        Ca & 0.07 & 0.05 & 0.08 \\
        Fe & 0.08 & 0.08 & 0.10 \\
        Ce & 0.11 & 0.14 & 0.15 \\
        Nd & 0.07 & 0.04 & 0.10 \\ \hline
    \end{tabular}
    \label{tab:stdev+syst}
    \tablefoot{The values of $\sigma_{[X/Fe]}$ are the square roots of the $\sigma^2_{[X/Fe]}$ given in Table \ref{tab:PlusMinus}.}
\end{table*}

\FloatBarrier
\clearpage

\section{Abundance ratios [X/Fe] versus effective temperature \texorpdfstring{$T_{eff}$}{Teff}\label{Teffmultiplot}}

\begin{figure*}[ht]
\centering
   \includegraphics[width=17cm]{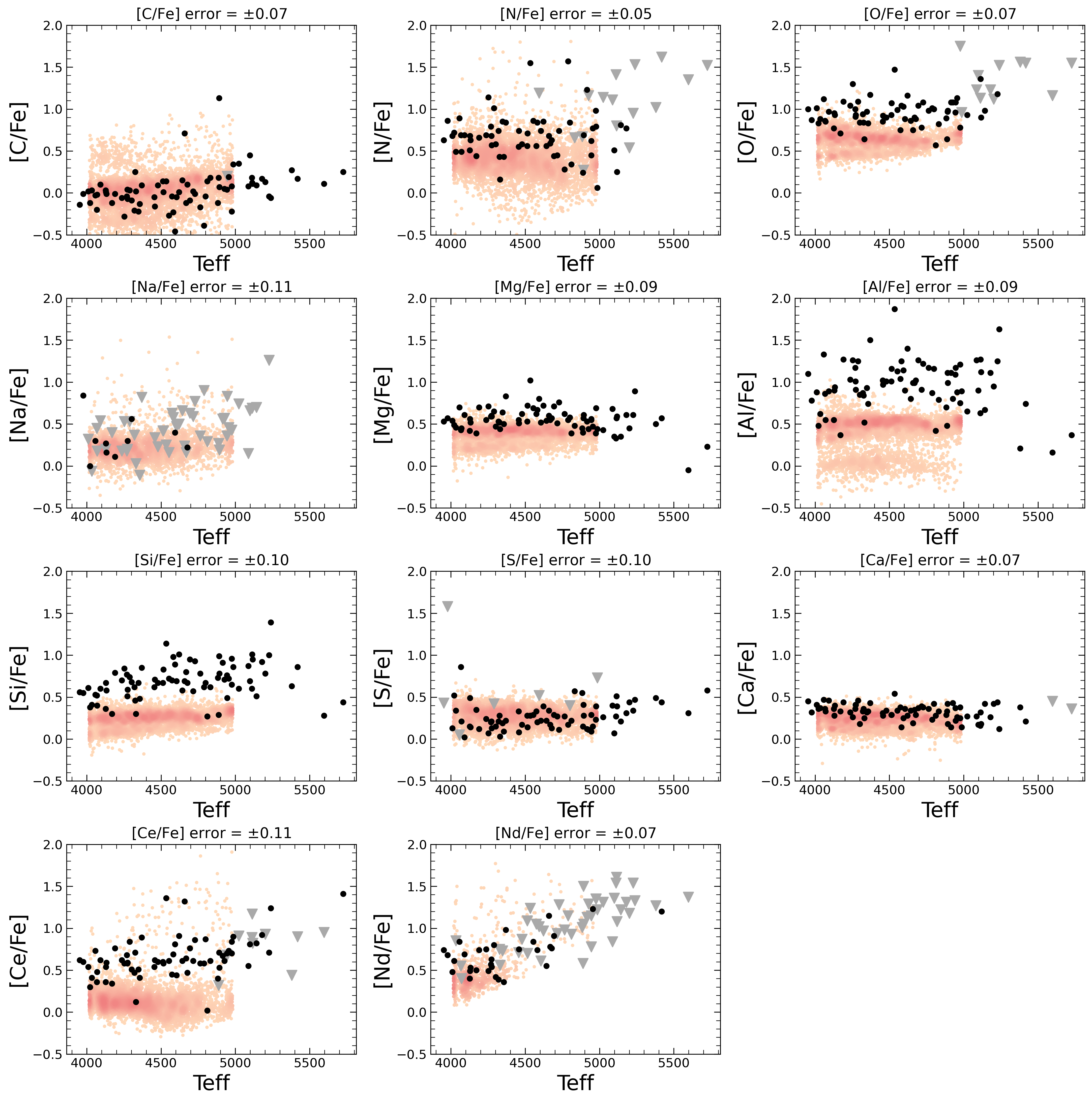}
     \caption{Abundance ratios [X/Fe] vs. effective temperature $T_{eff}$, displaying one element X in each panel. Gray downward triangles indicate upper limits. The background sample was selected and the corresponding abundances were calculated following the procedure described in Sects. \ref{Sample} and \ref{LineSelecProcessing}. In the case of S, the background stars have been selected out of the VAC from \citet{hayes22}. The color-coding reflects the number density of the stars. For clarity, error bars are suppressed. The errors adopted for each element are given in the title of each panel and correspond to the average standard deviation listed in Table \ref{tab:stdev+syst}. Here we assume that the error of the effective temperature is the average ASPCAP $T_{eff}$ error over the 78 sample stars ($\pm \unit[14.32]{K}$).}
     \label{fig:TeffMulti}
\end{figure*}

\FloatBarrier
\clearpage

\section{Zoom into [X/Fe] versus [Fe/H] and correlations\label{ZoomPlots}}

\begin{figure*}[ht]
  \resizebox{\hsize}{!}{
  \includegraphics{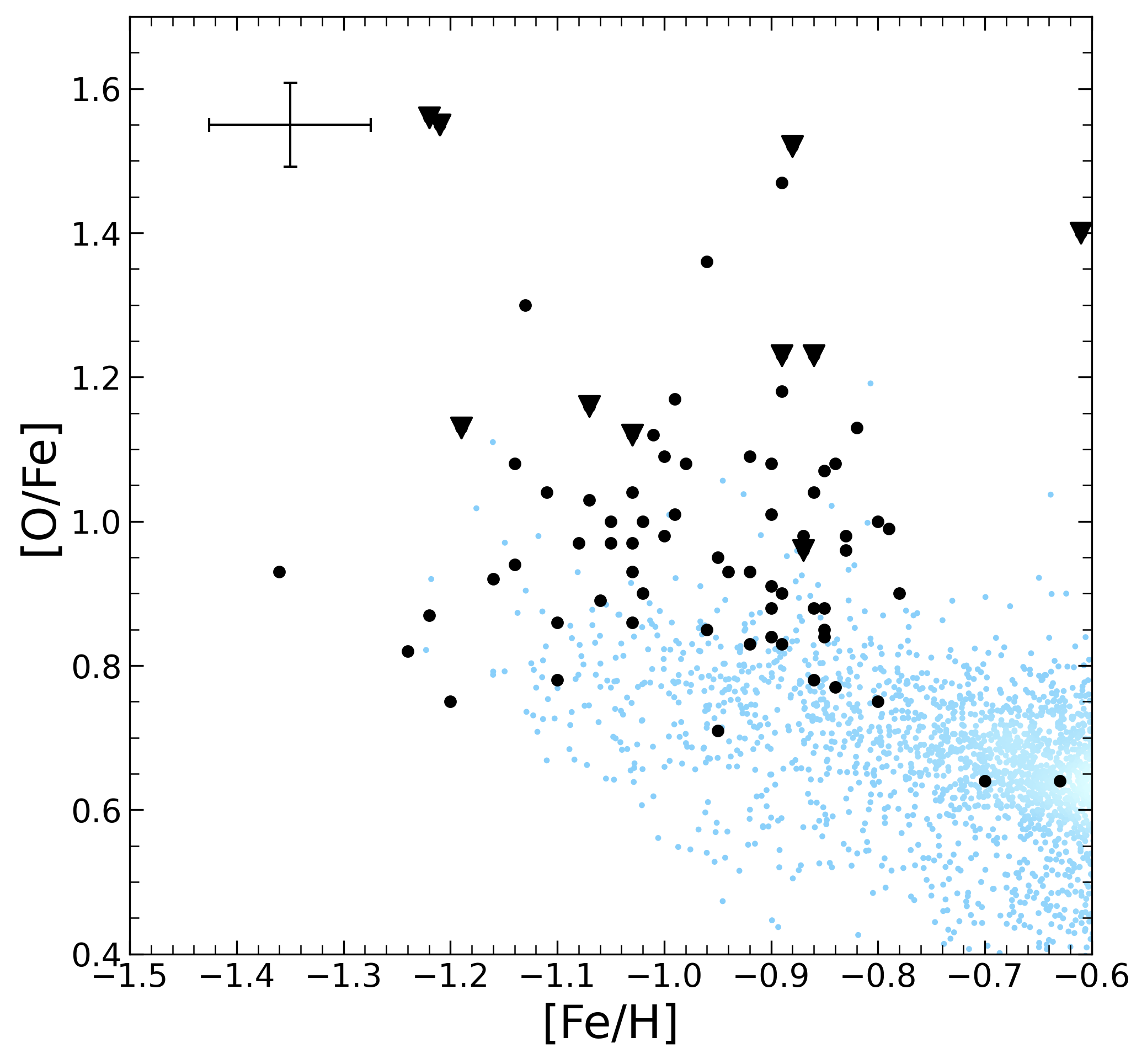}
  \includegraphics{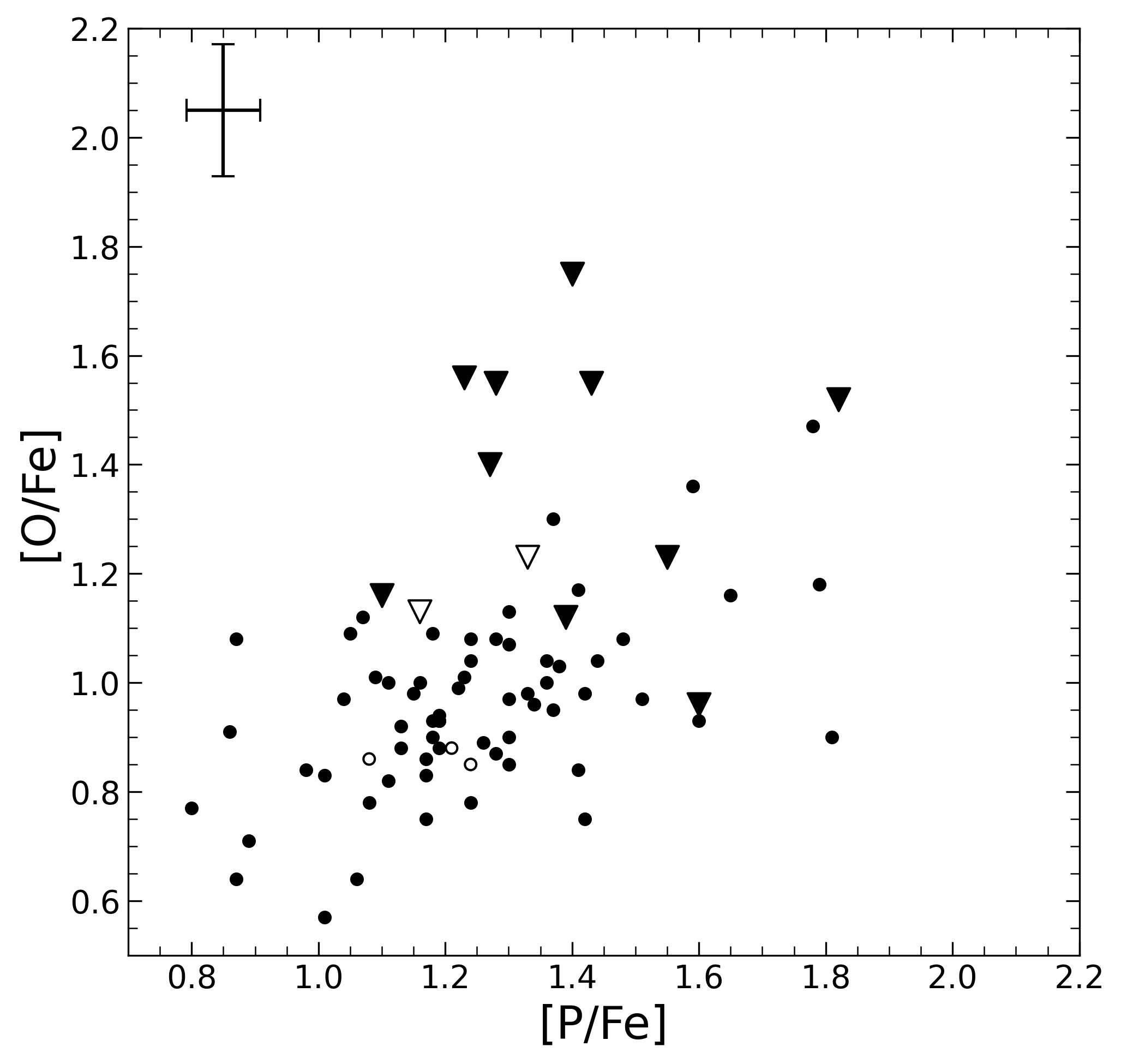}} 
  \caption{Detailed representation of the [O/Fe] abundance ratio. Left panel: Zoom into the [O/Fe] vs. [Fe/H] plot of Fig. \ref{fig:MetalMulti} with upper limits (solid triangles). Right panel: [O/Fe] vs. [P/Fe] to illustrate the correlation between P and O. Solid circles correspond to real measurements in both O and in P, empty circles to upper limits in P, solid triangles to upper limits in O, and empty triangles denote upper limits in O and P. Error bars are defined to reflect the typical standard deviation given in Table \ref{tab:stdev+syst}}
  \label{Fig:O}
\end{figure*}

\begin{figure*}[ht]
  \resizebox{\hsize}{!}{
  \includegraphics{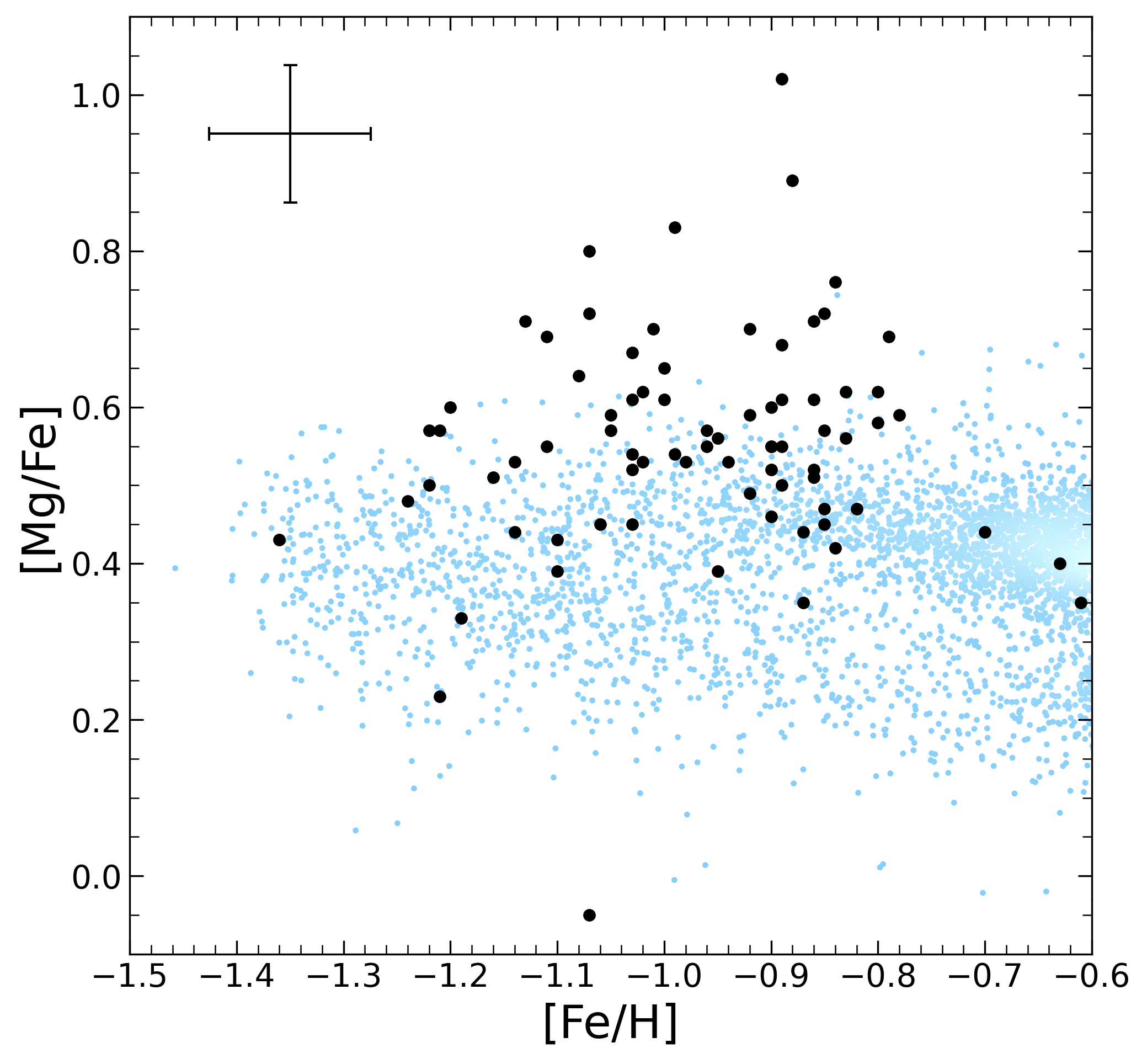}
  \includegraphics{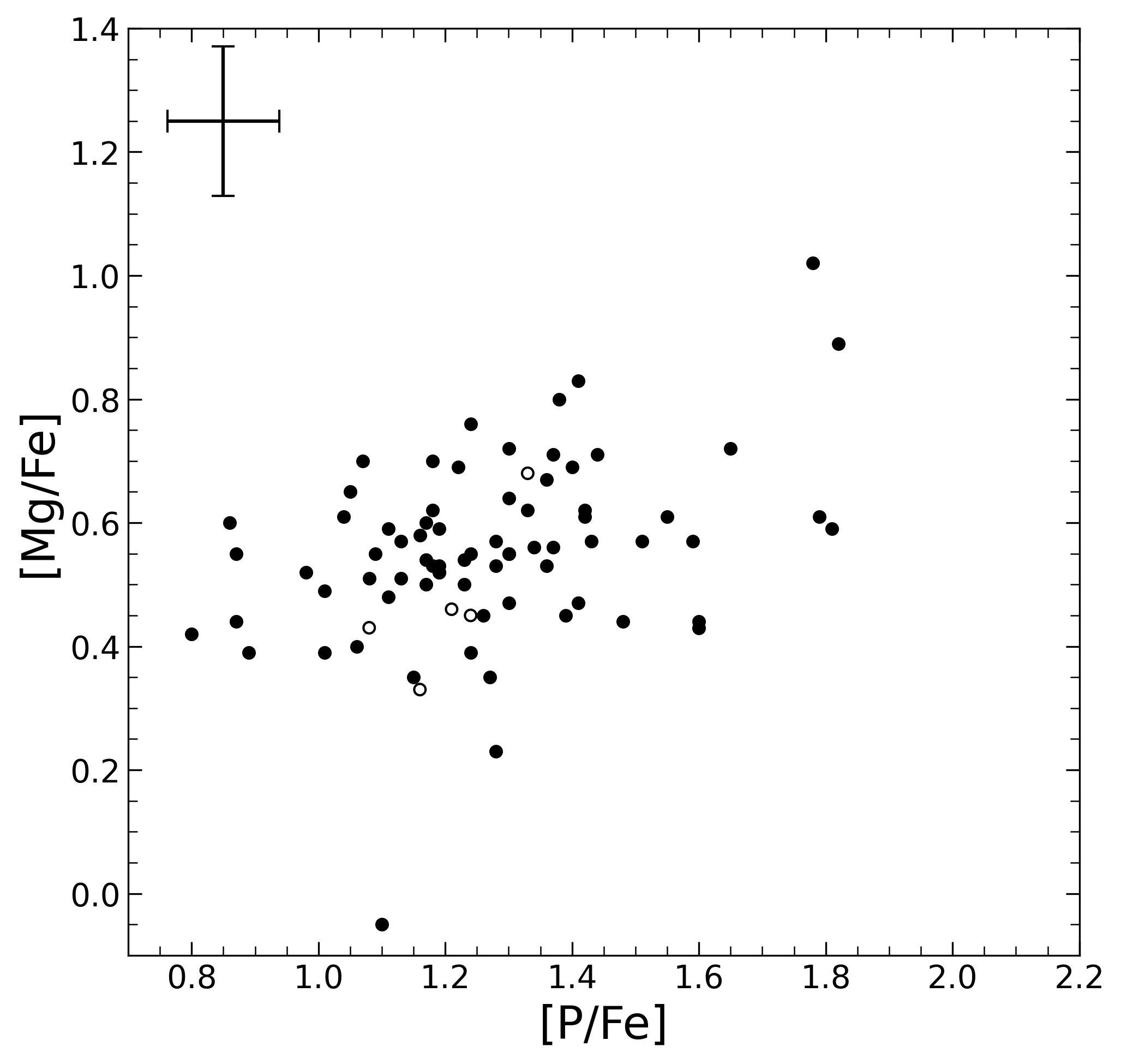}} 
  \caption{Same as Fig. \ref{Fig:O} for Mg.}
  \label{Fig:Mg}
\end{figure*}

\begin{figure*}[ht]
  \resizebox{\hsize}{!}{
  \includegraphics{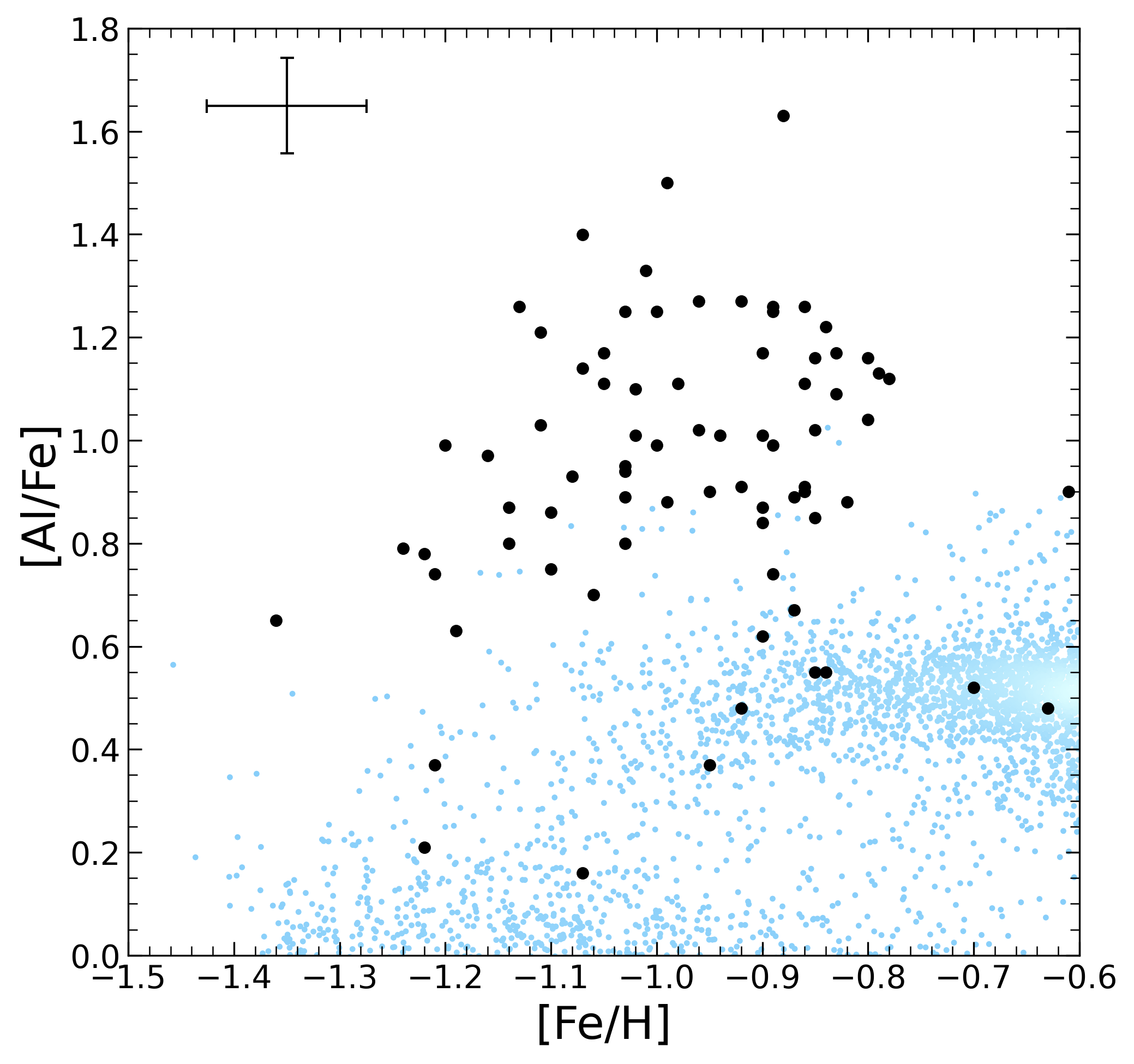}
  \includegraphics{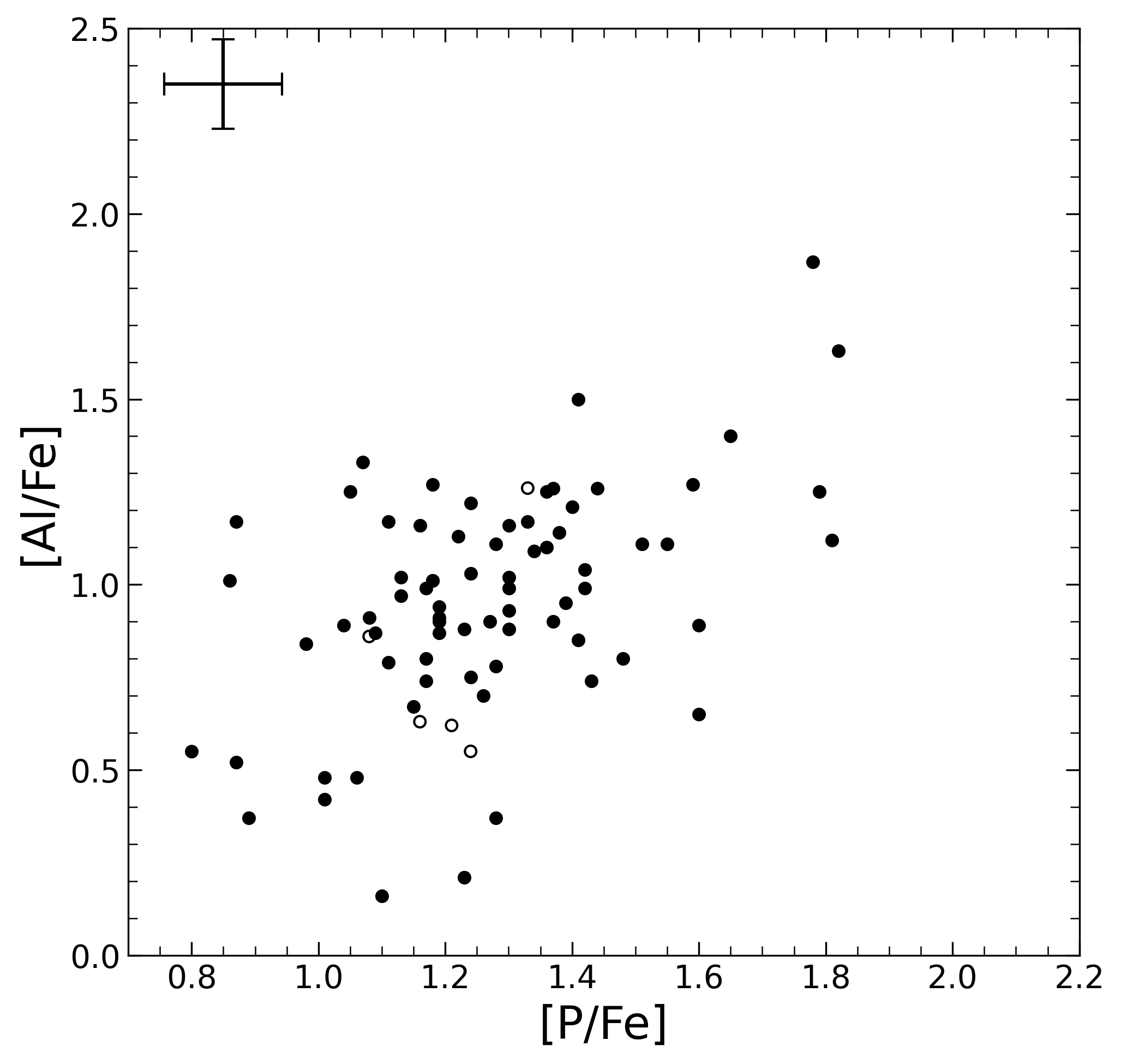}} 
  \caption{Same as Fig. \ref{Fig:O} for Al.}
  \label{Fig:Al}
\end{figure*}

\begin{figure*}[ht]
  \resizebox{\hsize}{!}{
  \includegraphics{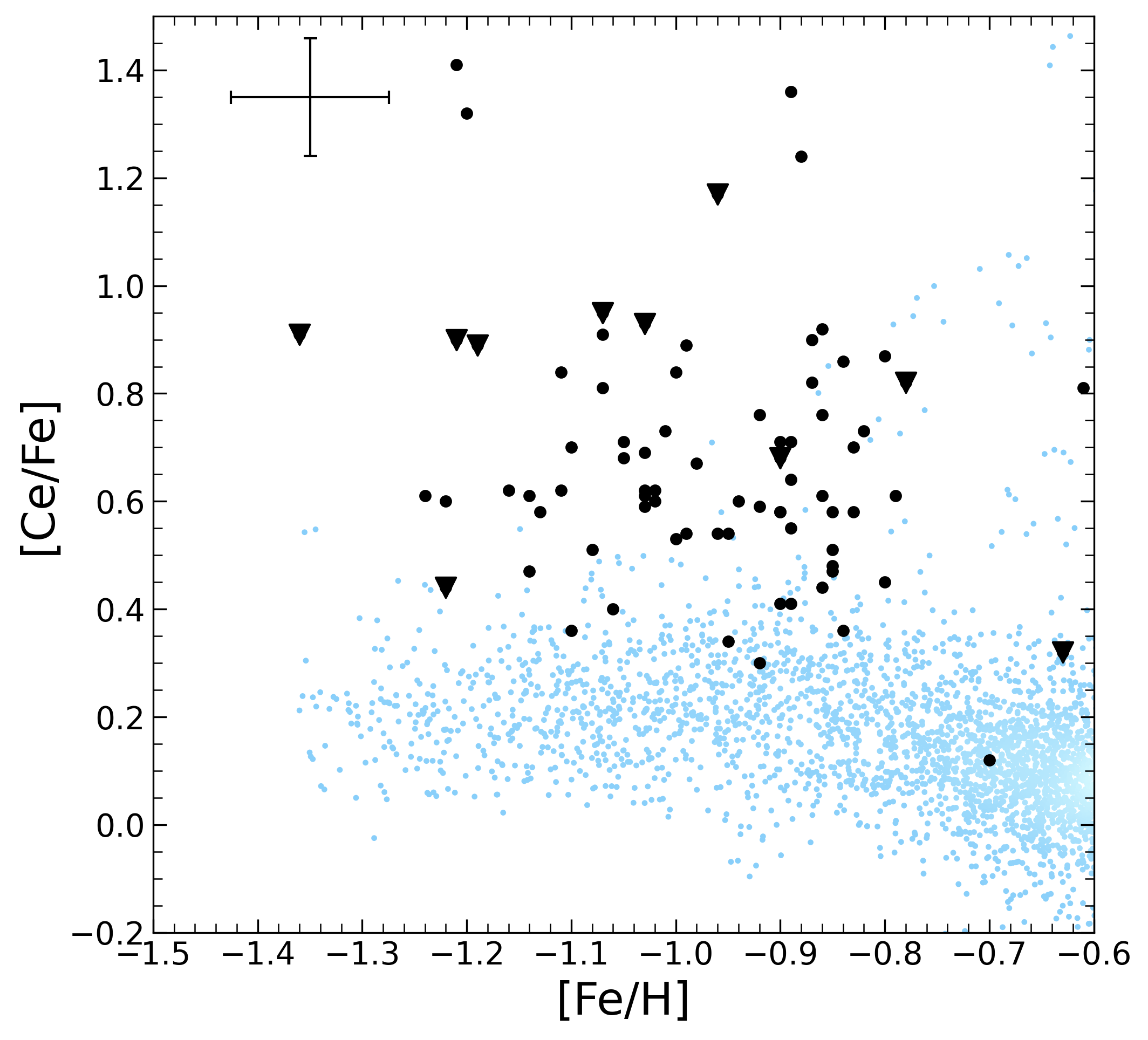}
  \includegraphics{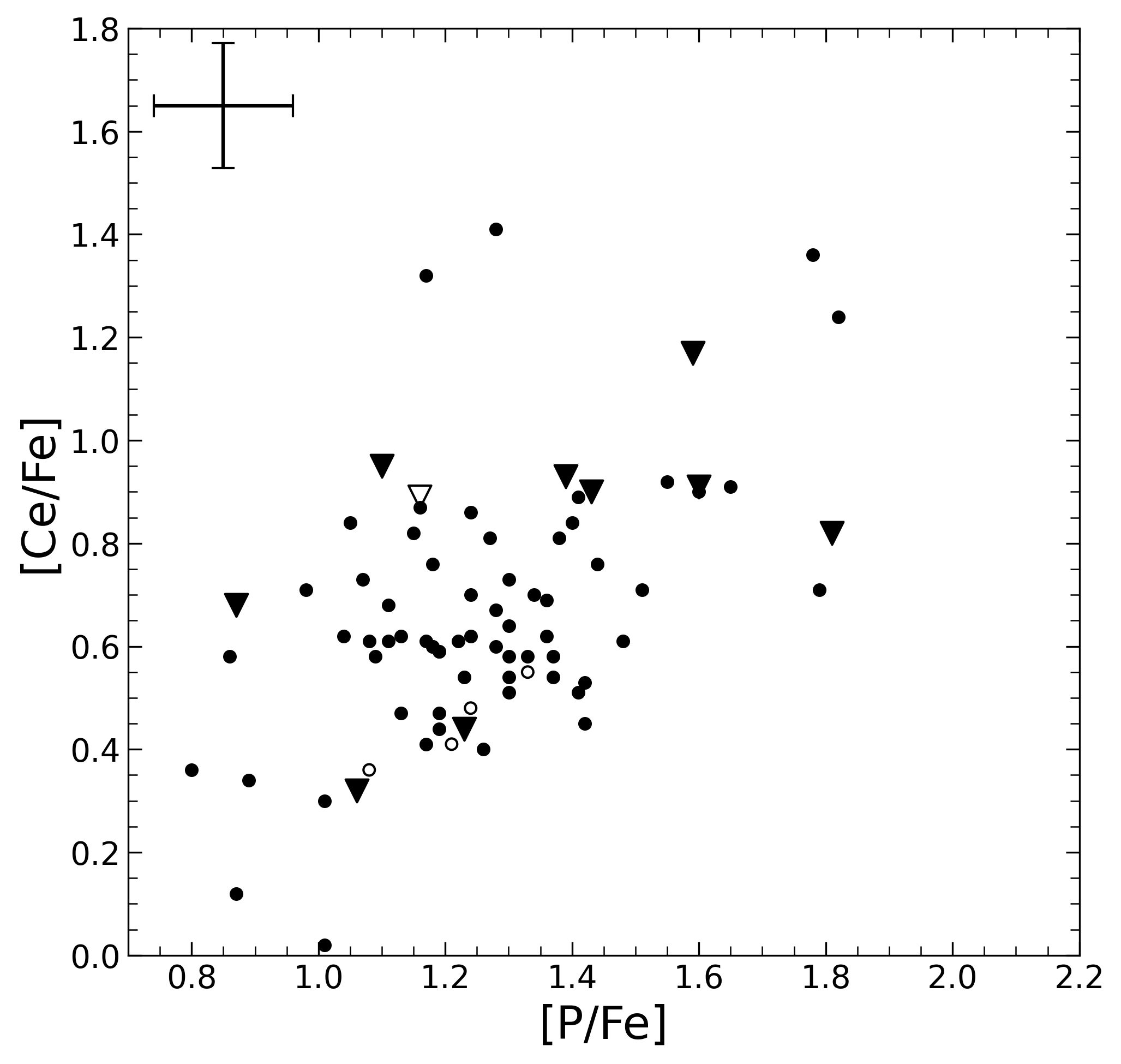}} 
  \caption{Same as Fig. \ref{Fig:O} for Ce.}
  \label{Fig:Ce}
\end{figure*}

\begin{figure*}[ht]
  \resizebox{\hsize}{!}{
  \includegraphics{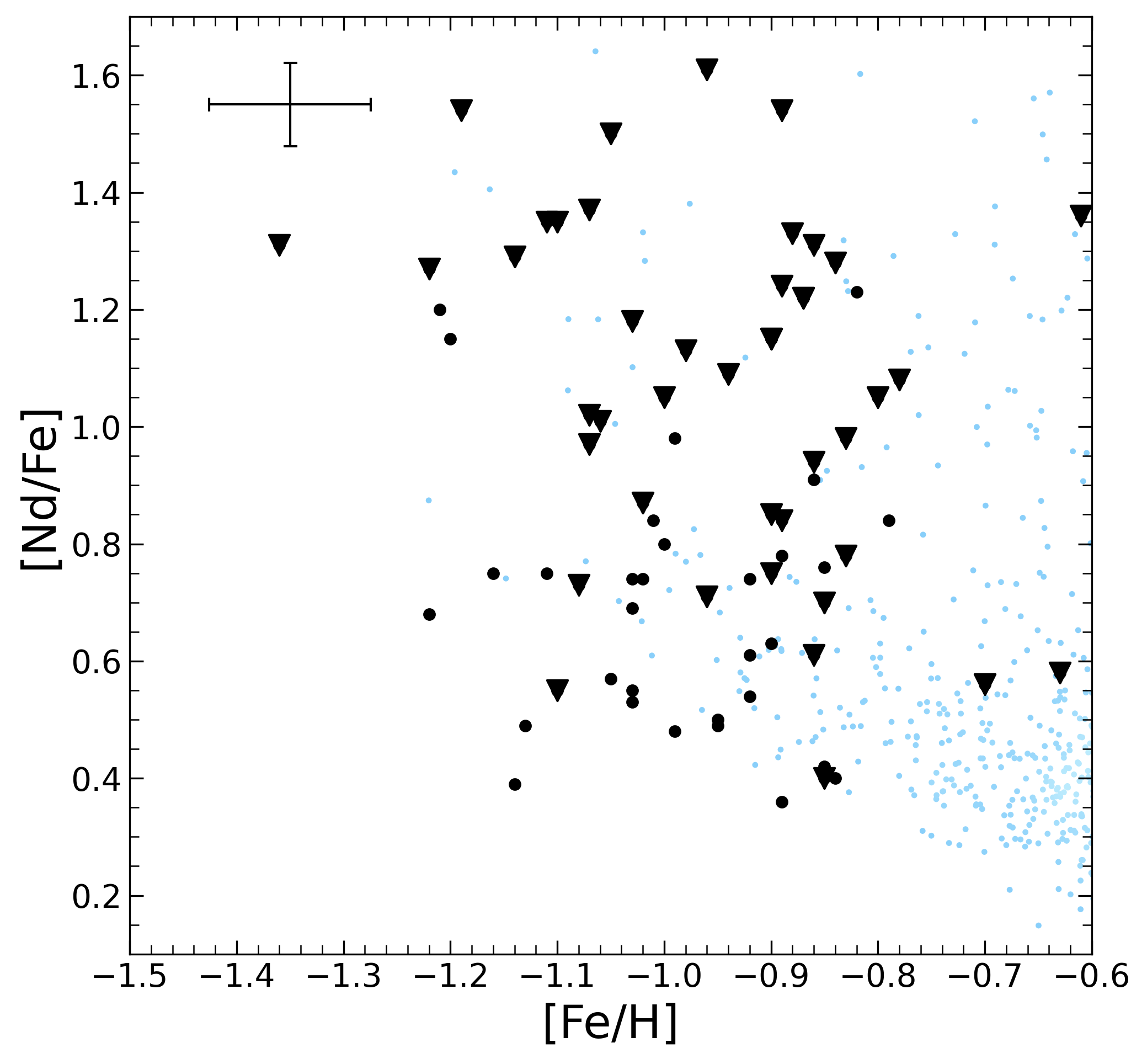}
  \includegraphics{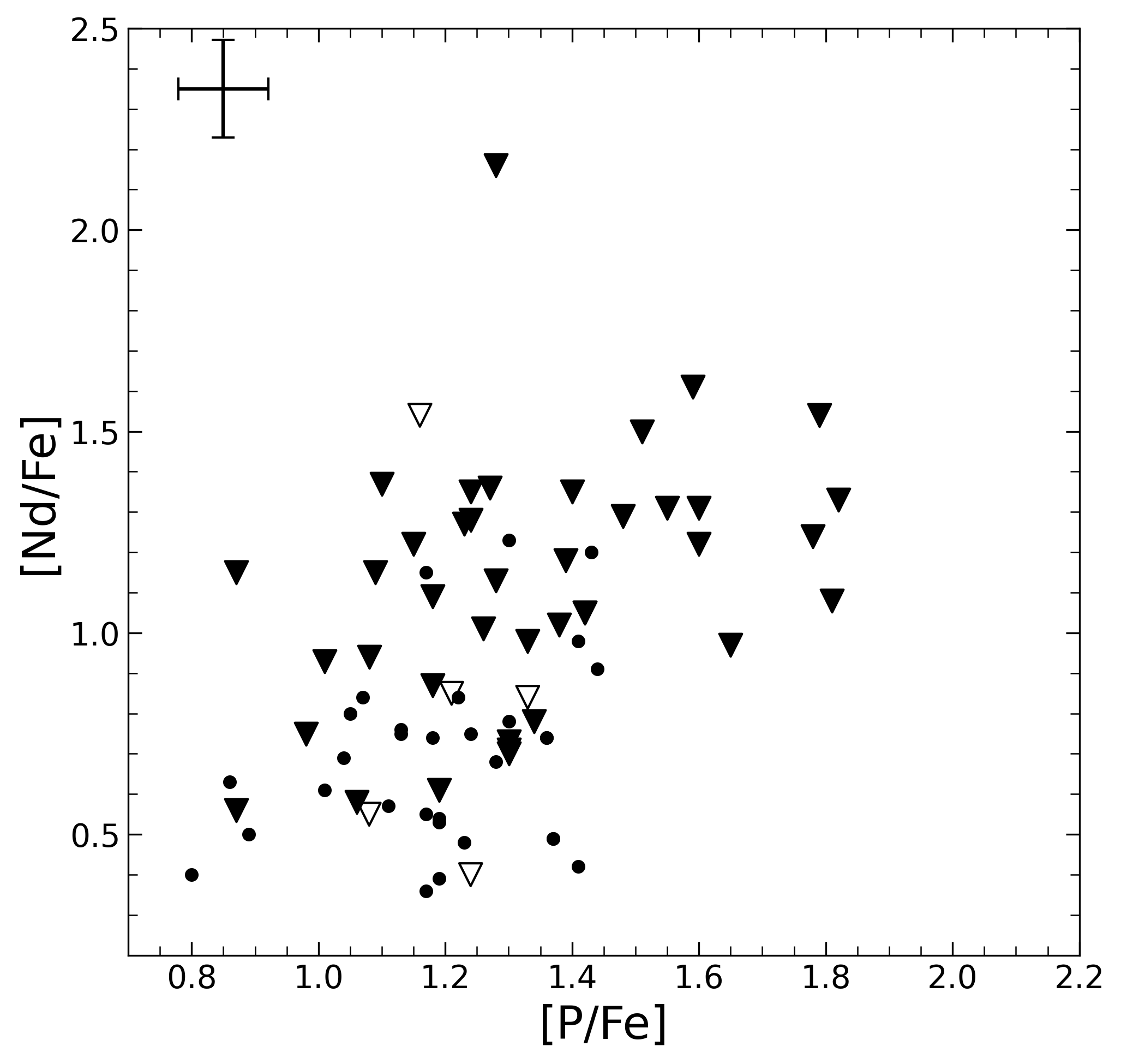}} 
  \caption{Same as Fig. \ref{Fig:O} for Nd.}
  \label{Fig:Nd}
\end{figure*}

\end{appendix}

\end{document}